\documentclass{article}
\usepackage[dvips]{graphicx}
\usepackage{latexsym}
\usepackage[hypertex]{hyperref}

\def\d{{\rm d}}
\def\i{\ifmmode{\rm i}\else\char"10\fi}

\def\e{\mathop{\rm e}\nolimits}

\def\Tr{\mathop{\rm Tr}\nolimits}
\def\Det{\mathop{\rm Det}\nolimits}
\def\usigma{\underline{\sigma}}

\def\kt{k_{\rm B}T}
\def\okappa{\overline{\kappa}}
\renewcommand{\theequation}{\arabic{section}.\arabic{equation}}

\title{Amphiphilic membranes}
\author{Luca Peliti\\
Dipartimento di Scienze Fisiche and Unit\`a INFM\\
Universit\`a ``Federico II'', Mostra d'Oltremare, Pad.~19\\
I-80125 Napoli (Italy)}

\date{January 17, 1995}

\begin{document}

\maketitle
\tableofcontents
\newpage

\section{Introduction}
Amphiphilic molecules (the word derives from the Greek
$\alpha\mu\phi\grave{\iota}$ $\phi\iota\lambda\acute{\iota}\alpha$,
meaning ``love on both sides'') are molecules
which both love and hate water. They are formed
by two parts with very different tastes, which are
covalently bound together: one, the {\em
hydrophilic head}, is polar or even ionized,
and tends therefore to be close to the small,
polar water molecules; the
other, the {\em hydrophobic tail}, is usually
a hydrocarbon chain, which perturbs the high
order of water, and has therefore the tendency
to pack close to similar chains.

Amphiphilic molecules in solvents (water and/or oil)
may form several structures (micelles, hexagonal phases,
cubic phases\dots), but we shall discuss mostly the cases
in which they form a {\em bilayer}, i.e., a sheet
made up of two layers of amphiphilic molecules: in water,
the hydrophilic heads stem out of the bilayer on both
sides, while the corresponding tails remain at the
interior. {\em Membranes\/} can be formed by an isolated
bilayer or by several bilayers stuck one on top of
another: in this case one speaks of {\em multilayers.}

Amphiphilic membranes are a physical realization of fluctuating
surfaces. They are thin sheets (50--100\AA) of amphiphilic molecules
immersed in a fluid, usually water or brine (water and salt).
They can be of natural or artificial origin: the most important
example of natural membranes is the {\em cell membrane},
which separates the interior of all living cells from its
exterior. Living cells, and in particular eucaryotic ones,
possess a large number of membranes, like the nuclear membrane,
which separates the nucleus from the rest of the cell,
allowing, e.g., mRNA to pass from the inside to the
outside, and several chemical signals in the reverse direction,
or the Golgi apparatus, which acts as a sort of ``chemical factory''
for the cell. Mitochondria are also organelles essentially
formed by a membrane, folded on itself several times.

Artificial amphiphilic membranes have recently become
a lively research field stimulated by their applications
in the industry,
in medicine and in cosmetics.
One can form with them tunable or ``active''
filters, simulating, as it were, the action of the
cell membrane. They are also able to close
on themselves, forming vesicles (small closed surfaces),
which may act as drug carriers, designed to open up and
release their load when the ``correct'' physico-chemical
conditions are found. Several other applications can also be envisaged.

The physics of amphiphilic membranes is a wide
subject, and it is out of question to review it fully in this series of
lectures. I shall mostly dwell on the aspects which fit more closely
the scope of the School, namely those involving shape fluctuations.
It will be necessary to review briefly the basic
physical chemistry involved to understand why membranes form
at all, which features govern their shape, and their equilibrium or
dynamical behavior. General introductions to the statistical
mechanics of amphiphilic membranes can be found
in ref.~\cite{Jerusalem}.
A good introduction to the basic
properties of biological membranes is found in \cite[Chap.~12]{Stryer}.
References~\cite{Silver},\cite{Cevc93} contain introductions to the physical
chemistry of amphiphilic molecules.

Therefore, in the next section I shall briefly dwell on the structure
of their basic components, the amphiphilic molecules, and give an overview
of the structures they form in the presence of water and/or oil.
In the following section I shall discuss the free energy of
an isolated fluid membrane as a function of its shape. The
corresponding hamiltonian (due to W. Helfrich) lies at the basis
of the current understanding of vesicle shapes and of membrane
fluctuations. The following section contains a brief review of recent
theoretical and experimental work on vesicle shapes.
Then the effects of fluctuations on amphiphilic membranes
will be discussed: these involve on the one
hand the characteristic flicker phenomenon in
vesicles, and on the other hand the renormalization
of the elastic parameters appearing in the Helfrich hamiltonian.
All these aspects are reflected in the phase
behavior of interacting fluid membranes to which the
last section is dedicated. A few technical
points are discussed in the four Appendices.

\section{Amphiphilic molecules and the phases they form}
\begin{figure}\begin{center}
\includegraphics[width=6cm]{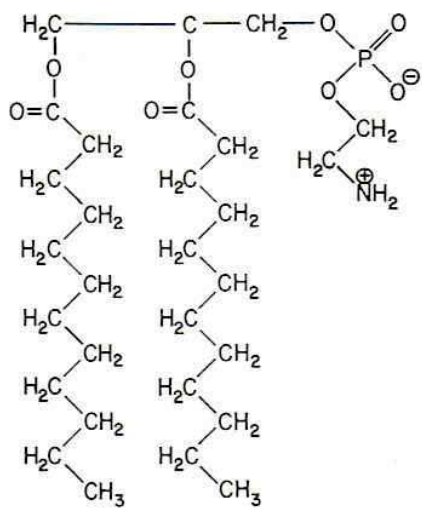}
\end{center}\caption{Chemical structure of
a typical phosphoglycerid:
1,2~dilauryl-DL-phosphatidylethanolamine (DLPE). From
B.~L.~Silver, {\em The physical chemistry of
membranes\/} (Boston: Allen \& Unwin, 1988) p.~2.}
\label{DLPE-fig}
\end{figure}

\setcounter{equation}{0}
Biological membranes are formed by a bilayer
of amphiphilic molecules, the most common of which
are {\em phospholipids}.
Their hydrophilic head
is a phosphate, and their tail is formed
by one or two fatty acids. Most often the
two parts are connected by a glycerine hinge,
and the molecule is therefore called a {\em phosphoglycerid.}
The chemical structure of
a~typical phosphoglycerid is represented in fig.~\ref{DLPE-fig}.

The glycerine molecule which forms the hinge of the structure
is shown
on the top. On one side, it is ester-linked
to a phosphate group \hbox{--POOH--O--R} carrying the
ethanolamine residue \hbox{R=--CH$_2$--CH$_2$--NH$_2$}.
In physiological conditions the head is almost
always ionized, yielding \hbox{--PO$^-$--CH$_2$--CH$_2$--NH$_3^+$.}
Via the other two carbons (1 and 2) the glycerine is ester-linked
to two fatty acid chains, of the general structure
\hbox{CH$_3$--(CH$_2$)$_n$--COOH.} In our case
one has $n=10$, and the acid is called
acid $n$-dodecanoic, or lauric acid.
The example helps to clarify the terminology.
Other phosphoglycerids may differ from DLPE
either by having just one fatty acid chain ({\em
monoglycerids\/}), or by the nature of the fatty
acid tail. For example, if the
two chains have $n=14$ (palmitic acid)
one has {\em dipalmitoylphosphatidylethanolamine,}
(DPPE). If we change now
the residue R to cho\-line R$_1=$--CH$_2$--CH$_2$--N$^+$--(CH$_3$)$_3$,
we have {\em dipalmi\-toyl\-phosphatidylcholine\/} (DPPC),
also called lecitin, which is an important
component of natural membranes.

\begin{figure}\begin{center}
\includegraphics[width=8cm]{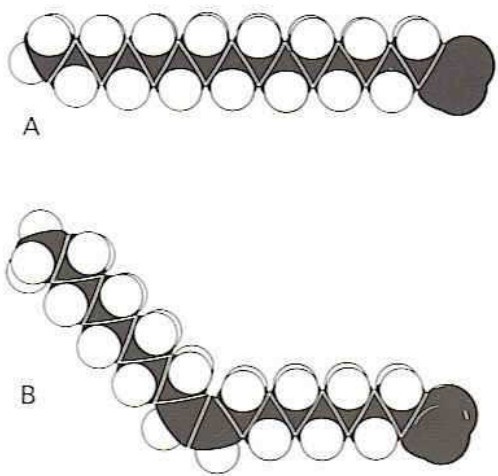}
\end{center}\caption{Space filling models of (A)
palmitate (C$_{16}$, saturated) and (B) oleate
(C$_{18}$, unsaturated). The {\em cis\/} double
bond in oleate produces a bend in the
hydrocarbon chain. From L.~Stryer, {\em Biochemistry\/}
(New York: Freeman, 1988) p.~285.}\label{unsat-fig}
\end{figure}

In general, the two fatty acids of natural
phosphoglycerids are different, and one of them
is quite often unsaturated. The knee which appears
because of the insaturation, like that shown in
fig.~\ref{unsat-fig},
helps in keeping the fluidity of the membrane.

\begin{figure}\begin{center}
\includegraphics[width=8cm]{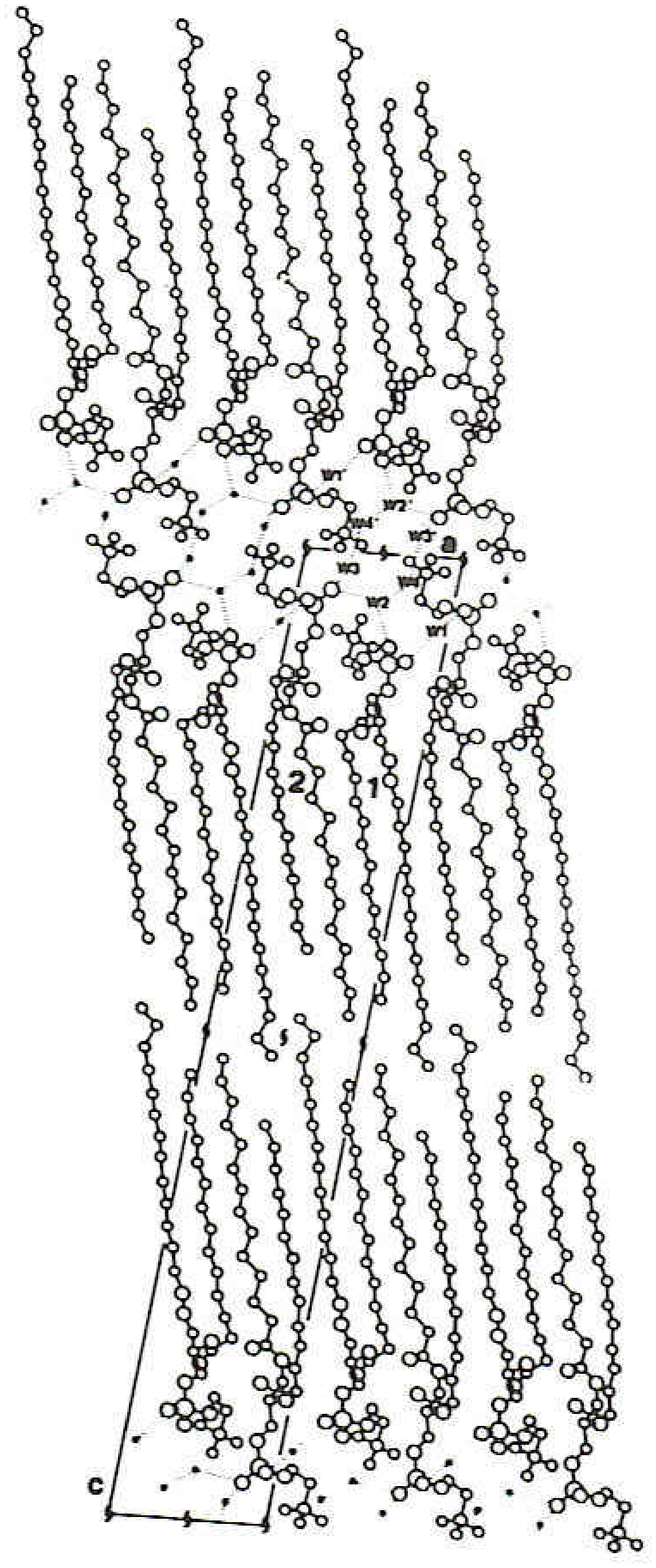}
\end{center}\caption{Crystal structure of DMPC projected
onto the crystallographic {\em a-c\/} plane. From R.~H.~Pearson,
I.~Paschen, {\em Nature\/} {\bf 281} 499 (1979).}
\label{crystal-fig}
\end{figure}

\begin{figure}\begin{center}
\includegraphics[width=8cm]{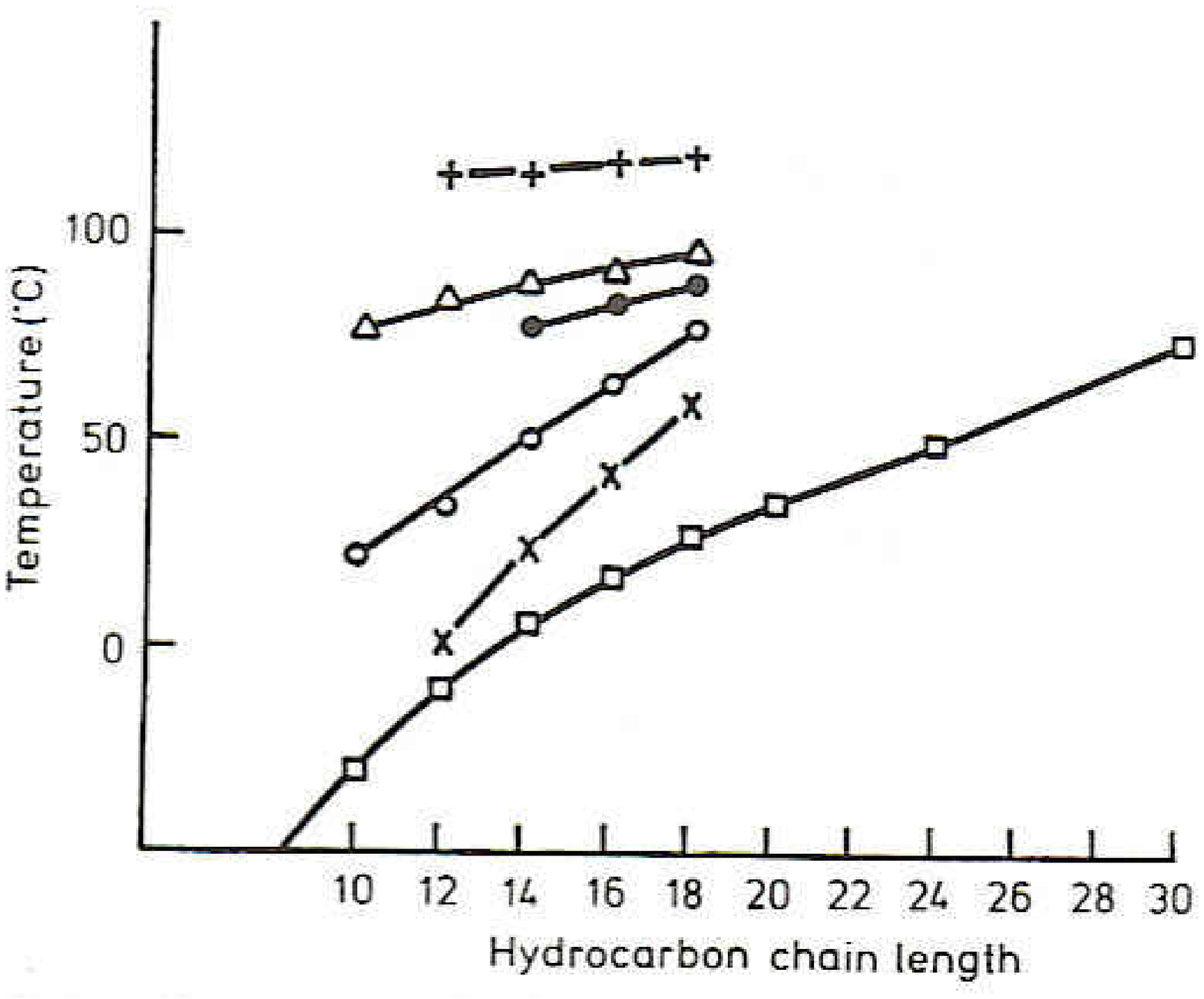}
\end{center}\caption{Chain melting temperatures $T_{\rm t}$ for
some amphiphiles as a function of chain length.
$+$ anhydrous 1,2-diacyl-DL-phosphatidylethanolamines;
$\bullet$, 1,2-diacyl-DL-phosphatidylethanolamines in water;
$\triangle$, anhydrous 1,2-diacyl-L-phosphatidylcholine
monohydrates;
$\circ$, 1,2-diacyl-L-phosphatidylcholine
monohydrates;
$\times$, 1,2-diacyl-L-phosphatidylcholine
monohydrates in water.
For comparison, the melting points of some normal
paraffines, $\Box$, are given. From D.~Chapman, et al.,
{\em Chem.~Phys.~Lipids\/} {\bf 1} 445 (1967).}
\label{trtemp-fig}
\end{figure}

\begin{figure}\begin{center}
\includegraphics[width=8cm]{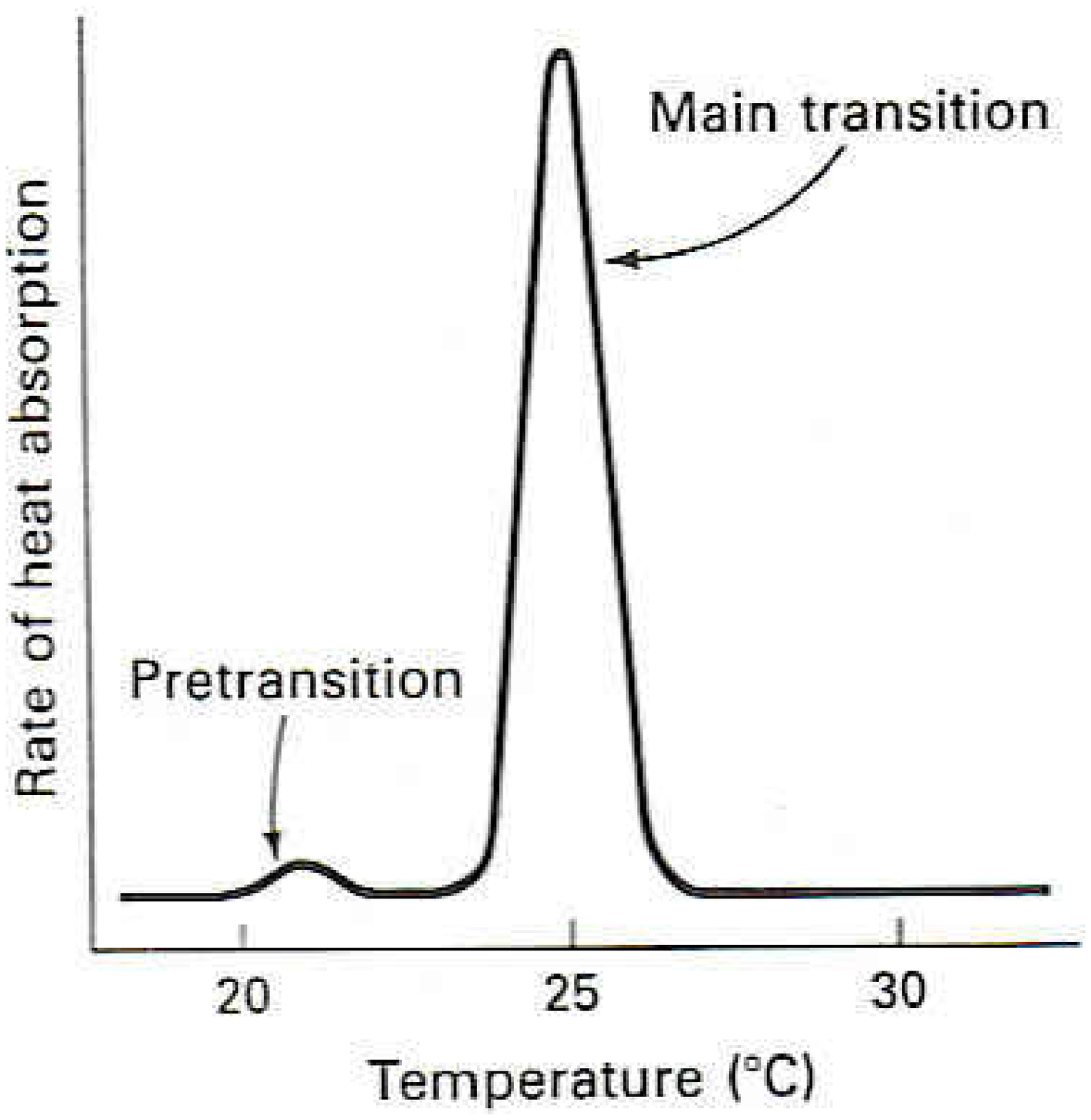}
\end{center}\caption{Differential scanning calorimetry of phosphatidyl choline.
The small peak (the pretransition) comes from a change in the
tilt of fatty acyl chains with respect to the bilayer.
The major peak arises from the phase
transition in which crystalline fatty
acyl chains become disordered because
of the introduction of kinks. From L.~Stryer, {\em Biochemistry\/}
3rd ed.\ (New York: Freeman, 1988) p.~297.}
\label{trans-fig}
\end{figure}

Pure phospholipids may be made to crystallize, and the crystal
structure of several of them have been determined by
X-ray or electron diffraction. In the crystalline state
they are stuck in bilayers, their tails all {\em trans},
with their heads folded approximately parallel to the
bilayer surface and linked (mostly by hydrogen bonds)
into a firm network. This organization
is schematically represented in fig.~\ref{crystal-fig}.
This order is disrupted, as the
temperature increases, via a two-step process:
\begin{figure}\begin{center}
\includegraphics[width=8cm]{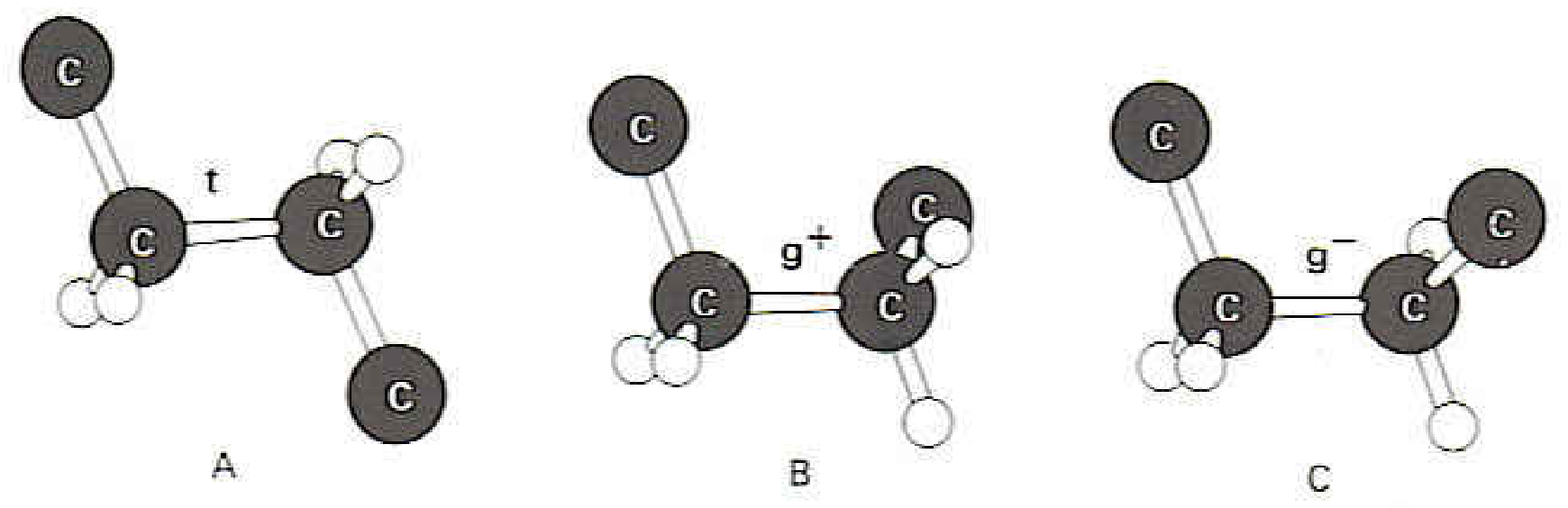}
\end{center}\caption{Conformation of \hbox{C--C} bonds in fatty
acyl chains: (A) {\em trans\/} (t) conformation;
(B and C) a 120$^{\circ}$ rotation yields a {\em gauche\/}
(g) conformation, which can be either g$^+$ (clockwise
rotation) or g$^-$ (counterclockwise rotation).
{}From L.~Stryer, {\em Biochemistry\/} (New York: Freeman, 1988) p.~297.}
\label{gauche-fig}
\end{figure}
\begin{itemize}
\item At a temperature $T_{\rm t}$ called the {\em
transition temperature\/} the order of the chains breaks down:
i.e., a significant
fraction of the carbon links goes over to the {\em gauche\/}
configurations (cf.~fig~\ref{gauche-fig}),
providing a gain in entropy against
a loss in van der Waals attraction
among the chains. This is sometimes called
``chain melting'' or ``premelting''.
\item The actual melting temperature $T_{\rm m}$ is reached
when the ionic lattice formed by the hydrophilic heads
eventually breaks down.
\end{itemize}
The melting temperature $T_{\rm m}$ of most
pure phosphoglycerids
is comparatively high ($\sim 200^{\circ}{\rm C}$),
and does not depend strongly on the length of
the fatty acid tail.
On the other hand the transition temperature is closer to
room temperature and increases with the hydrocarbon
chain length: a plot is shown in fig.~\ref{trtemp-fig}.
We expect therefore, in most phospholipids, to find
a phase in which a relatively fluid hydrocarbon
core is sandwiched between two relatively rigid
polar sheets. This feature makes possible to
sustain fluid bilayers in an aqueous medium.
In isolated bilayers in a solvent one observes
a sharp anomaly of the specific heat at a transition
temperature $T_{\rm t}$ somewhat lower than
that observed for pure anhydrous phospholipids (see fig.~\ref{trans-fig}).
This anomaly is accompanied by a fast variation
of the mechanical properties of
the bilayer, and is known as the ``main transition''.

The idea to keep in mind is that amphiphilic
molecules are complex ones, with a number of
internal degrees of freedom in the tails. The transition
takes place essentially in their tails. The disorder in the tails
eventually induces a liquid-like disorder in the
location of the molecules on the sheet.
This is quite different from a melting
transition in a sheet of bead-like molecules.
The coupling between crystallization
and chain ordering has not yet been
satisfactorily treated~\cite{Bloom-QRB}.
Its relevance can be grasped from the observation
that most biological membranes appear to work
very close, but slightly above the
transition temperature. Some suggested
explanations of this fact can be found in~\cite{Bloom-QRB}.

\begin{figure}\begin{center}
\includegraphics[width=8cm]{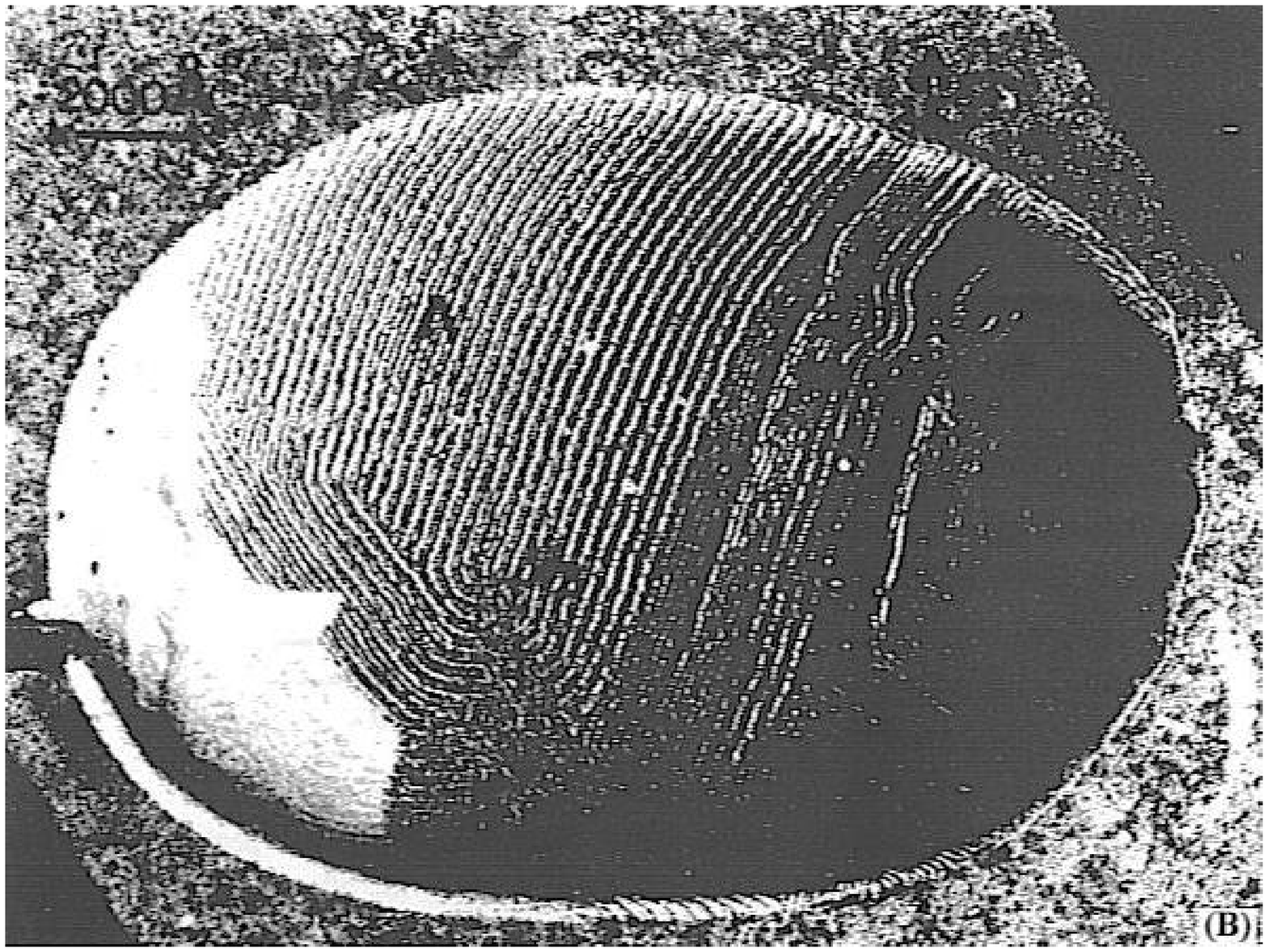}
\end{center}\caption{Electron microscope photograph
of ``rippled'' phases in DPPC, produced by cooling large
vesicles from the P$_{\beta'}$ phase. From E.~Sackmann,
et al., {\em Liquid crystals of one- and two-dimensional
order,} (Berlin: Springer, 1980) p.~314.}\label{pbeta}
\end{figure}

As soon as water enters the picture,
it makes it much more complicated. With a low
water content, the system maintains by and large
the lamellar organization characteristic of the
pure crystal. The layers can
exhibit a number of different phases, which
have different stability domains. For example
in a mixture of dimyristoylphosphatidylcholine (DMPC)
and water, one can find at least three phases
at high DMPC concentration \cite{DMPC-old,DMPC-new,DMPC-2}.
They are called
the L$_{\alpha}$, L$_{\beta}$ and P$_{\beta}$ phases.
In the L$_{\alpha}$ phase the bilayers are fluid and
flat on average. If one lowers the temperature $T$,
or decreases the water content $\phi$, one goes to an
ordered, ``solid like'' phase L$_{\beta}$, in which
the hydrocarbon chains are ordered and the molecules
do not diffuse freely. In these phases, the order
of the hydrocarbon chains implies a larger
thickness of the bilayers. One can also observe an intermediate
``rippled'' phase P$_{\beta}$ (see fig.~\ref{pbeta}),
 in which the bilayers
exhibit an undulated structure and almost solid-like
diffusion properties~\cite{DMPC-old}. Analysis
of X-ray experiments on these ``rippled'' phases
strongly suggests that they are characterized
by a modulation of the bilayer
thickness \cite{X-ray}. The hydrocarbon chains are often tilted
with respect to the bilayer: one then denotes the
phases as L$_{\beta'}$ or P$_{\beta'}$. In fact,
there are several different L$_{\beta'}$ phases,
distinguished by the relationship between the tilt
and in-plane bond orientational order \cite{lbeta}.

\begin{figure}\begin{center}
\includegraphics[width=8cm]{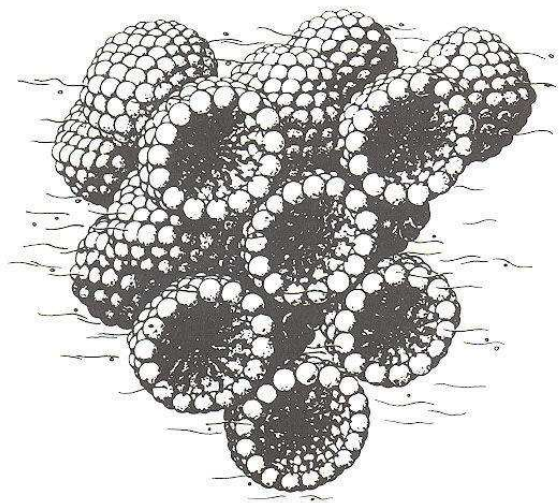}
\end{center}\caption{Scheme of micellar structure. The spheres indicate
the head groups and the wiggly lines the lipid
chains of phospholipids. From  C.~L.~Khetrapal, et al.,
{\em Lyotropic liquid crystals} (Berlin: Springer, 1975).}
\label{micelle-fig}
\end{figure}

\begin{figure}\begin{center}
\includegraphics[width=8cm]{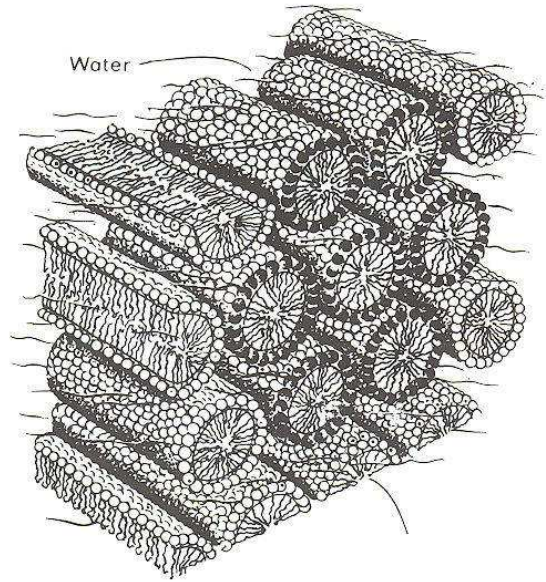}
\end{center}\caption{Scheme of hexagonal close-packed structure.
{}From C.~L.~Khetrapal, et al., {\em Lyotropic liquid crystals}
 (Berlin: Springer, 1975).}\label{hex-fig}
\end{figure}

The ``rippled'' P$_{\beta}$ phases may be considered as
an intermediary structure between the lamellar ones found
at low water concentration and the ones
found at low amphiphile concentration. If we add amphiphilic
molecules to pure water, the molecules first
go preferentially to the air-water interface,
forming a monolayer, with their heads toward
water as long as the concentration does
not exceed the {\em critical
micelle concentration\/} (cmc), which is of the
order of $10^{-10}\,{\rm mol}$. Below the
cmc the amphiphilic molecules are overwhelmingly in the
monomer form, at higher concentration added monomers
appear almost exclusively in aggregates, mainly of globular
form with the hydrophilic heads on the surface.
These aggregates are called {\em micelles}. They form more
readily for single-chain amphiphiles (e.g., monoglycerids)
and are favored by the presence of large head groups.
The structure of micelles is depicted in fig.~\ref{micelle-fig}.
In this figure, the concentration of amphiphiles is high
enough to let the micelles arrange in a close-packed bcc
lattice: this is the simplest example of the remarkable {\em cubic
phases\/} formed by amphiphiles.
As the amphiphile concentration is increased, one observes
the appearance of nonspherical micelles, and eventually
of cylindrical rods. These rods behave as ``living polymers''
at low concentrations, and organize as a close packed,
hexagonal phase, like that shown in fig.~\ref{hex-fig},
at higher concentration.

\begin{figure}\begin{center}
\includegraphics[width=10cm]{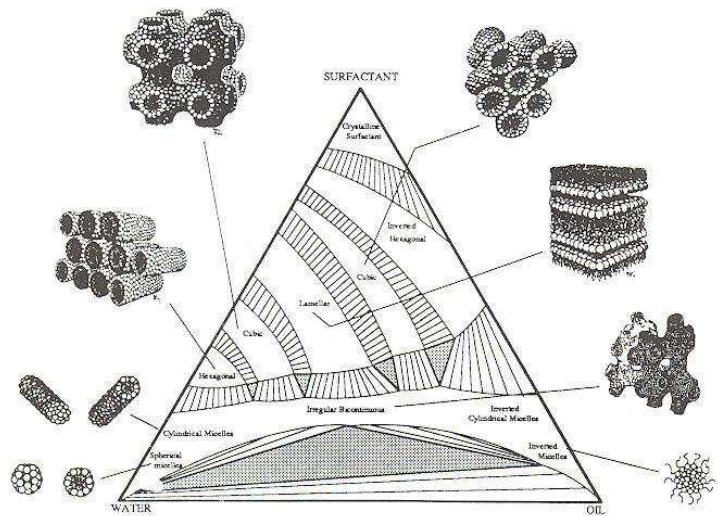}
\end{center}\caption{Schematic oil-water-surfactant phase diagram
with microstructures depicted. From H.~T.~Davis et al.,
in J.~Meunier, D.~Langevin, N.~Boccara (eds.),
{\em  Physics
of amphiphilic layers\/} (Berlin: Springer, 1987) p.~311.}
\label{phasediag-fig}
\end{figure}

If we add paraffine oil, close in composition to the
hydrocarbon tails of phospholipids, we may stabilize some
new phases. A schematic picture of the resulting phase
diagram is found in ref.~\cite{phase-diag} (fig.~\ref{phasediag-fig}).
The new phenomena are determined by the fact that,
if the oil content is large enough, amphiphilic
molecules can form a {\em monolayer,}
with their tails towards the oil-rich phase
and their heads towards the water-rich one.

This opens the way to {\em bicountinuous\/} phases,
in which both the oil-rich and the water-rich phases
percolate. Some of these phases have a periodic, cubic
structure, similar to that represented by the cartoon
in fig.~\ref{phasediag-fig}: the actual structures are
however more complex and difficult to draw. They are
collectively known as ''plumber's nightmare'' phases,
since they may be considered as pipeworks in which
the interior and the exterior look the same. Of great
interest, both theoretical and experimental, are
the {\em irregular bicontinuous\/} phases. As shown
from the phase diagram, they can be obtained by adding
oil to a water-amphiphile solution (with a high enough
amphiphile content), without crossing any phase barrier.
Moreover they can coexist with the water-rich and the oil-rich
phases, having a water-oil ratio close to one. In this coexistence
they will remain between the water and oil phases because
of their density, and are known therefore as middle-phase
microemulsions. The name microemulsions intimates that
they are formed of almost equal proportions of oil and water,
like ordinary emulsions: however, whereas emulsions are
{\em nonequilibrium} structures obtained by suspending droplets,
say, of oil in water by means of intense mixing, microemulsions
are {\em equilibrium phases}. In nonequilibrium emulsions, the droplet
size ranges from the micrometer to a fraction of millimeter,
whereas in microemulsions one observes irregularities in the local
composition at scales of the order of a few hundred \AA ngstr\"oms,
much smaller than the wavelength of visible light. As a consequence,
microemulsions usually appear transparent.

There is also the rather paradoxical possibility
that the bilayer forms an interface separating
two percolating domains occupied
by the {\em same\/} solvent. This is known
as the {\em sponge phase\/}~\cite{sponge}.

The structure of these phases can be studied by the ``traditional''
means of light or X-ray diffraction (depending on the
characteristic size of the structures), but can be also
directly exhibited by freeze fracture.
In this technique the sample is rapidly frozen to the
temperature of liquid nitrogen. The frozen
sample is then fractured by means of a microtome knife. Cleavage
usually occurs in the middle of bilayers. The exposed
regions can then be shadowed with carbon or platinum,
which produces a replica of the interior of the bilayer.
In order to expose the exterior of the membrane,
one can combine freeze fracture with etching. First, the
interior of a frozen membrane is exposed by fracturing; then, the ice
that covers one of the adjacent membrane surfaces is
sublimed away: this process is called
deep-etching. The combined technique, called
{\em freeze-etching electron microscopy,} provides a view
of the interior of a membrane and of both its surfaces.
The structure of emulsion and microemulsions can
also be exhibited in this way.
\begin{figure}\begin{center}
\includegraphics[width=6cm]{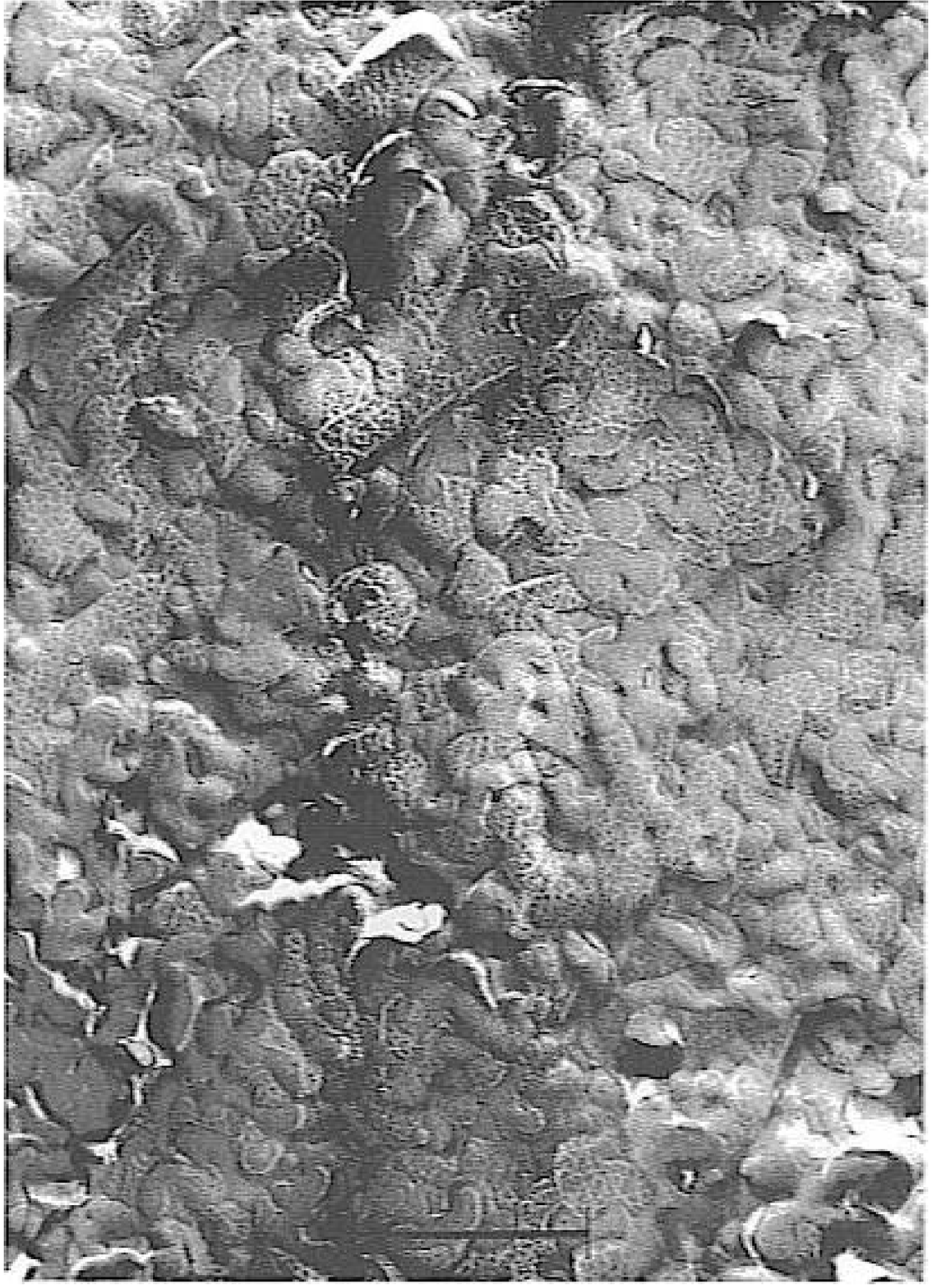}
\end{center}\caption{Freeze-fracture of a bicontinuous microemulsion
containing equal volumes of water
and oil. The shadow material decorates
the oil-rich parts of the fracture face. Bar: 500 nm. From
W.~Jahn, R.~Strey, in J.~Meunier, D.~Langevin, N.~Boccara (eds.),
{\em Physics of amphiphilic layers,} (Berlin: Springer, 1987)
p.~354.}\label{micro-fig}
\end{figure}
\begin{figure}\begin{center}
\includegraphics[width=6cm]{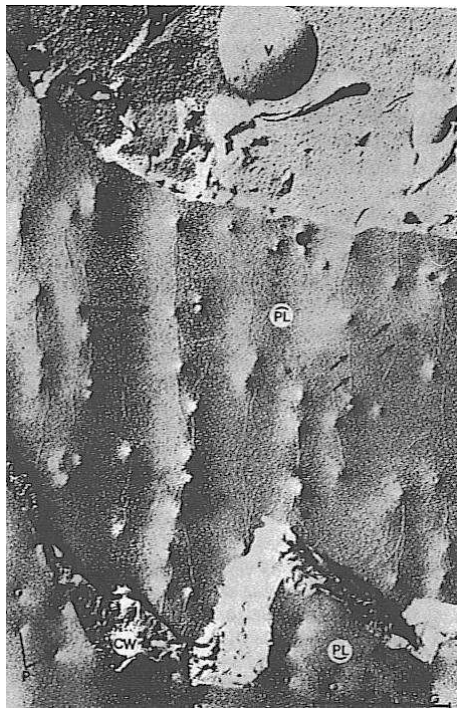}
\end{center}\caption{Freeze-fracture of a cell membrane from an onion root tip.
The uppermost area in the photograph is the cell cytoplasm with a vacuole (V)
evident. The center area shows the fracture face of the cell
membrane (plasmalemma, PL).
The fracture has also ruptured the cell wall (CW).
Plasmodesmata (P), fine threads connecting
the cytoplasm of adjacent cells, have also
been broken off. From D.~Branton, D.~W.~Deamer, {\em Membrane structure\/}
(Berlin: Springer, 1972).}
\label{memb-fig}
\end{figure}

By use of these techniques it has been possible
to prove the general validity of the {\em fluid mosaic model\/} of
cell membranes, proposed by S.~J.~Singer and~G.~Nicholson in 1972
\cite{Singer}.
It may be schematically represented by the cartoon in fig.~\ref{memb-fig}.
The main part of the membrane is formed by a bilayer,
which is a mixture of several kinds of amphiphilic molecules,
in particular phospholipids and glycolipids
(sugar-containing lipids, like in particular
sphingomyeline and cerebroside), plus smaller ones
like cholesterol. Integral membrane proteins are
dissolved in the bilayer: they can freely diffuse laterally,
but cannot move out of the surface. Other proteins
carry hydrophilic tails (sometimes containing
sugar: one then speaks of glycoproteins) which extend in the exterior
of the cell. In the interior of the cell there is
often a network of filaments, suitably anchored
to the bilayer. This is the case in particular of the red
blood cells (erythrocytes), whose skeleton
is formed by filaments of spectrin bound to
``buoys'' formed by other proteins, like
ankyrin and ``protein~4.1''.

\begin{figure}\begin{center}
\includegraphics[width=10cm]{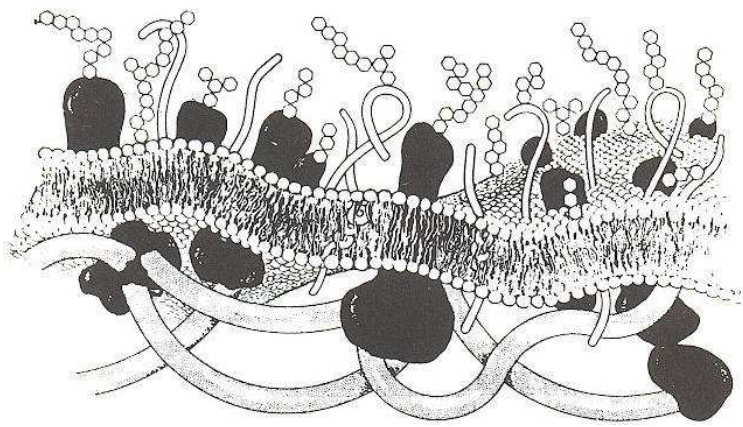}
\end{center}\caption{Schematic illustration of a biological membrane (courtesy
of Ove Broo S\o rensen of the Danish Technical University).
The membrane itself is a mixture of several kinds of
phospholipids, plus smaller amphihilic molecules, among
which (in animals) cholesterol has a prominent role.
The dark objects traversing the membranes
are ``integral membrane proteins''. Other, more flexible,
membrane proteins carry carbohydrates (represented as small polygons)
in their exterior. The rubber-like cytoskeleton attached to
the inner side of the membrane exerts a strong influence
on the mechanical properties of the composite membrane.}
\label{biomemb-fig}
\end{figure}
The fluidity of biological membranes
can be exhibited by fluorescence photobleaching
recovery experiments in intact cells. One first attaches
a fluorescent dye to a specific membrane component.
One then looks at a small region ($\sim 3\mu{\rm m}^2$)
through a fluorescence microscope.
One then destroys the fluorescent molecules in this
region with a very intense light pulse from a laser.
The fluorescence of this region is then monitored as
a function of time. The rate of recovery is related
to the diffusion coefficient $D$ of the fluorescent-labeled
molecules. The order of magnitude of $D$ turns
out to be $10^{-8}\,{\rm cm}^2{\rm s}^{-1}$,
implying that a molecule can diffuse about $2\mu{\rm m}$
in $1\,{\rm s}$. On the other hand, the characteristic times
for a molecule to pass from one membrane layer to the other
(flip-flop) are of the order of several hours. These times
can be measured, although with some difficulty, by NMR~\cite{flip-flop}.
Therefore any asymmetry of the lipid bilayer can be preserved for
long periods.

Of course a biological membrane is extremely complex to characterize
physically. A strong tendency has therefore developed towards
the study of model membranes, containing just one (or two)
phospholipids, and in case a controlled concentration of
impurities. For example, one can suspend the amphiphile in
an aqueous medium, and the agitate the mixture by high-frequency
sound waves. This procedure is called {\em sonication}.
Alternatively, on can dissolve the lipid in ethanol,
and then inject the solution via a fine needle into water.
In this way one obtains aqueous compartments closed
by a lipid bi- or (more often) multi-layer. These structures
are called {\em lipid vesicles\/} or {\em liposomes}.
It is of course possible to prepare the liposomes in
a solution containing some interesting drug, and then separate
them from the surrounding solution by dialysis or by gel filtration.
This technique provides in principle  a way to control the
delivery of drugs to target cells.

\section{Isolated membranes: the Helfrich hamiltonian}\setcounter{equation}{0}
In this section I shall discuss the fundamentals of the physical
description of isolated membranes: I shall neglect therefore
the interaction between membranes, and I shall not consider the
microscopic mechanism leading to their formation.

We can write the free energy of a bilayer composed of $2N$
amphiphilic molecules immersed in water as a function of the area $a=A/N$
per molecule \cite{deGennes}:
\begin{equation}
F=2N\phi(a)=2N\left[\phi_{\rm phob}(a)+\phi_{\rm phil}(a)+
\phi_{\rm int}(a)\right].
\end{equation}
In this equation, $\phi(a)$ is the chemical potential
of molecules in the bilayer, $\phi_{\rm phob}$ and $\phi_{\rm phil}$
are the effective attractive and repulsive parts, respectively,
due to their interaction with water, while $\phi_{\rm int}$
is a direct interaction between amphiphiles which will
generically be repulsive. In general, $\phi(a)$
will have a minimum at some preferred area-per-molecule
$a=a_0$. If the membrane does
not exchange molecules with the reservoir, and
if it can freely adjust its total area, the equilibrium
will be reached when
\begin{equation}
\left.\frac{\partial F}{\partial a}\right\vert_{a_{\rm eq}}
=0,
\end{equation}
implying
\begin{equation}
a_{\rm eq}=a_0.
\end{equation}
Therefore, the surface tension $\gamma(a)$ of the
membrane vanishes at equilibrium:
\begin{equation}
\gamma\left(a_{\rm eq}\right)=\left.\frac{\partial F}{\partial A}\right\vert
_{a_{\rm eq}}=\left.\frac{\partial \phi}{\partial a}\right
\vert_{a_0}=0.
\end{equation}
This simple thermodynamic argument is usually evoked to explain
why the surface tension of amphiphilic membranes can be very
small.

If we consider a
{\em fluctuating\/} amphiphilic membrane,
it is necessary to distinguish between
its {\em total\/} area $A$ and its {\em projected\/} area
$A_{\rm p}$~\cite{tension}.
To fix one's ideas, suppose that the membrane spans a planar frame
of area $A_{\rm p}.$ If the compressibility of the amphiphilic
monolayers is low, the total area $A$ is proportional to the
total number $2N$ of molecules forming the membrane: $A=a_0 N.$
It can only vary through changes in the number of molecules $N.$
Then the two quantities $A$ and $A_{\rm p}$ can be considered as
{\em independent\/} thermodynamic variables~\cite{deGennes}.
They are both extensive: their thermodynamic conjugates represent
{\em distinct\/} physical quantities. The {\em area coefficient\/} conjugate
to the total area $A,$ which we denote by $\gamma,$ is for incompressible
fluids directly proportional to the chemical potential $\mu$
of amphiphilic molecules. The {\em film tension\/} conjugate
to the projected area $A_{\rm p},$ which we denote by $\tau,$
corresponds to the physical ``surface tension''. One can then consider
four different thermodynamical ensembles:
\begin{enumerate}
\item $(A,A_{\rm p})$-ensemble: isolated, framed membranes.
\item $(A,\tau)$-ensemble: isolated, unframed membranes.
\item $(\gamma,A_{\rm p})$-ensemble: open, framed membranes.
\item $(\gamma,\tau)$-ensemble: open, unframed membranes.
\end{enumerate}
Experimentally, the most important situations are the
{\em open, framed\/} systems (which can be experimentally
realized in black lipid films or monolayers) and the
{\em isolated, unframed\/} systems, in which the projected
area can fluctuate, which correspond to the
case of lipid vesicles (at least as long as exchange
of lipids with the surrounding solution can be neglected).

\smallskip\leftline{\em Open, framed systems.}
\noindent In this ensemble the projected area $A_{\rm p}$ is fixed and
the total area $A$ fluctuates. The fluctuations are governed
by a hamiltonian of the form
\begin{equation}
{\cal H}=\gamma A+{\cal H}_{\rm el},
\end{equation}
where ${\cal H}_{\rm el}$ contains the contribution of elastic internal forces
(bending energy, shear modulus, etc.). The partition function
is written as the sum over all film configurations ${\cal C}$
with fixed $A_{\rm p}$:
\begin{equation}
Z_O=\sum_{\cal C}\exp\left\{-\frac{{\cal H}({\cal C})}{\kt}
\right\}.
\end{equation}
The free energy in this ensemble is
\begin{equation}
G_O(\gamma,A_{\rm p})=-k_{\rm B}T\ln Z_O(\gamma,A_{\rm p}),
\end{equation}
and the film tension is simply defined as the free energy per unit projected
area:
\begin{equation}
\tau=\lim_{A_{\rm p}\to\infty}\frac{G_O(\gamma,A_{\rm p})}{A_{\rm p}}.
\end{equation}

\smallskip\leftline{\em Isolated, unframed systems}
\noindent In this ensemble the total area $A$ is fixed while the
projected area may fluctuate. The thermodynamic potential is obtained
from $G_O$ by first going to the isolated, framed
$(A,A_{\rm p})$ ensemble where both $A$ and $A_{\rm p}$ are
fixed, and the thermodynamic potential is obtained by a Legendre
transform
\begin{equation}
F_I(A_{\rm p},A)=G_O(A_{\rm p},r)-\gamma A,
\qquad A=\left.\frac{\partial G_O}{
\partial \gamma}\right|_{A_{\rm p}}.\label{tension:framed}
\end{equation}
Then one goes to the isolated, unframed film ensemble by a second
Legendre transform which defines the associated thermodynamic
potential
\begin{equation}
G_I(\tau,A)=F_I(A_{\rm p},A)-\tau A_{\rm p},
\end{equation}
where the surface tension $\tau$ is defined by
\begin{equation}
\tau=\left.\frac{\partial F_I}{\partial A_{\rm p}}\right|_A=
\left.\frac{\partial G_O}{\partial A_{\rm p}}\right|_r.\label{tension}
\end{equation}
It is clear that in the thermodynamic limit, $A\to\infty,$
the surface tension defined in this ensemble by eq.~(\ref{tension})
coincides with the surface tension defined for the open, framed
system by eq.~(\ref{tension:framed}).

\begin{figure}
\begin{center}
\includegraphics[width=8cm]{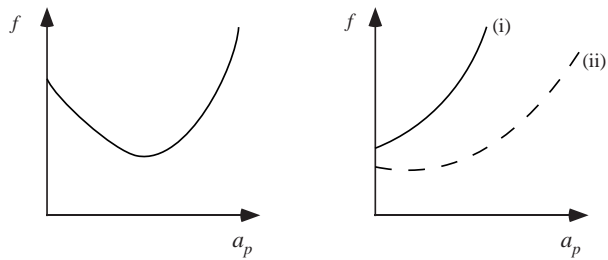}
\end{center}
\caption{The free energy density $f$ as a function of the
area ratio $a_{\rm p}$. (a) $f$ has a minimum
for $a_{\rm p}>0$ and the tension $\tau$ vanishes.
(b) $f$ has its minimum at $a_{\rm p}=0$ and the
tension $\tau$ may either be positive (i)
or zero (ii).}\label{tension:fig}
\end{figure}

We can now analyse the meaning of the tension $\tau$ for
isolated, unframed systems with fixed total area $A.$
Let us assume that the projected area fluctuates around
its mean value $\left<A_{\rm p}\right>.$ This mean value
can be obtained by minimizing $F_I(A_{\rm p},A)$ with
respect to $A_{\rm p},$ while $A$ is fixed.
In the thermodynamic limit it is useful to consider the free
energy density
\begin{equation}
f=\frac{F_I}{A},
\end{equation}
as a function of the area ratio
\begin{equation}
a_{\rm p}=\frac{A_{\rm p}}{A}.
\end{equation}
Two situations are then possible, as depicted in fig.~\ref{tension:fig}.
\begin{itemize}
\item $f(a_{\rm p})$ has its minimum for a nonzero ratio
$0<a_{\rm p}<1$ (fig.~\ref{tension:fig}(a)). The membrane
is then said to be {\em flat}, since, although shrunk due to thermal
fluctuations, it still keeps the global structure
of a two-dimensional object. In this case, eq.~(\ref{tension})
implies that the surface tension vanishes.
\item $f(a_{\rm p})$ has its minimum at $a_{\rm p}=0$
(fig.~\ref{tension:fig}(b)). The membrane is then said to be
{\em crumpled,} since it is so shrunk by thermal fluctuations that
its extension in space does not scale linearly with its internal tension.
In this case it is obvious from fig.~\ref{tension:fig}(b) that the
tension $\tau$ has in general no reason to vanish:
however, there might be marginal situations in which,
although the minimum of $f(a_{\rm p})$ is at $a_{\rm p}=0,$
its slope at this point vanishes, and one has therefore
$\tau=0.$
\end{itemize}

Let us consider a {\em fluid\/} membrane freely fluctuating
in a solvent. We assume that its thickness
is much smaller than the length scales describing
its shape and its undulations. Its fluidity implies that all
the internal degrees of freedom, related, e.g., to the
hydrocarbon chain conformation, the local molecular density,
etc., reach equilibrium on a fast time scale. We are thus
led to the conclusion that the free energy of the membrane depends on its
shape alone. In our hypotheses, we may describe its
shape as a geometrical surface.

\begin{figure}
\begin{center}
\includegraphics[width=8cm]{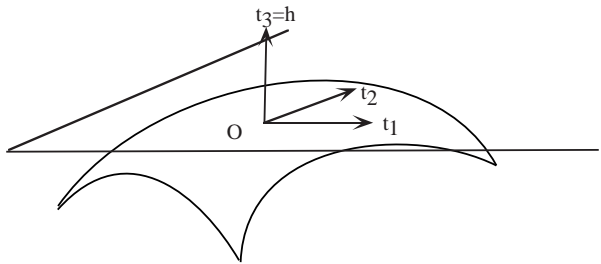}
\end{center}
\caption{The tangent plane at a point O. The distance of
a generic point P from it is given by
$h={{1}\over{2}}\sum_{ij}\Omega_{ij}t^it^j$. The principal
directions on the surface at the point O are also
drawn.}
\label{surf-fig}\end{figure}

The curvature of a surface can be described by two quantities, the
{\em mean curvature\/} $H$, and the {\em Gaussian curvature\/}
$K$. Consider a portion of the surface around an
arbitrary point O, like the one
depicted in~fig.~\ref{surf-fig}.
Draw the tangent plane through O and choose the
local coordinate axes $(t^1,t^2,t^3)$ so that O is their
common origin,
$t^1$ and $t^2$ lie on the plane, and $t^3=h$ is
normal to it. The surface is locally represented by
a quadratic form:
\begin{equation}
h=\frac{1}{2}\sum_{i,j=1}^2\Omega_{ij}t^it^j+\ldots
\label{quad-eq}
\end{equation}
The matrix $\Omega=\left(\Omega_{ij}\right)$
obviously vanishes for a locally flat surface.
This expression of the distance $h$ from the tangent plane
in terms of Cartesian coordinates on it, is also
called the {\em second fundamental form\/} of the surface.
The curvature is defined via the invariants of the second fundamental form,
namely:
\begin{itemize}
\item The {\em mean curvature\/} $H$ is given by $H={\rm Tr}\,\Omega;$
\item The {\em Gaussian curvature\/} $K$ is given by
$K={\rm Det}\,\Omega.$
\end{itemize}
The eigenvalues $c_1$, $c_2$ of $\Omega$ are called the principal
curvatures. They are the inverse of the curvature radii of
the intersections of the surface with two planes containing the
normal to the surface, and mutually perpendicular. The intersections
of these planes with the tangent plane define the principal curvature
directions. The principal curvatures are the extreme values of the
curvature of any intersection of the surface with a normal plane.
In the exceptional case where $c_1=c_2$, the point O is called an {\em
umbilical point,} and the principal directions are not defined.
The mean curvature $H=c_1+c_2$ is positive (negative) if the
surface is locally mostly above (below) the tangent plane, going
in the positive direction along the normal. The Gaussian curvature
$K=c_1c_2$ is positive if the surface is locally on one side
of the tangent plane (elliptic point), and negative if it is
on both sides (hyperbolic point).

The curvatures $H$ and $K$ have a simple geometrical interpretation.
Consider a small portion $S$ of the surface, and
let $\Delta A$ be its area. Construct
now a new surface $S'$ by displacing each point
of $S$ of a distance $\delta$ along the normal,
in the direction of positive $h$.
The new surface will have the area
\begin{equation}
\Delta A'=
\Delta A\,\left(1+\delta H+\delta^2 K+o(\delta^2)\right).
\label{eq:curv}
\end{equation}

We now make the following assumptions on the free energy
of membranes:
\begin{itemize}
\item The surface is smooth, and can be locally represented
by the parametric equation $\vec{r}=\vec{r}(\underline{\sigma})$,
where $\vec{r}$ denotes a point in three-dimensional
ambient space, and $\underline{\sigma}=(\sigma^1,\sigma^2)$
are local coordinates of the surface. The function
$\vec{r}(\underline{\sigma})$ is differentiable an arbitrary
number of times.
\item The free energy can be expressed as a local
functional of $\vec{r}(\underline{\sigma})$ and its derivatives.
This assumption rules out, for the time being, the effects
of the interaction of the membrane with itself,
due for example to close contacts when a part
of the membrane folds on the rest.
\item The free energy must be invariant under
Euclidean transformations applied to $\vec{r}$
and under reparametrization transformations
like $\underline{\sigma}
\to\underline{\sigma}'=\underline{\sigma}'(\underline{\sigma})$.
\end{itemize}

It was noticed by Canham~\cite{Canham} and then by
Helfrich~\cite{Helfrich73} that these hypotheses imply
that, considering only the contributions
of derivatives of $\vec{r}$ up to second
order, one obtains the following general expression of the free energy
of a membrane:
\begin{equation}
F=\int_S \d A\,\left[\gamma+\frac{1}{2}\kappa
\left(H-H_0\right)^2+\overline{\kappa}K\right].
\label{eq:Helfrich}
\end{equation}
Here $\d A$ is the area element and $\gamma$ is the
area coefficient. The integral is extended over the
membrane surface $S$.
The coefficients $\kappa$ and $\overline{\kappa}$
are known as the {\em rigidity\/} and {\em Gaussian rigidity\/}
respectively. The quantity $H_0$ is called the {\em spontaneous
curvature.}

The free energy (\ref{eq:Helfrich}) has been derived only on
the basis of geometrical considerations. However it is necessary
to gain some insight on the terms appearing in this expression by
considering in more detail the elasticity of a fluid membrane.
One finds several discussions of this subject in the
literature \cite{Bivas84,Szleifer,Helfrich81},\cite[App.~B]{Xavier}.

The simplest derivation starts by considering a monolayer.
Let us assume that the elastic energy per molecule in a
given configuration is the sum of the energy ${\cal E}_{\rm h}$ pertaining to
the heads and the energy ${\cal E}_{\rm t}$ pertaining to the tails.
We take as a reference the neutral surface,
where the moments of the elastic forces on heads and tails cancel.
We assume for definiteness that the
positive direction is towards the heads, and denote by $\delta_{\rm h}$
and $\delta_{\rm t}$ the distance of the heads and the tails, respectively,
from the reference surface.

\begin{figure}
\begin{center}
\includegraphics[width=8cm]{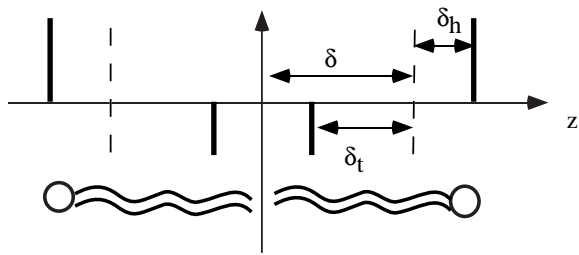}
\end{center}
\caption{Simple model of a bilayer in an aqueous medium. One
assumes that the stresses are localized on the heads and tails
of the amphiphilic molecules. The position of the neutral
surfaces is denoted by the broken lines.}\label{model:fig}
\end{figure}

The areas $a_{\rm h}$ of the heads and $a_{\rm t}$ of the tails are given by:
\begin{eqnarray}
a_{\rm h}&=&a\left(1+H\delta_{\rm h}+K\delta_{\rm h}^2+\ldots\right),\\
a_{\rm t}&=&a\left(1-H\delta_{\rm t}+K\delta_{\rm t}^2+\ldots\right),
\end{eqnarray}
where $a$ is the area of the molecule on the neutral surface
(cf.~fig.~\ref{model:fig}).
We can now write, assuming a simple (Hooke-like) elasticity:
\begin{eqnarray}
{\cal E}_{\rm h}&=&\frac{1}{2}k_{\rm h}
\left(\frac{a_{\rm h}-a_{{\rm h}0}}{a_{{\rm h}0}}\right)^2,\\
{\cal E}_{\rm t}&=&\frac{1}{2}k_{\rm t}
\left(\frac{a_{\rm t}-a_{{\rm t}0}}{a_{{\rm t}0}}\right)^2,
\end{eqnarray}
where $a_{{\rm h}0}$ and $a_{{\rm t}0}$
are the equilibrium values of head and
tail areas, and $k_{\rm h}$ and $k_{\rm t}$
are elastic constants (with the
dimensions of an energy per molecule).
The elastic energy ${\cal E}$ per unit molecule can be written
\begin{equation}
{\cal E}={\cal E}_{\rm h}+{\cal E}_{\rm t}.\label{ener:eq}
\end{equation}
For a {\em flat} membrane we can obtain from it the
expression of the elastic energy of compression {\em per unit area\/}:
\begin{equation}
{\cal H}^{({\rm m})}_{\rm comp}=\frac{{\cal E}}{a_0}=\frac{E_0}{a_0}+
\frac{1}{2}k^{({\rm m})}\left(\frac{a-a_0}{a_0}\right)^2,\label{comp:eq}
\end{equation}
where $E_0$ is a constant, and where
the compression modulus $k^{({\rm m})}$ and the equilibrium
area per molecule $a_0$ are respectively given by
\begin{eqnarray}
k^{({\rm m})}&=&\frac{k_{\rm h}}{a_{{\rm h}0}}+
\frac{k_{\rm t}}{a_{{\rm t}0}},\\
a_0&=&k^{({\rm m})}\bigg/\left(\frac{k_{\rm h}}{a_{{\rm h}0}^2}+
\frac{k_{\rm t}}{a_{{\rm t}0}^2}\right)=
\frac{k^{({\rm m})}}{\beta_{\rm h}+\beta_{\rm t}}.
\end{eqnarray}
The index $({\rm m})$ reminds us that we are dealing
with a monolayer.
The condition that
the moments of the compression forces
vanish on the neutral surface implies
$\delta_{\rm h}\beta_{\rm h}=\delta_{\rm t}\beta_{\rm t}.$
We can now let the expression of the areas
$a_{\rm t}$ and $a_{\rm h}$
in the expression of the elastic energy, obtaining
\begin{eqnarray}
{\cal F}^{({\rm m})}&=&{\cal H}^{({\rm m})}_{\rm comp}
+{\cal H}^{({\rm m})}_{\rm  bend}=\nonumber\\
&&=\sigma^{({\rm m})}+\frac{1}{2}k^{({\rm m})}
\left(\frac{a-a_0}{a_0}\right)^2+
\frac{1}{2}\kappa^{({\rm m})}\left(H-H_0^{({\rm m})}\right)^2+
\overline{\kappa}^{({\rm m})} K,
\end{eqnarray}
where
\begin{eqnarray}
\sigma^{({\rm m})}&=&\frac{E_0}{a_0},\\
\kappa^{({\rm m})}&=&(\delta_{\rm h}+\delta_{\rm t})^2k^{({\rm m})}
\frac{\beta_{\rm h}\beta_{\rm t}}{(\beta_{\rm h}
+\beta_{\rm t})^2}
=\delta_{\rm h}\delta_{\rm t}\,k^{({\rm m})},\\
\overline{\kappa}^{({\rm m})}&=&(a_{{\rm h}0}-a_{{\rm t}0})
\frac{\beta_{\rm h}\beta_{\rm t}
(\beta_{\rm h}-\beta_{\rm t})}{(\beta_{\rm h}+\beta_{\rm t})^2},\\
H_0^{({\rm m})}&=&\frac{\overline{\kappa}}{\kappa}
\frac{1}{\delta_{\rm h}-\delta_{\rm t}}.
\end{eqnarray}
We see that $\kappa^{({\rm m})}>0,$ while
$\overline{\kappa}^{({\rm m})}$ can be
of either sign.
We can now estimate the order of magnitude
of the rigidities $\kappa^{({\rm m})}$, $\overline{\kappa}^{({\rm m})}$.
For pure phospholipids
the measured compressibility is
$k\sim50\, {\rm dyn}/{\rm cm}=5\,10^{-1}\,{\rm
Jm}^{-1},$ and the distances $\delta_{\rm h},$ $\delta_{\rm t}$ are of the
order of $10\,{\rm \AA}.$ We obtain therefore
$\kappa^{({\rm m})}\sim 5\,10^{-20}
{\rm J},$ which must be compared with the termal energy
$k_{\rm B}T\sim 4.2\, 10^{-21}J$ at $T=300\,{\rm K}.$
The fact that typical rigidities
are only slightly larger than the thermal energy makes fluctuations
important in understanding the behavior of
amphiphilic membranes.

We can now consider a {\em bilayer\/}, by putting two monolayers,
called ``internal'' (i) and ``external'' (e)
on the top of each other. We assume that there are no interactions
between the monolayers, so that
\begin{equation}
{\cal F}_{\rm b}
={\cal F}^{({\rm m})}_{{\rm i}}+{\cal F}^{({\rm m})}_{{\rm e}},
\end{equation}
where the suffixes refer to the two monolayers. We denote by
$a_i$ the area per molecule of the monolayer $i={\rm i},\,{\rm e},$
measured on its neutral surface. We can thus define
the particle density $\rho_i$ of each monolayer,
measured along a bilayer reference surface, placed midway
between the two neutral surfaces of the
monolayers. Let us denote by $\delta$
the distance between the bilayer reference surface
and that of the monolayers. For a bilayer with heads outside,
for example, we have $\delta=\delta_{\rm t}.$ We choose the positive direction
on the
normal, in going, e.g., from the internal to the external layer.
We then have
\begin{eqnarray}
\frac{1}{\rho_{{\rm i}}}&=&a_{{\rm i}}\,
(1-H\delta+K \delta^2+\ldots),\\
\frac{1}{\rho_{{\rm e}}}&=&a_{{\rm e}}\,
(1+H\delta+K \delta^2+\ldots).
\end{eqnarray}
This allows us to define the mean density $\overline{\rho}$
and the density difference $\tilde{\rho}$:
\begin{eqnarray}
\overline{\rho}&=&\frac{\rho_{{\rm i}}+\rho_{{\rm e}}}{2},\\
\tilde{\rho}&=&\frac{\rho_{{\rm i}}-\rho_{{\rm e}}}{2}.
\end{eqnarray}
In the case where the spontaneous curvature $H_0$ of each
monolayer vanishes, one obtains the following expression
of the elastic energy density per unit area of the bilayer:
\begin{eqnarray}
{\cal F}_{\rm b}&=&\sigma_{\rm b}\overline{\rho}a_0
+\frac{1}{2}k_{\rm b}\left(a_0
\overline{\rho}-1\right)^2+\frac{1}{2}
k_{\rm b}(a_0\tilde{\rho})^2\nonumber\\
&&+a_0k_{\rm b}D\tilde{\rho}H+\left(2\kappa_{\rm b}
+\frac{k_{\rm b}\delta^2}{2}\right)H^2
+\overline{\kappa}_{\rm b}K.\end{eqnarray}
In this equation, the bilayer parameters (denoted by
the suffix (b)) are simply the
sum of the corresponding parameters of
the monolayers.

If the spontaneous curvature of the monolayers does not vanish, but
the two monolayers are made of the same material, one obtains the
same expression, but with slightly renormalized values of the
elastic parameters. In either case, the spontaneous curvature $H_0$
of the bilayer vanishes.

The only difference with respect to the
Helfrich hamiltonian lies in the terms depending on the
particle densities $\overline{\rho},$ $\tilde{\rho},$ which
we now discuss.
The first term, $\sigma_{\rm b}\overline{\rho}a_0,$ yields a contribution
proportional to the total number $2N$ of molecules in the membrane,
but independent of its geometry. It corresponds to
a shift in the chemical potential of the amphiphile. As long as one considers
vesicles over short time scales, so that amphiphilic molecules
are not exchanged with the solvent, this term plays no role.
The second term represents the energy cost necessary to induce
local density fluctuations. As we have seen, the rigidity $\kappa_{\rm b}$
is of the order of $k_{\rm b}\delta_{\rm h}\delta_{\rm t}$: therefore,
it costs about the same energy to bend the membrane with
a radius of the order $10^3\delta\sim 1\,\mu{\rm m}$ as to
produce a change of order $10^{-3}$ in the relative area $(A-A_0)/A_0,$
where $A_0=Na_0.$
{}From a practical point of view, therefore, the membrane can
be considered incompressible and its area can be set equal to $A_0.$
The third term represents a coupling between the local density
difference between the two monolayers and the mean curvature $H.$
Indeed, if we try to bend the membrane, the molecules contained
in the interior layer are compressed with respect to
those contained in the exterior layer, and therefore tend to escape.

Let us now consider a closed vesicle, formed by a bilayer with
identical monolayers. If the observation times are not
too long (of the order of several hours, or even a few
days) the total number of molecules
contained in the bilayer remains constant: on the other hand,
as we have seen, the ``flip-flop'' times are also quite
long, and therefore the number of molecules contained in each
monolayer can be considered constant. Therefore
the integrals of both $\overline{\rho}$ and $\tilde{\rho}$
over the whole bilayer surface are constant.
In particular, that means that their average values
are given by
\begin{eqnarray}
\overline{\rho}_0&=&\frac{N_{\rm i}+N_{\rm e}}{2A},\\
\tilde{\rho}_0&=&\frac{N_{\rm i}-N_{\rm e}}{2A},
\end{eqnarray}
where $N_{\rm e,i}$ are the number of molecules in the
two monolayers, and $A$ is the area of the vesicle.
Therefore the term $\oint\d A\gamma\overline{\rho}a_0$
is independent of the bilayer shape.
It is reasonable to assume, in view of the fast lateral
diffusion times of the amphiphiles, that $\tilde{\rho}$
locally equilibrates faster than the vesicle shape.
It will then assume locally the  value dictated by
the optimization of $F_{\rm b}$
when the curvatures are kept fixed.
One has
\begin{equation}
\tilde{\rho}(\vec{r})=\tilde{\rho}_0+
\delta \tilde{\rho}(\vec{r}),
\end{equation}
where $\delta\tilde{\rho}$ satisfies
\begin{equation}
\delta\tilde{\rho}=-\tilde{\rho}_0-\frac{\delta}{a_0}H.
\end{equation}
We can use again eq.~(\ref{eq:curv}) to show that
this equation implies
\begin{equation}
\oint_S\d A \delta\tilde{\rho}=0.
\end{equation}
If we assume that $\overline{\rho}=a_0^{-1}$
locally, and let the result for $\delta\tilde{\rho}$ into the
expression for ${\cal F}_{\rm b}$ we obtain
\begin{equation}
F_{\rm b}=\oint_S\d A{\cal F}_{\rm b}=
\oint_S\d A\left\{\frac{1}{2}\kappa_{\rm b} H^2
+\overline{\kappa}_{\rm b} K
\right\},
\end{equation}
up to a term independent of the bilayer geometry.

These arguments lead thus to the conclusion that
vesicles formed by symmetric bilayers are described by
the Helfrich hamiltonian with a vanishing spontaneous curvature.
A small asymmetry between the monolayers can
be induced by a difference in the solvent: e.g., if
a higher concentration in ions screens more effectively
on the interior the electrostatic
interaction among the heads. One is safe to assume
that any induced spontaneous curvature will be rather small.
However, the surface must satisfy the additional constraints, fixing the
value of (i) the enclosed volume $V$, (ii) the total area $A$, and
(iii) the area difference between the two monolayers,
which can be expressed in terms of the integral of the mean curvature $H$:
\begin{equation}
\Delta A=2 \delta \oint_S\d A H=2 \delta M.
\end{equation}

This model of a vesicle, due to
Svetina et al.~\cite{Svetina82}, is known as the
{\em bilayer coupling\/}~(BC) model. The earlier work on vesicle
shapes had considered the spontaneous curvature as
an independent parameter, and had neglected
the constraint on the integral of the mean curvature.
This model has is known as the {\em spontaneous
curvature\/}~(SC) model. Although
the spontaneous curvature $H_0$ can be
considered as the conjugate variable to the integral of the
mean curvature $M$, the two descriptions
are not equivalent. The point to keep in mind is that
the Helfrich curvature energy does not grow with the size of the system
as an extensive free energy, in the way
one is accustomed to find in thermodynamical systems.
In fact, it is scale invariant: consider
a surface $S$ defined by some parametric equation
of the form $\vec{r}=\vec{r}_0(\overline{\sigma})$. Now define the new
surface $S'$ by the parametric equation
$\vec{r}=\lambda \vec{r}_0(\overline{\sigma})$,
where $\lambda>0$. In this transformation, we have
\begin{eqnarray}
H&\to &H'=\lambda^{-1}H;\\
K&\to &K'=\lambda^{-2}K;\\
\d A&\to &\d A'=\lambda^2\d A.
\end{eqnarray}
Therefore the elastic energy remains locally invariant:
on the other hand, since $H$ (and therefore $M$) is multiplied by
$\lambda^{-1}$, the corresponding conjugate field $H_0$ should be multiplied
by $\lambda$ in order to keep the ``scaled'' constraint.
Since the differential ${\cal F}_{\rm b}\d A$
of $F_{\rm b}$ remains locally invariant
upon a scale transformation, it will be invariant upon all
transformations which locally reduce to a scale transformation
(plus translations and rotations)~\cite{Willmore}. This form the class
of {\em conformal transformations,} which in three dimensions
are obtained as the group formed by translations, rotations
and a three-parameter family of {\em special conformal transformations,}
obtained as the combination of an inversion, a translation, and
another inversion. Of course, the constraints mentioned above are
not invariant in general upon such a transformation.
{}From now on, we shall only refer to bilayers, and
correspondingly drop the (b) suffix on the rigidity moduli.

\section{Vesicle shapes}\setcounter{equation}{0}
The theory of the equilibrium shapes of phospholipid
vesicles started in 1970, when Canham~\cite{Canham} showed
that the characteristic discoidal shape of red blood cells
(erythrocytes) can be obtained by the minimization
of the curvature elastic energy for a particular
value of the area and enclosed volume costraints.
Actually the case of the red blood cells, although
of great interest, is somehow misleading, since
the spectrin-ankyrin network present on the interior
of the red blood cell membrane makes its properties
somehow different from those of a fluid
membrane~\cite{Stokke,Boal92}.

An extensive study of vesicle shapes was performed by Deuling and
Helfrich in the seventies \cite{Deuling,Deuling2}. The problem
was reconsidered more recently, in particular because of the
experiments of the Sackmann group \cite{Berndl,Kaes91,Kaes93},
which exhibited shape transitions of single
vesicles induced by changes in temperature. This prompted
the research on the identification of the phase
diagram of vesicle shapes (for recent reviews, see, e.g., \cite{Lipowsky91},
and the Proceedings contained in \cite{Lipowsky92}). The interest
has been further enhanced by the
observation of toroidal vesicles by Mutz and Bensimon~\cite{Mutz91},
which has stimulated several investigations on vescles with higher
genus \cite{Seifert91,Seifert91b,Michalet94,Xavier}.

In the spirit of mean field theory, the shape of a vesicle is obtained
by minimization of the elastic free energy~(\ref{eq:Helfrich})
\begin{equation}
F=\oint_S\d A\,\left\{\frac{1}{2}
\kappa H^2+\overline{\kappa}K\right\}.
\end{equation}
For phospholipids, $\kappa\sim 10^{-19}$ J \cite{Duwe,Mutz}.
This expression of the energy also yields a nice
argument to explain why isolated membranes form
vesicles at all.
Indeed the elastic free energy $F$ of any given shape
is independent of its scale: e.g., for a sphere one has
\begin{equation}
F=8\pi \kappa+4\pi \overline{\kappa}.
\end{equation}
If the membrane is not closed, but has a free edge of
length $\ell$, the free energy of the edge will be
proportional to $\ell$, times some line tension $\tau,$
which is of the order of $10^{-20}$--$10^{-19}$ J nm$^{-1}$
\cite{Harbich,Fromherz}.
As soon as the lateral size of the membrane becomes
larger than $\kappa/\tau,$ i.e., 1--10 nm, it becomes energetically
favorable for the membrane to get rid of its open edge
and form a closed vesicle. A detailed study of
this transition, analytic at zero temperature, and
via simulations at $T>0,$ is contained in ref.~\cite{Boal-Rao}.

The bilayer is relatively permeable to water~\cite{Cevc93},
much less to ions: the permeability ratio is of order
$10^9$. On a time scale of several hours
the ions enclosed in the vesicle at its formation
remain there. As a consequence, any variation in the
enclosed volume---due to the permeation of some amount of
water---would lead to the apperance of an osmotic pressure.
Equilibrium is reached when the osmotic pressure balances
the membrane tension. On the other hand, the time needed
to exchange amphiphilic molecules between the two
layers, or between the bilayer and the solution,
are of the order of several hours. Therefore, as long
as we consider shorter times, we can
argue that the enclosed volume $V$, the total bilayer
area $A$, and the number of molecules contained
in each bilayer are constant. The last constraint,
as we have seen, is equivalent to a constraint on the total
mean curvature $M=\oint \d A\,H.$
These constraints
can be taken into account by Lagrange multipliers, which
correspond to the pressure difference $p$ between the interior and the
exterior, the lateral tension $\gamma,$ and a quantity $\mu$
proportional to a variation of the spontaneous curvature.
In principle, one cannot rule out the
possibility that a small spontaneous curvature $H_0$
is present.
The Euler-Lagrange equations read
\begin{equation}
\delta\left(F+pV+\gamma A+\mu M\right)=0.
\end{equation}

It is a consequence of the Gauss-Bonnet theorem that the
Gaussian curvature term does not play any role in this
variational problem. According to this theorem,
the integral $\oint \d A\,K$
is a topological invariant:
\begin{equation}
\oint_S\d A K=4\pi\chi_{\rm E},
\end{equation}
where $\chi_{\rm E}$ is the Euler-Poincar\'e characteristic, equal
to one minus the number of ``handles'' of the surface:
i.e., it is equal to 1 for the sphere, to 0 for the torus,
to -1 for the torus with two holes\dots \ The last term in (\ref{eq:Helfrich})
remains therefore constant
for all continuous deformations of a given surface.
The variational equation depends only
on the parameters $p$, $\gamma$, $\kappa$ and $H_0.$
The solution of the Euler-Lagrange equations
will be characterized by the values $V$ of
the volume, $A$ of the area, and $M$ of the total
mean curvature. If we now perform the scale
transformation $\vec{r}\to\lambda \vec{r}$,
we have
\begin{eqnarray}
V&\to&V'=\lambda^3V;\\
A&\to&A'=\lambda^2A;\\
M&\to&M'=\lambda M.
\end{eqnarray}
This shape will be a solution of the same
Euler-Lagrange equation, corresponding
to the same value of $\kappa$, but where
\begin{eqnarray}
p&\to&p'=\lambda^{-3}p;\\
\gamma&\to&\gamma'=\lambda^{-2}\gamma;\\
\mu&\to&\mu'=\lambda^{-1}\mu.
\end{eqnarray}
With these substitutions, the free energy
remains invariant.

We can use this scale invariance to draw
the phase diagram of vesicle shape as a function
of dimensionless variables. One naturally introduces
the radius $R_0$ as the radius of the sphere
having the same area $A$:
\begin{equation}
R_0=\left(\frac{A}{4\pi}\right)^{\frac{1}{2}}.
\end{equation}
One can thus define:
\begin{itemize}
\item The reduced volume:
\begin{equation}
v=\frac{V}{\frac{4}{3}\pi R_0^3}=6\sqrt{\pi}\frac{V}{A^{\frac{3}{2}}};
\end{equation}
\item The reduced total curvature $m$:
\begin{equation}
m=\frac{M}{R_0}.
\end{equation}
\end{itemize}

\begin{figure}
\begin{center}
\includegraphics[width=6cm]{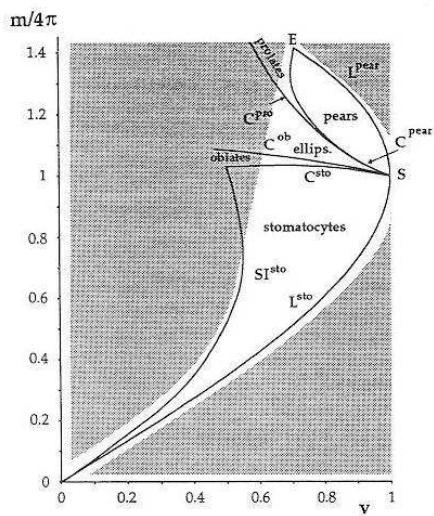}
\end{center}
\caption{Phase diagram of shapes in the BC model.
{}From K.~Berndl,
J.~K\"as, R.~Lipowsky, E.~Sackmann, U.~Seifert,
in L. Peliti (ed.) {\em Biologically Inspired Physics\/}
(New York: Plenum, 1991).}\label{vesicle-fig}
\end{figure}

The resulting Euler-Lagrange equations cannot
be analytically solved in general. If one looks for
{\em axisymmetric shapes\/} one can transform
these equations into a system of first-order ordinary differential
equations, which can be solved numerically
\cite{Deuling,Peterson}.
In this way one obtains the phase diagrams shown
in fig.~\ref{vesicle-fig}, as a function of $v$
and $\Delta a=m/4\pi$.
The continuous lines $C^{\rm pear}$ and $C^{\rm sto}$
denote lines of continuous transitions
at which the up/down symmetry of the vesicle
shape is broken. $L^{\rm pear}$,
$L^{\rm sto}$ and $L^{\rm dumb}$ denote limit shapes.
The dumbbell region contains for large $v$-values
prolate ellipsoids, and the discocyte region
oblate ellipsoids.

\begin{figure}\begin{center}
\includegraphics[width=10cm]{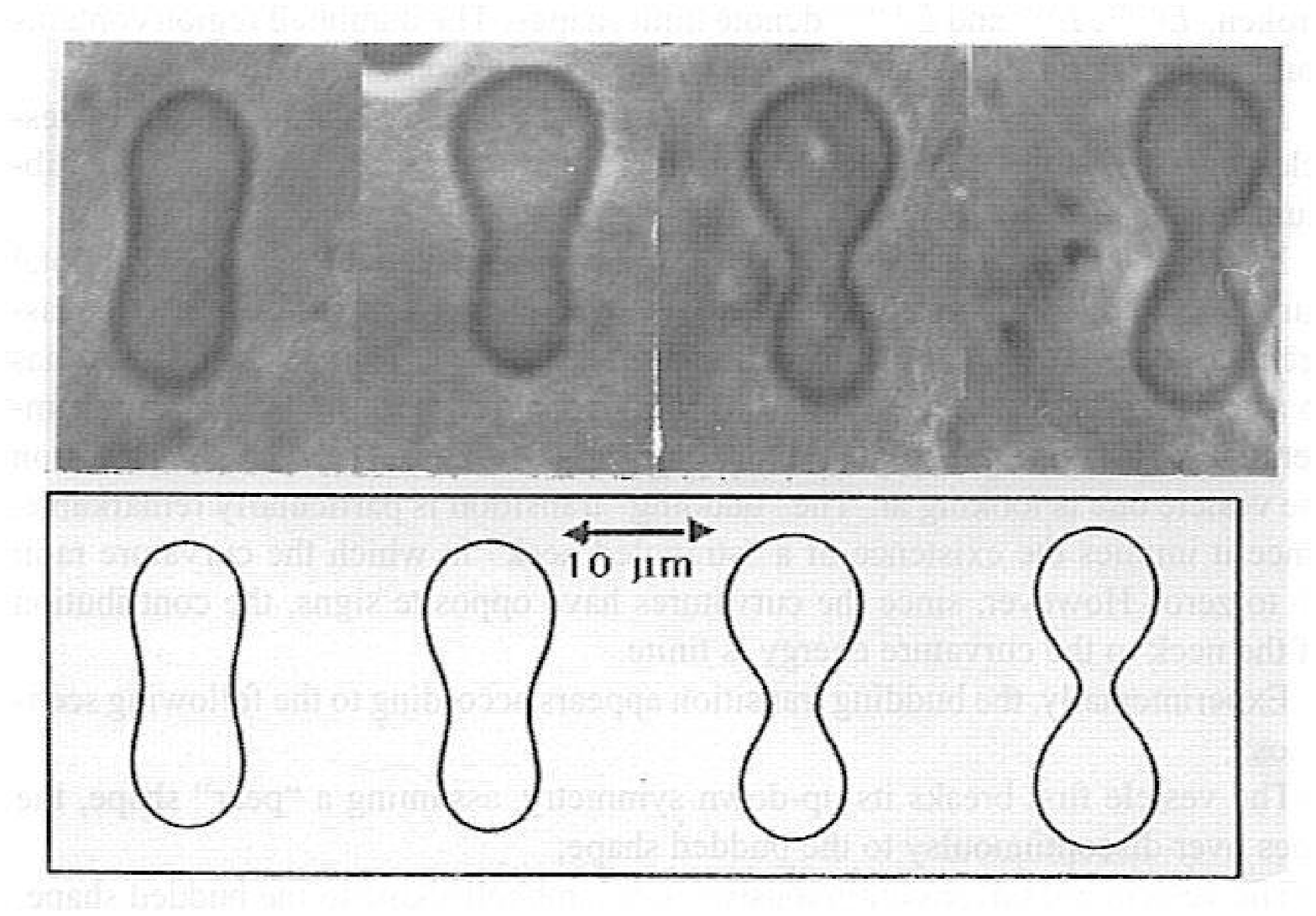}\\
\includegraphics[width=10cm]{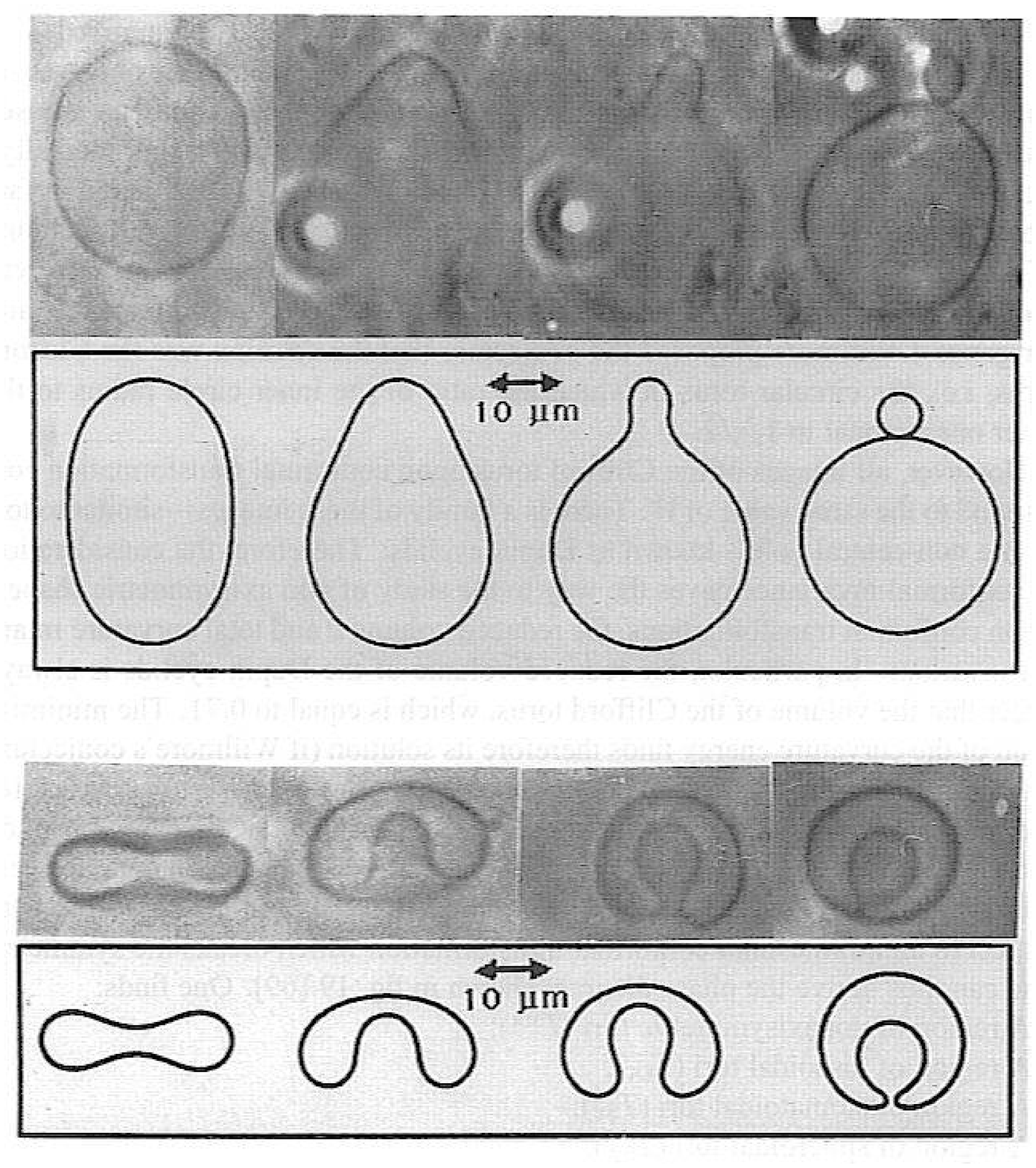}
\end{center}\caption{Demonstration (1) of a budding transition;
(2) symmetric-asymmetric reentrant transition;
(3) discocyte-stomatocyte transition.
Vesicles of DMPC in water: the reconstructed
trajectories in $(v,m)$ space correspond to
the dotted lines in the previous figure. From K.~Berndl,
J.~K\"as, R.~Lipowsky, E.~Sackmann, U.~Seifert,
in L. Peliti (ed.) {\em Biologically Inspired Physics\/}
(New York: Plenum, 1991).}
\label{vesicleshape-fig}
\end{figure}

The figure also contains pointed lines, labelled by numbers,
which correspond to trajectories observed in actual
phospholipid vesicles.
The photographs of these vesicles
are shown in fig.~\ref{vesicleshape-fig},
together with the calculated shapes.
In the interpretation of these experiments, the
value of the reduced total curvature $m$ is {\em inferred\/}
by the observed shape changes instead of being
measured. In order to explain the transitions observed
in a given vesicles, one has to assume that the inner and outer layer
expand at different rates when the
temperature is raised, introducing a phenomenological parameter
which depends on the vesicle one is looking at.
The ``budding'' transition is particularly
remarkable, since it implies
the existence of a ``strangled neck'' in which
the curvature radii go to zero. However, since
the curvatures have opposite signs, the contribution
of the neck to the curvature energy is finite.

Experimentally, the budding transition
appears according to the following scenarios:
\begin{itemize}
\item The vesicle first breaks its up-down
symmetry, assuming a ``pear'' shape, the
goes over discontinuously to the budded shape;
\item The vesicle goes over continuously
via a dumbbell shape to the budded shape.
\end{itemize}
While the second possibility agrees with the
calculated phase diagram of the BC model,
the first one has some difficulties:
the BC model predicts a continuous budding transition,
on the other hand the SC model does not predict
the intermediate ``pear'' shape. A model
based on area-difference elasticity
has been introduced to resolve this
difficulty~\cite{Miao93}. It seems to me
that the continuous elastic model of the membrane
breaks down near the strangled neck, which
may have an additional energy cost due to
the strain it imposes on the phospholipid tails:
the extra cost could turn a continuous transition into
a discontinuous one.

Mutz et al.~first observed toroidal vesicles
in partially polymerized vesicles~\cite{Mutz91}. Vesicles
of the same or higher genus have then been
observed in fluid vesicles made with the same phospholipid.
This observation has aroused great interest, because it
opened the possibility of investigating ``experimentally''
a classic mathematical problem posed by Willmore~\cite{Willmore}.
He had asked in fact which shapes minimized, for each topological genus,
the Willmore functional $W=\oint_S\d A H^2,$ which is none other
than the Helfrich Hamiltonian with zero spontaneous curvature.
For genus~0 the solution is shown to be the sphere,
and for genus~1 Willmore proposed the conjecture that
the solution was the Clifford torus, i.e., the
circular torus in which the ratio of the inner circle
radius to the outer one is equal to $1/\sqrt{2}.$

\begin{figure}
\begin{center}
\includegraphics[width=10cm]{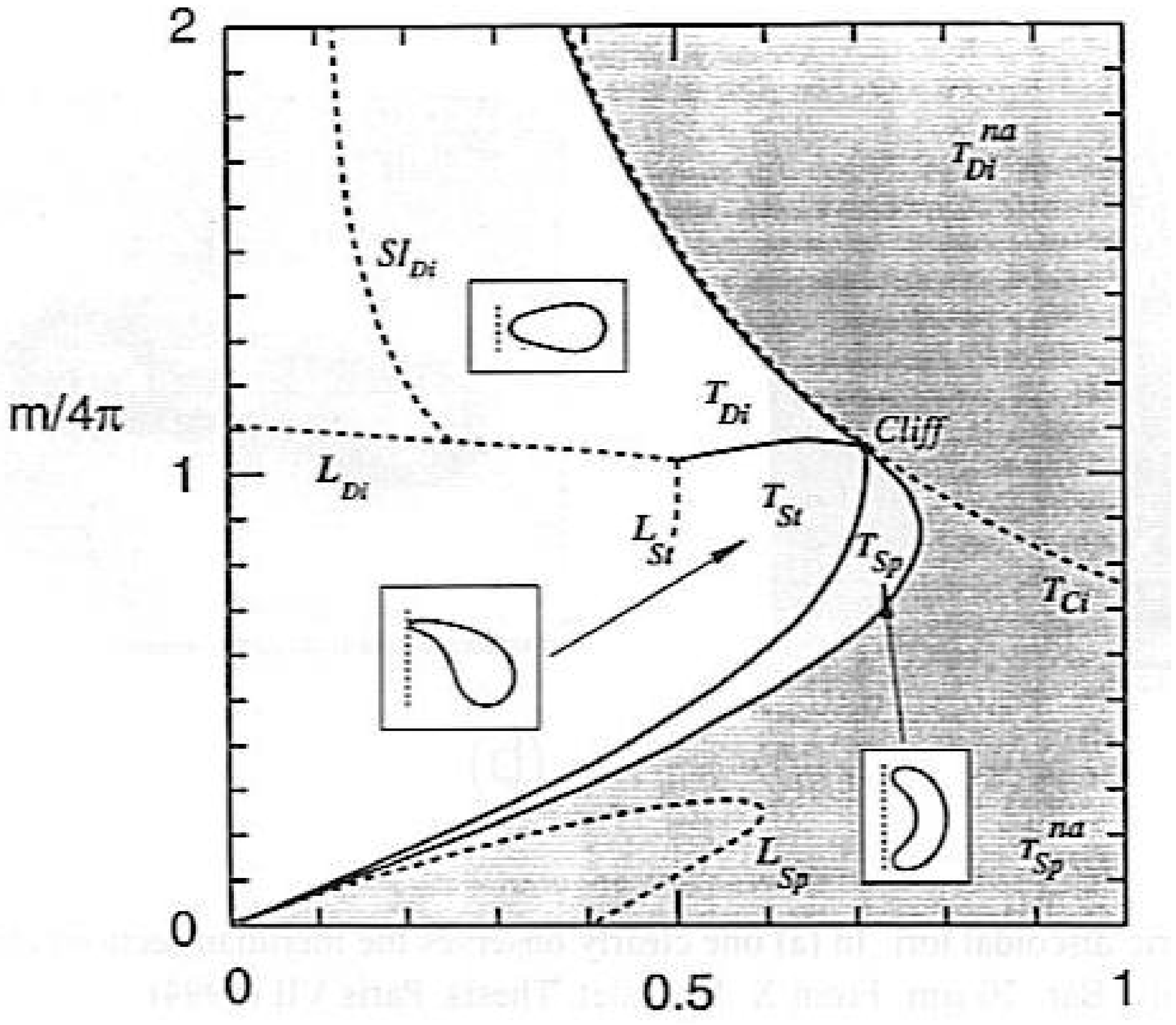}
\end{center}\caption{Phase diagram for vesicles of genus~1
in the BC model. Gray areas correspond to
non axisymmetric shapes. From F.~J\"ulicher,
U.~Seifert, R.~Lipowsky, {\em J.~Phys.~II France\/}
{\bf 3} 1681 (1993).}
\label{torus-fig}
\end{figure}

However, all images of the Clifford torus upon conformal
transformation correspond to the same value of $W.$
There is a family of these images---similar to
tori with a non-central
hole---known as Dupin cyclids. Therefore, the consideration
of conformal invariance paves the way to the study
of non axisymmetric shapes. Upon conformal transformations,
the reduced volume $v$ and total curvature $m$ are not invariant.
In particular, the reduced volume of the Dupin cyclids
is always larger that the volume of the Clifford
torus, which is equal to 0.71. The minimization
of the curvature energy finds therefore its
solution (if Willmore's conjecture is true)
if we consider the SC model (without the constraint
on the total curvature) and look for surfaces
with reduced volume larger that 0.71.
Other forms can be obtained by the solution
of the differential equations, as it had been done
for vesicles of genus 0.
One can check the stability of the axisymmetric
forms with respect to an infinitesimal conformal
transformation which breaks the symmetry.
One can thus derive the phase diagram shown in
fig.~\ref{torus-fig}~\cite{Julicher}.
One finds:
\begin{itemize}
\item A region of nonaxisymmetric tori ($T^{\rm na}$);
\item A region of discoidal tori ($T_{\rm Di}$);
\item A region of stomatoidal tori ($T_{\rm St}$);
\item A region of spheroidal tori ($T_{\rm Sp}$).
\end{itemize}
The last three zones have boundaries on the limit
curves: $SI_{\rm Di}$, where discoidal tori
auto-intersect, and $L_{\rm Di},$ $L_{\rm St},$
$L_{\rm Sp}$ where the diameter of axisymmetric
tori vanishes.
On the continuous lines which separate $T_{\rm Di}$
from $T_{\rm St}$ and this from $T_{\rm Sp}$ the shapes
cross over continuously from one family to the other.
All four families have the Clifford torus {\em Cliff\/}
in common.
Circular tori form the boundary between axisymmetric
and non axisymmetric shapes for $m>m_{\rm Cliff}=4\pi\times
2^{-\frac{3}{4}}\pi^{\frac{1}{2}}=4\pi\times 1.05.$
For smaller values of $m,$ this boundary is formed by
spheroidal tori.
Discoidal tori are equilibrium shapes only within the BC
model with zero spontaneous curvature.

\begin{figure}\begin{center}
\includegraphics[width=10cm]{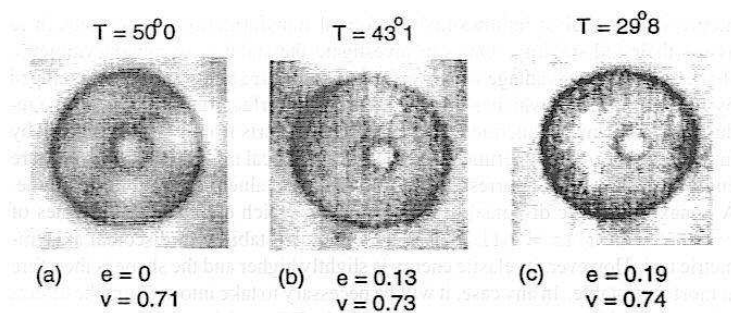}
\end{center}\caption{Breaking of axial symmetry of the Clifford torus
upon cooling. The transformation is reversible.
The parameter $e$ denotes the excentricity of the cyclid.
The bar corresponds to 10 $\mu$m. From X.~Michalet,
Thesis, Paris VII (1994).}
\label{Clifford}
\end{figure}

\begin{figure}\begin{center}
\includegraphics[width=10cm]{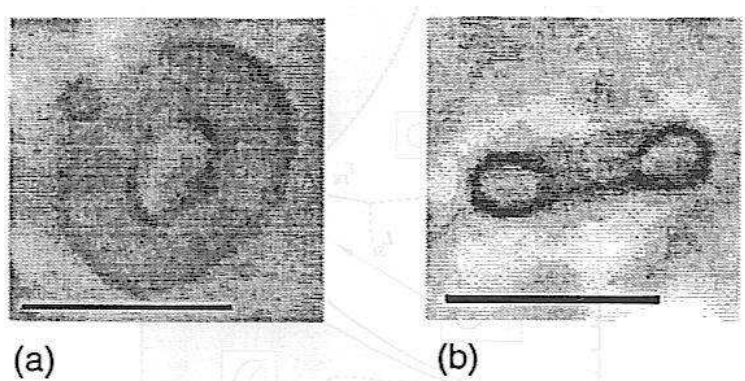}
\end{center}\caption{Axisymmetric discoidal tori. In (a) one clearly observes
the meridian sections characteristic of the discoidal family.
Bar: 10 $\mu$m. From X.~Michalet, Thesis, Paris VII (1994).}
\label{discoidal}
\end{figure}

Experimentally~\cite{Xavier} one finds indeed that vesicles
having the Clifford torus shape turn into non axisymmetric
shapes (close to Dupin cyclids) upon cooling (which induces
a decrease of the bilayer area, and thus an increase
in reduced volume) (see fig.~\ref{Clifford}).
One can also observe axisymmetric discoidal tori, supporting
the idea that the vesicles are described by the BC model---although
one could not rule out the possibility that these shapes are merely
metastable (see fig.~\ref{discoidal}).
However, one can also observe nonaxisymmetric discoidal tori,
which do not appear as stable shapes in the phase diagram of the
BC model with vanishing spontaneous curvature.
These shapes cannot be obtained as conformal transforms of
a stable axisymmetric shape.
This indicates that the stability of axisymmetric shapes
against infinitesimal conformal transformations is not
enough to assess their full stability.
One can investigate the stability of nonaxisymmetric
shapes by taking advantage
of powerful program {\tt Surface Evolver} developed by
K.~A.~Brakke~\cite{Evolver}, which, starting from a given
surface shape, lets it evolve reducing the Willmore functional
at each step. One starts from a shape obtained by image
analysis of the experiment and looks for the local minimum
of the curvature energy near to it---and corresponding to
the same value of the reduced volume. A nonaxisymmetric
discoidal torus is obtained, which corresponds to values
of $v=0.52$ and $m/4\pi=1.11,$ well in the region
of stability of discoidal axisymmetric tori. However,
its elastic energy is slightly higher and the shape is therefore
at most metastable. In any case, it will be necessary to take
into account the effects of a nonvanishing spontaneous curvature
in the BC model.

Vesicles of higher topology have more recently been observed.
The important new fact is that for some values of the parameters
$v$ and $m$ there is a one-parameter family of conformal
transformations which conserves the Willmore functional
and satisfies at the same time the constraints~\cite{Julicher}.

\begin{figure}\begin{center}
\includegraphics[width=10cm]{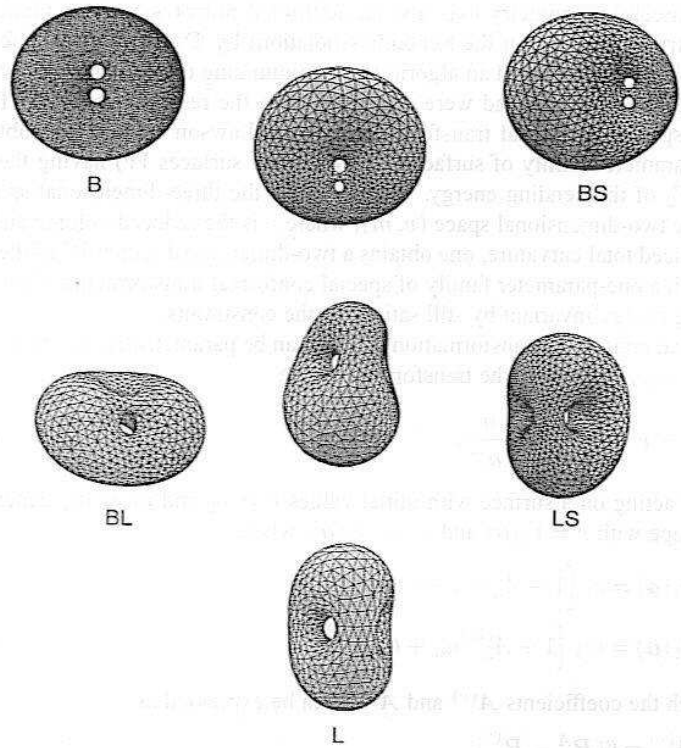}
\end{center}\caption{Willmore surfaces
with bending energies $F=F_2\simeq 1.75\times 8\pi \kappa.$
These shapes correspond to points
at the boundary of the region ${\cal W}$:
The ${\cal D}_{3h}$ symmetric Lawson surface L;
the ${\cal D}_{2h}$ symmetric button surface B;
and examples of ${\cal C}_{2v}$ symmetric shapes
along the lines $C_{BS},$ $C_{BL},$ and $C_{LS},$ respectively.
{}From F. J\"ulicher, U. Seifert, R. Lipowsky,
{\em Phys.\ Rev.\ Lett.}\ {\bf 71} 452 (1993).}
\label{Lawson-fig}
\end{figure}

By numerically minimizing a discretized version
of the Helfrich hamiltonian, Hsu, Kusner and Sullivan found that
the Lawson surface L shown in fig.~\ref{Lawson-fig} corresponds
to a minimum, with a value $F=F_2\simeq 1.742\times 8\pi\kappa$~\cite{Hsu}.
This surface has a threefold symmetry axis and an
additional mirror symmetry plane:
this symmetry is denoted (in the Sch\"onflies notation) by ${\cal D}_{3h}.$
J\"ulicher, Seifert and Lipowsky~\cite{Julicher} developed an algorithm
for minimizing the bending energy $F$ for a triangulated surface,
and were able to confirm the result of ref.~\cite{Hsu}.
By applying special conformal transformations to the Lawson surface,
one obtains a three-parameter family of surfaces
(the Willmore surfaces ${\cal W}$) having the same value $F_2$ of the bending
energy. By projecting the three-dimensional space ${\cal W}$
onto the two-dimensional space $(v,m),$ where $v$ is the reduced volume
and $m$ is the reduced total curvature, one
obtains a two-dimensional region $W$ of the plane in which a
one-parameter family of special conformal transformations leaves
the bending energy invariant by still satisfying the constraints.

Special conformal transformations (SCT) can be parametrized by a vector
$\vec{a}=(a_x,a_y,a_z),$ defined by the transformation rule
\begin{equation}
\vec{r}\to \vec{r}'=\frac{\vec{r}/r^2+
\vec{a}}{\left(\vec{r}/r^2
+\vec{a}\right)^2}.
\end{equation}
A SCT acting on a surface with initial values
$v=v_1$ and $m=m_1$ generates a new shape with
$v=v_1(\vec{a})$ and $m=m_1(\vec{a}),$ where
\begin{eqnarray}
v_1(\vec{a})&=&v_1\left[1+A^{(v)}_{\alpha}a_{\alpha}+O(\vec{a}^2)\right],\\
m_1(\vec{a})&=&m_1\left[1+A^{(m)}_{\alpha}a_{\alpha}+0(\vec{a}^2)\right],
\end{eqnarray}
in which the coefficients $\vec{A}^{(v)}$
and $\vec{A}^{(m)}$ can be expressed as
\begin{eqnarray}
\vec{A}^{(v)}&=&6(\vec{R}^A-\vec{R}^V),\\
\vec{A}^{(m)}&=&2(\vec{R}^A-\vec{R}^M),
\end{eqnarray}
in terms of the center of volume $\vec{R}^V=\oint\d V\vec{R}/V,$
the center of area $\vec{R}^A=\oint\d A
\vec{R}/A,$ and the center of mean curvature
$\vec{R}^M=\oint\d A H\vec{R}/M.$ Thus, the conformal mode which
conserves both $v$ and $m$ can be identified as the
SCT with $\vec{a}$ obeying the differential equation
\begin{equation}
\frac{\d\vec{a}}{\d s}=\vec{A}^{(v)}\times\vec{A}^{(m)},
\end{equation}
where $s$ parametrizes the path in the space ${\cal W}.$

The boundaries of the region $W$ can be determined
by first introducing the button surface B shown in ~\ref{Lawson-fig}.
This surface is conformally equivalent to the
Lawson surface L and corresponds to
$v=0.66,$ and $m=1.084\times 4\pi.$ It has three
orthogonal symmetry planes, i.e., symmetry ${\cal D}_{2h}.$
Choose the $(x,y)$ plane to be the midplane of the disk,
with the centers of the two holes along the $x$ axis.
By applying SCT one can define a line $C_{BL}$
of conformally equivalent surfaces of symmetry ${\cal C}_{2v}$ which connect
B to L. This line is defined by $\vec{a}=(0,0,a_z)$ with $0\le a_z\le 3.4/R_0.$
A further increase in $a_z$ breaks the threefold symmetry of L, generating
the line $C_{LS}$ with ${\cal C}_{2v}$-symmetric shapes:
for $a_z=15.5/R_0,$ the shape along this
approaches a sphere with two infinitesimal handles at
$(v,m)=(1,4\pi).$ If on the other hand we break the
$x$-$z$ symmetry plane of B by a SCT transformation with
$\vec{a}=(0,a_y,0),$ we obtain the line $C_{BS}$
which also approaches a sphere at $(v,m)=(1,4\pi).$
The three lines $C_{BL},$ $C_{LS}$ and $C_{BS}$ form the boundary
of the region $W$.

\begin{figure}\begin{center}
\includegraphics[width=10cm]{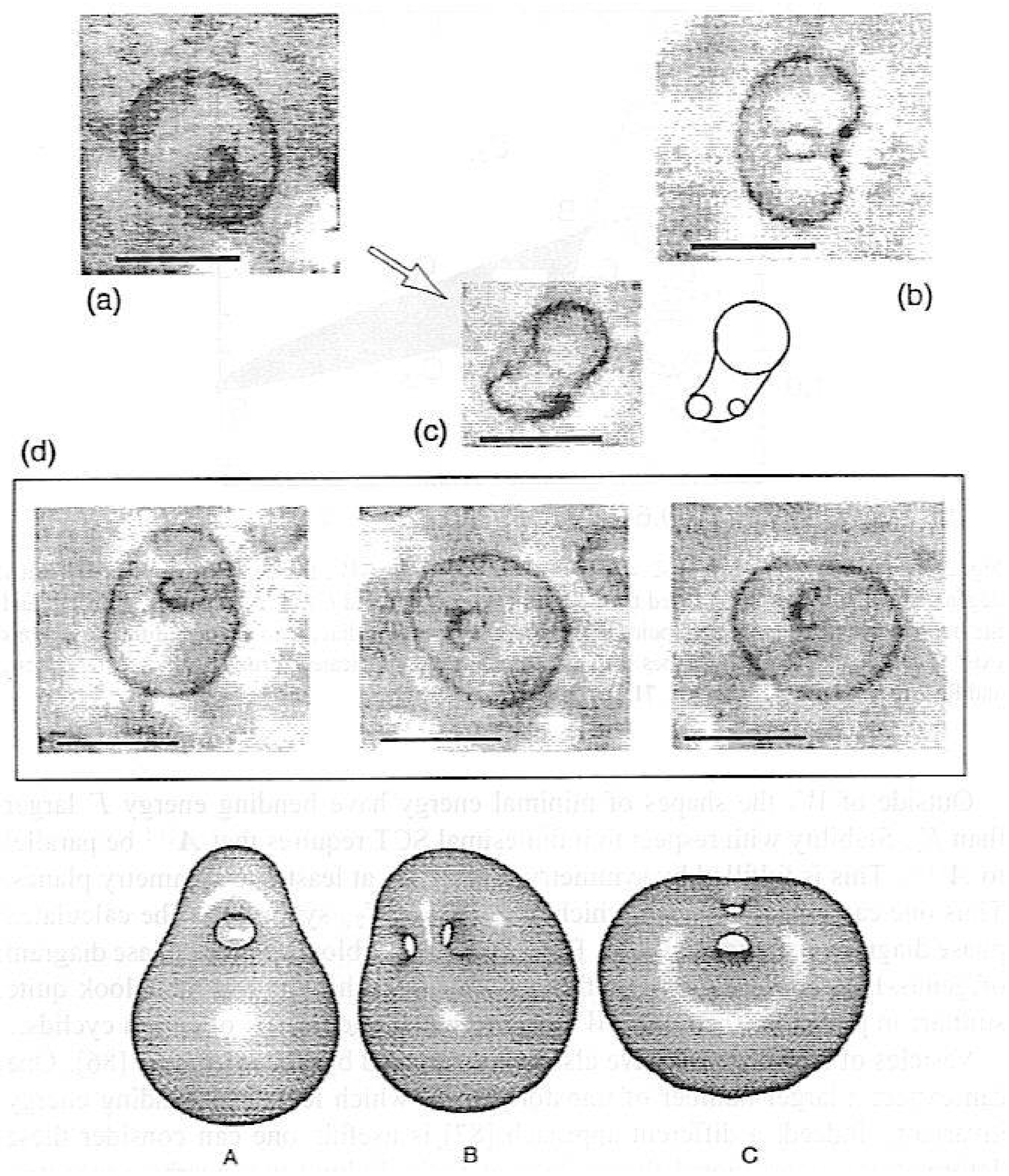}
\end{center}\caption{Example of a vesicle of type BS undergoing
conformal diffusion. Bar: $10\mu{\rm m}.$
(a): View from ``above'', showing the absence
of symmetry planes normal to the focal plane.
A small rotation leads to view (b) and eventually to view (c).
The series of three views (d) shows the vesicle at intervals
of a few seconds. From X.~Michalet, Thesis,
Paris VII (1994).}
\label{diffusion-fig}
\end{figure}

Within this region one should be able to observe fluctuations
of the shape of the vesicle among conformal transforms:
this phenomenon has been named {\em conformal diffusion\/}
and has first been observed
by Michalet and collaborators~\cite{Xavier}.
Examples of this behavior are shown in fig.~\ref{diffusion-fig}.
There is of course the delicate experimental problem of
discriminating between conformal diffusion and ordinary thermal
fluctuations.

\begin{figure}\begin{center}
\includegraphics[width=8cm]{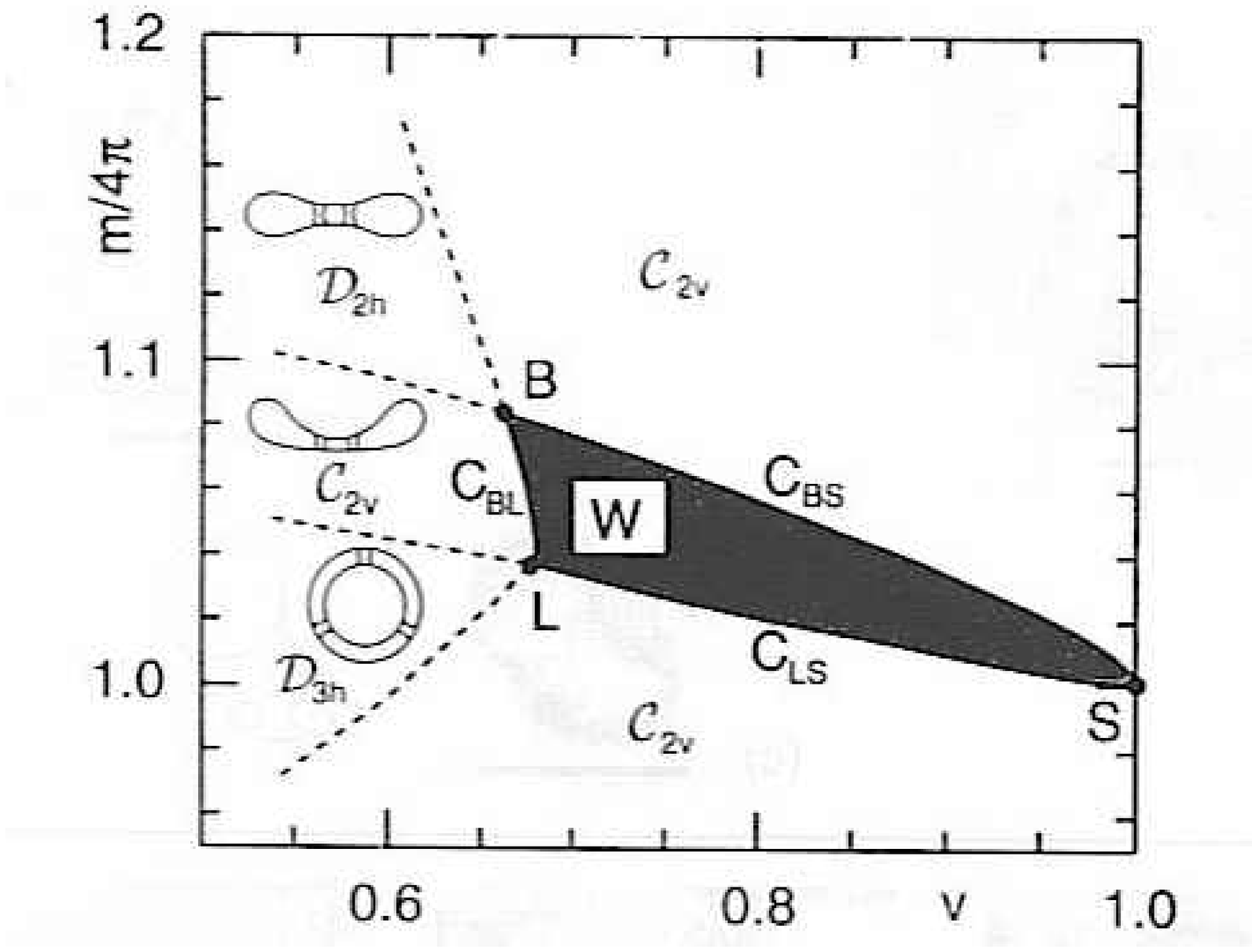}
\end{center}\caption{Phase diagram of genus-2 vesicles.
Within the region $W$, the ground state is conformally
degenerate.
This region is bounded by the lines $C_{BS},$ $C_{BL},$ and
$C_{LS}.$ The Lawson surface
L and the button surface B
are special point at the boundary of $W.$ Adjacent to
$W,$ five different regions exist. The symmetries
of the shapes within these regions are
indicated. From
F.~J\"ulicher, U.~Seifert, R.~Lipowsky,
{\em Phys.\ Rev.\ Lett.}\ {\bf 71} 452 (1993).}
\label{Julicher-fig}
\end{figure}

\begin{figure}\begin{center}
\includegraphics[width=8cm]{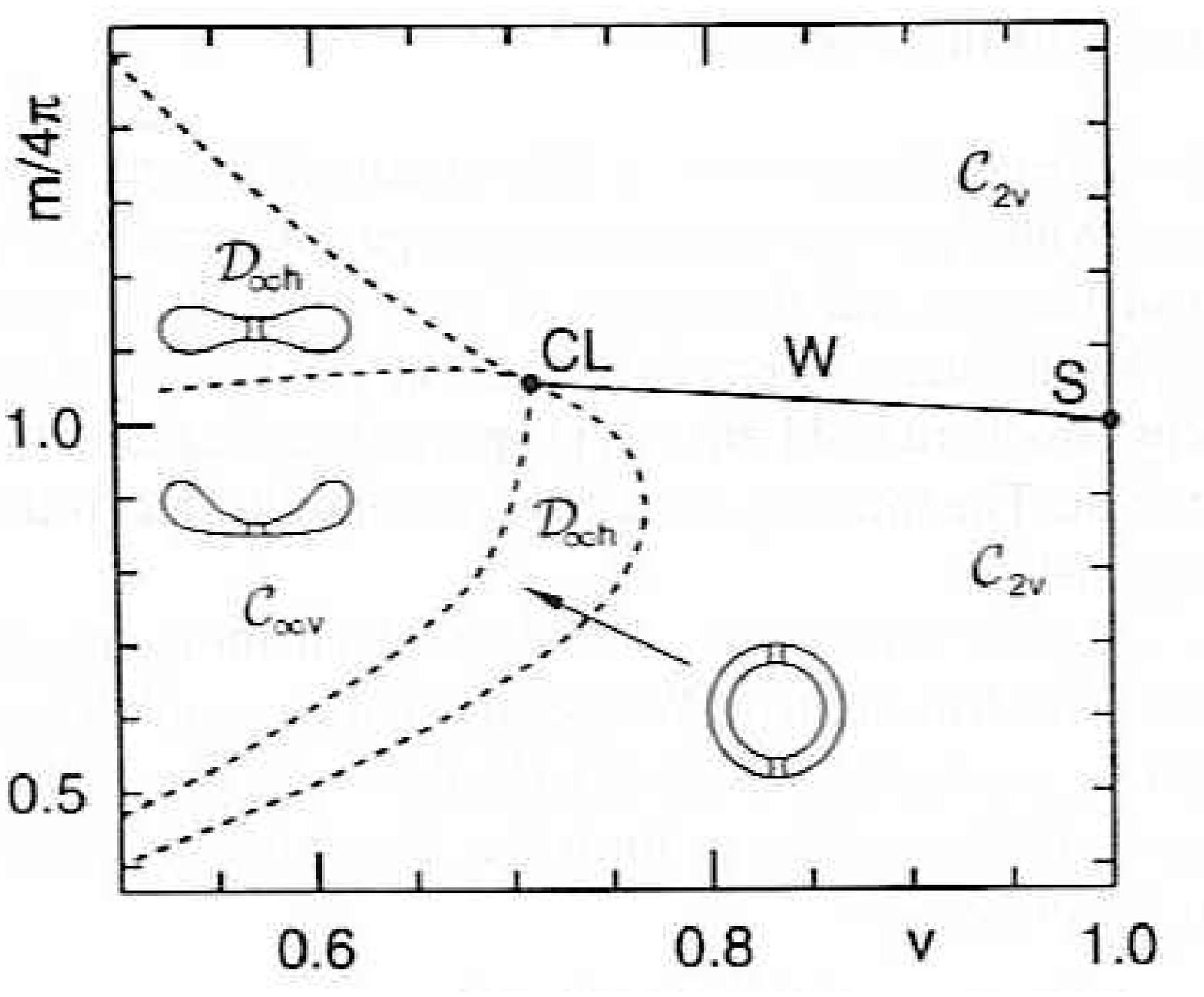}
\end{center}\caption{Phase diagram for genus-1 vesicles. The line
$W$ denotes the Willmore surfaces (Dupin cyclids) starting
at the Clifford torus CL. Compare
with the previous figure.}
\label{Julicher2-fig}
\end{figure}

Outside of $W$, the shapes of minimal energy have bending
energy $F$ larger than $F_2.$ Stability with
respect to infinitesimal SCT requires that $\vec{A}^{(v)}$
be parallel to $\vec{A}^{(m)}.$ This is fulfilled
by symmetry if there are at least two symmetry
planes. Thus one can look for shapes which have
at least ${\cal C}_{2v}$ symmetry.
The calculated phase diagram is shown in fig.~\ref{Julicher-fig}.
For comparison, a blow up of the phase diagram of genus-1
vesicles is shown in fig.~\ref{Julicher2-fig}.
One sees that the diagrams look quite similar: in
particular the region $W$ corresponds to the line $W$
of Dupin cyclids.

Vesicles of higher genus have also
been observed by Michalet et al.~\cite{Xavier}.
One can expect a larger number of transformations which leave
the bending energy invariant. Indeed, a different
approach~\cite{Michalet94} is useful: one can consider
these deformations as positional fluctuations
of necks linking two nearby concentric
membranes. The strategy is to consider the shape of
a neck of radius $a$ linking two square parallel
pieces of membrane of size $L\times L$
with periodic boundary conditions. The problem
breaks into a {\em inner\/} problem, in which
the surface can be assimilated to a {\em minimal surface\/}
(with vanishing mean curvature), and an {\em outer\/} problem
which can be solved via an electrostatic
analogy. The result is that
the necks behave as a gas of free particles
with a hard core repulsion of range $\ell_{\rm c}\simeq \sqrt{aL}.$
Therefore vesicles with a low density of necks will
fluctuate freely: only when two necks come nearby will they feel
the hard core repulsion.

\begin{figure}\begin{center}
\includegraphics[width=10cm]{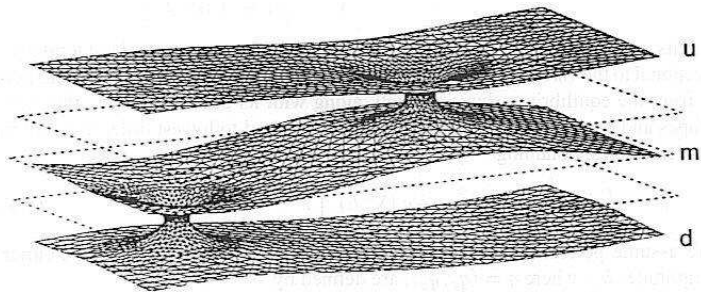}
\end{center}\caption{Example of a structure with $M=3$ membranes and $N=2$
necks connecting them. Periodic boundary conditions are
understood. From X.~Michalet, D.~Bensimon, B.~Fourcade,
{\em Phys.~Rev.~Lett.}\ {\bf 72} 168 (1994).}\label{double-fig}
\end{figure}

The approach can be generalized to the case of $M$ membranes
connected by $N$ necks: an example with $M=3$ and
$N=2$ is shown in fig.~\ref{double-fig}.
In the limit
where $M\to\infty$ and $N\to\infty$ one recovers the sponge phase.

We have thus seen that the intimations of the Helfrich
hamiltonian on vesicle shapes, even at the mean field level,
reveal an unsuspected richness.

\section{Shape fluctuations in vesicles}\setcounter{equation}{0}
Red blood cells suspended
in solution exhibit a remarkable {\em flicker
pheno\-men\-on}, which is best seen with a phase
contrast microscope and appears as a shimmering at the junction of
the rim and the center of the cells. The origin
of this phenomenon was debated since its first observation
in 1890~\cite{Browicz} and until the quantitative
analysis by Brochard and~Lennon~\cite{Brochard}
proved beyond doubt that it was due to Brownian
motion. The intensity of the phenomenon is due
to the fact that the surface tension vanishes.

Indeed, let us estimate the amplitude of
shape fluctuations in a membrane described by the Helfrich
hamiltonian. We assume that the equilibrium shape of
the membrane is planar, and we take its plane to be
the $(x,y)$ plane. We represent the shape of the membrane
in the Monge form, i.e., by giving the third coordinate $z,$
as a function of the other two:
\begin{equation}
z=h(x,y).
\end{equation}
We set $\underline{\sigma}=(x,y).$ The hamiltonian then
takes the form:
\begin{equation}
F=\int \d^2\underline{\sigma}\sqrt{1+\left(\nabla h\right)^2}
\left[\gamma+\frac{1}{2}\kappa\left(\nabla\cdot\frac{\nabla h}{\sqrt{1
+\left(\nabla h\right)^2}}\right)^2\right].
\end{equation}
In this equation $\nabla$
denotes the two-dimensional nabla operator,
and the term proportional to the Gaussian curvature is
understood. We assume that the deformation $h$ from
 the equilibrium shape is small, along with its derivatives:
 i.e., that both slopes and curvatures are small.
We can then expand to lowest order
in $h$ and in its derivatives, obtaining
\begin{equation}
F\simeq\int \d^2\underline{\sigma} \left[\frac{1}{2}\gamma\left(\nabla h
\right)^2+\frac{1}{2}\kappa\left(\nabla^2 h\right)^2\right].
\end{equation}
We assume periodic boundary conditions
on a square of side $L$.
The Fourier amplitudes
$h_{\underline{q}},$ where $\underline{q}=(q_x,q_y),$ are defined by
\begin{equation}
h_{\underline{q}}=\int\d^2\underline{q}{\rm e}^{{\rm i}
\underline{q}\cdot\underline{\sigma}}
h({\underline{\sigma}}).
\end{equation}
In terms of these amplitudes, the
Helfrich hamiltonian becomes
\begin{equation}
F=\frac{1}{2L^2}\sum_{\underline{q}}\left[\gamma q^2+\kappa q^4\right]\left|
h_{\underline{q}}\right|^2.
\end{equation}
By the equipartition theorem we have
\begin{equation}
\left<\left|h_{\underline{q}}\right|^2\right>=
L^2\frac{k_{\rm B}T}{\gamma q^2
+\kappa q^4}.
\end{equation}

We see that, if the surface tension vanishes,
the fluctuation amplitude diverges, at small $q,$ like
$q^{-4}.$ The finite size of the membrane
imposes a low $q$ cutoff
at $\pi/L.$ We can then integrate the former expression to
get an estimate of the fluctuation amplitude:
\begin{equation}
\left<h^2\right>=\frac{1}{(2\pi)^2L^2}
\int_{\pi/L}^{\infty}\d^2\underline{q}
\left<\left|h_{\underline{q}}\right|^2\right>=
L^2\frac{k_{\rm B}T}{4\pi^3\kappa}.
\end{equation}
The square amplitude diverges like $L^2$. If the
surface tension had not vanished, we would have had a much weaker
(logarithmic) divergence. As an order of magnitude, since
$\kappa\sim 10^{-19}$--$10^{-20}\,{\rm J}$, taking $L\sim 8\mu{\rm m}$
as for human red blood cells, we obtain $\sqrt{\left<h^2\right>}\sim
0.4$--$0.7 \,\mu{\rm m}.$

More generally, let us consider the fluctuations of a $D$-dimensional
surface in the Monge representation, governed by a generic Hamiltonian
of the form
\begin{equation}
{\cal H}\propto \int \d^D\underline{\sigma}
\left(\nabla^{\omega}h\right)^2.
\end{equation}
For the case of the rigidity-dominated
membranes we have $\omega=2,$
whereas for the case of an ordinary interface with a
nonvanishing surface tension we would have $\omega=1.$
The {\em wandering exponent\/} $\zeta$ describes
the behavior of the excursions of the membrane as a function of
its lateral size $\left<h^2\right>\propto L^{2\zeta}$ or,
equivalently, the behavior of the height-height correlation function:
\begin{equation}
\left<\left(h(0)-h(\underline{\sigma})\right)^2\right>\propto \left|
\underline{\sigma}\right|^{2\zeta}.
\end{equation}
We obtain
\begin{equation}
\zeta=\frac{1}{2}\left(2\omega-D\right).
\end{equation}
We obtain therefore $\zeta=1$
for the case of membranes (whereas $\zeta=0$
for interfaces with nonvanishing surface tension),
which implies that the aspect
ratio of the fluctuations over lateral size does not decrease
with increasing size. This means that membranes should appear
``wobbled'' in the same way even at the largest scales. As we shall
see, the situation is even worse.

Helfrich~\cite{Helfrich85} recognized the important effects
of membrane fluctuations. On the one hand,
the projected area $A_{\rm p}$ of the membrane
is reduced with respect to its true area $A$ (see eq.~(\ref{area:eq})
below).
Moreover, the wandering
of the membrane from its equilibrium
configuration
implies the existence of a long-range
repulsion between undulating membranes, as it had already been
pointed out in 1978
by Helfrich~\cite{Helfrich78}. To fix one's ideas, let us consider a
membrane constrained by two parallel, planar walls, set at a distance
$2d$ apart. The membrane is on average in the
middle, but it collides with either wall from time to time.
Collisions are separated by a typical length $L_{\rm coll}\sim d^{1/\zeta},$
where $\zeta$ is the wandering exponent. Therefore the free energy
per unit area will have a repulsive contribution proportional
to the density of the collisions, i.e., to $L_{\rm coll}^{-D}
\sim d^{-D/\zeta}.$
This ``steric repulsion'' term decreases like $d^{-2}$ for membranes,
i.e., just like the van der Waals attraction at short distances.
One can thus expect in principles regimes in which van der Waals forces
dominate, and parallel membranes are bound to each other, or
in which steric repulsion dominates, and they repel each other.
The transition between the two regimes is called {\em unbinding\/}:
although it is not described by this simple argument, as we shall see later,
it has been observed in actual membranes.
Finally, he recognized that the rigidity modulus $\kappa$ itself
is renormalized by the fluctuations, and becomes smaller and
smaller for larger and larger membranes~\cite{Helfrich85}.

A more rigorous analysis of shape fluctuations in vesicles requires one
to take into account the fact that the equilibrium shape is not planar.
The necessary techniques were developed by Peterson~\cite{Peterson85}.
These techniques have been recently applied to a reappraisal of the
flicker phenomenon in red blood cells by
Peterson, Strey, and Sackmann~\cite{Peterson92}.
In general one would like to compute the variation of
the elastic free energy $F$ under a slight deformation
of the vesicle shape, starting from a reference {\em equilibrium\/}
shape $S_0.$ It is assumed that the deformation satisfies the
constraints that define the ensemble.
There are many ways to parametrize the deformation: the
most convenient choice is the {\em normal gauge,} in which
the deformation is described by the distance $h$ of the deformed
surface $S$ from the reference one $S_0,$ measured along the
normal to $S_0$ at each point.
One then computes the variation of $F$ to second order in
the deformation:
\begin{equation}
\delta^2F[h]=\frac{1}{2}\int\d\underline{\sigma}
\d\underline{\sigma}'
\left.\frac{\delta^2F}{\delta h(\underline{\sigma})
\delta h(\underline{\sigma}')}\right|_{h=0}
h(\underline{\sigma})h(\underline{\sigma}').\label{Peterson:eq}
\end{equation}
The probability of a given deformation is proportional, in the Gaussian
approximation, to $\exp[-\delta^2 F/k_{\rm B}T].$
The constraints can be taken into account in the following way.
The deformation $h$ is splitted into a first-order term $h_1,$
which satisfies the constraints to first order,
and a second-order term $h_2,$ which is chosen to enforce
the constraints to second order. The first-order
term $h_1$ parametrizes the deformation, whereas
the variation of the elastic energy is given by the
second-order variation of $F$ times $h_1^2$ plus
the first order variation multiplied by the second-order
term $h_2.$ Equivalently, one may consider adjusting
the Lagrange multipliers in order to keep the constraints
satisfied to second order.

\begin{figure}\begin{center}
\includegraphics[width=10cm]{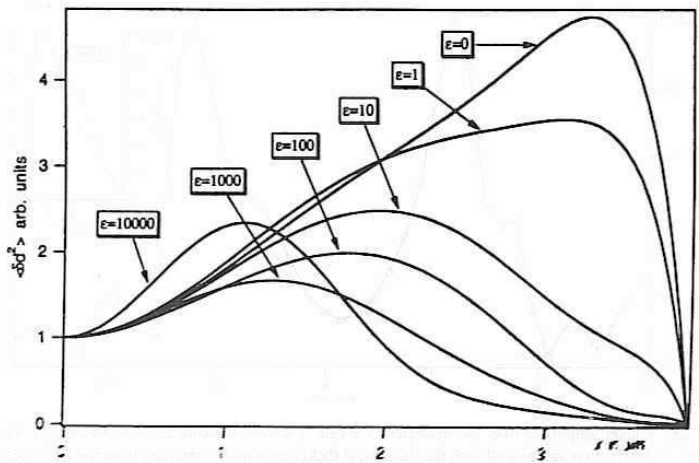}
\end{center}\caption{Thickness fluctuation profile for various values
of the ratio $\epsilon=\mu R^2_0/\kappa.$ From M.~A.~Peterson,
H.~Strey, E.~Sackmann, {\em J.~Phys.~II France\/} {\bf 2} 1273 (1992).}
\label{blood:fig}
\end{figure}

\begin{figure}\begin{center}
\includegraphics[width=10cm]{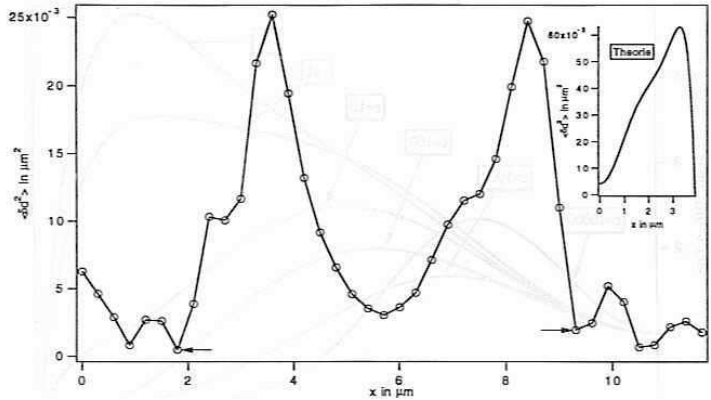}
\end{center}\caption{Flicker amplitude over the diameter of a cell.
(Arrows indicate the cell boundary.) The observed distribution agrees
well with the theoretical flicker amplitudes
obtained from the BC model. From M. A. Peterson, H.~Strey,
E. Sackmann, {\em J.~Phys.~II France\/}
{\bf 2} 1273 (1992).}
\label{blood:exp}
\end{figure}

Once the quadratic variation $\delta^2F$ of $F$ is obtained, one
diagonalizes it by taking advantage of its symmetries:
in the case of the red blood cells, the axial symmetry
and the up-down symmetry. It is an extremely stringent
check on the calculation that all the rigid Euclidean
motions emerge as zero energy modes. The amplitudes
of the fluctuating modes are obtained from the
equipartition theorem, and are then summed up
to obtain the fluctuation profile.
Since one cannot rule out in principle the existence
of a shear elasticity modulus $\mu$ in red blood cells,
due to the loose ankyrin-spectrin network present on
the interior of the membrane, the formalism
requires a slight generalization~\cite{Peterson85,Peterson92}.
The fluctuation profile then depends on the dimensionless
parameter $\epsilon=\mu R_0^2/\kappa.$ In fig.~\ref{blood:fig}
the thickness fluctuation profile is shown as a function of this parameter.
We see that, as the shear elasticity is reduced, the fluctuation
maximum moves toward the rim. At small values of the shear modulus,
this effect is more pronounced in the BC model than in the SC model.
Experimental results are shown in fig.~\ref{blood:exp}. They are
compatible with a very small value of the shear modulus $\mu,$
with the predictions of the BC model, and with a value
of $\kappa=1.4\times 10^{-19}$ J. This value is of the
order of that of pure lipid bilayers~\cite{Duwe},
but a factor two smaller than that of a mixture
of a physiological lipid and cholesterol~\cite{Duwe,Evans90}.
It is possible that  the discrepancy is due to the presence of small
proteins~\cite{Leibler86}, since red blood cell membranes
contain about 50\% proteins by weight, covering about 20\% of the area.

In order to be able
to go beyond the Gaussian approximation,
we now need to pay a closer look at fluctuations.
The most convenient framework introduces the
{\em effective potential\/} $\Gamma[\vec{r}_0]$
and the renormalization group \cite{Peliti85,Kleinert86,%
David89,Guitter89}.
Let us consider an fluctuating membrane, whose
instantaneous configuration is parametrically represented
by $\vec{r}=\vec{r}(\usigma),$ $\vec{r}\in{\bf R}^d$
and $\usigma\in{\bf R}^2.$
The free energy ${\cal F}$ of the membrane is given
in terms of the partition function
\begin{equation}
{\cal F}=-\kt\log\int{\cal D}\vec{r}
\exp\left\{-\frac{F[\vec{r}]}{\kt}\right\},
\end{equation}
where $F[\vec{r}]$ is the Helfrich hamiltonian
\begin{equation}
F=\oint_S\d A\left[\tau_0+\frac{\kappa_0}{2}H^2
+\overline{\kappa}_0K\right].
\end{equation}
If we want to assign the membrane a given average shape
$\vec{r}_0(\underline{\sigma}),$
we can introduce a fictitious external field $\vec{\lambda},$
coupling to the position $\vec{r},$ such that
$\left<\vec{r}(\usigma)\right>_{\lambda}=
\vec{r}_0(\usigma).$
The effective potential $\Gamma[\vec{r}_0]$
is the defined
as the Legendre transform of ${\cal F}$
with respect to $\lambda$:
\begin{equation}
\Gamma[\vec{r}_0]={\cal F}-\int\d A\,
\vec{\lambda}\cdot\vec{r}_0(\usigma).
\end{equation}
(A precise definition needs to take into account
reparametrization invariance.)

Euclidean and reparametrization invariances
dictate the form of the effective potential, just
as they do for the Helfrich hamiltonian~\cite{Peliti85}. Therefore
\begin{equation}
\Gamma[\vec{r}_0]=\int \d\bar{A}\left[
\tau_{\rm eff}+\frac{\kappa_{\rm eff}}{2}\bar{H}^2+
\overline{\kappa}_{\rm eff}\bar{K}+\ldots\right],
\end{equation}
where $\bar{A}$ is the area element of the
{\em average\/} membrane shape,
and $\bar{H},$ $\bar{K}$ its curvatures.
In this formula we have neglected
terms which depend on higher derivatives of the
average shape.
This expression defines the effective parameters
$\tau_{\rm eff},$ $\kappa_{{\rm eff}}$ and
$\overline{\kappa}_{{\rm eff}}.$
If the equilibrium shape is flat, all curvature
terms vanish at equilibrium, so that
\begin{equation}
\Gamma_{\rm eq}=\tau_{\rm eff}A_{\rm p}.
\end{equation}
Therefore $\tau_{\rm eff}$ is the effective frame
tension of the membrane.

Now, the effective frame tension $\tau_{\rm eff},$
must be equal to the ``$q^2$ coefficient,'' $\gamma,$
of the two-point height correlation
function of the membrane,
defined by
\begin{equation}
\left<|h_{\underline{q}}|^2\right>=\frac{1}{\gamma q^2+O(q^4)}.
\end{equation}

It is sufficient to recall that the effective
potential $\Gamma$ is also the generating functional
of the vertices, and in particular its second derivative
yields the inverse of the propagator. By going over
to the Monge representation of $\vec{r}_0$
we obtain
\begin{equation}
\left<|h_{\underline{q}}|^2\right>^{-1}=\left.
\frac{\delta^2\Gamma}{\delta h_{-\underline{q}}\delta
h_{\underline{q}}}\right|_{h=0}.\end{equation}
We obtain therefore
$\gamma=\tau_{\rm eff}.$
In a similar way, the effective rigidity
$\kappa_{\rm eff}$ can be related to
the coefficient of $\left(\nabla^2\bar{h}\right)^2$ in the
effective potential.

If we wish to investigate the behavior of the membrane at
larger and larger distances, we can apply the renormalization
group approach. The effective Wilson hamiltonian
${\cal H}_{\rm eff}(s)$ is obtained by
integrating out all fluctuations whose
wavenumbers $\underline{q}$ are
contained in a shell $\Lambda/s<\left|\underline{q}
\right|<\Lambda,$ where $s>1.$
Again by Euclidean symmetry, ${\cal H}_{\rm eff}(s)$
can be expanded in the form
\begin{equation}
{\cal H}_{\rm eff}(s)=\int\d A\left[
\tau_{\rm eff}(s)+\frac{\kappa_{\rm eff}(s)}{2}H^2+
\overline{\kappa}_{\rm eff}(s)K+\ldots\right].
\end{equation}
We then rescale the lengths in order
to bring the upper cutoff back to $\Lambda$:
\begin{equation}
\vec{r}\to s\,\vec{r},\qquad\underline{q}\to s^{-1}\underline{q},
\end{equation}
yielding the renormalized Hamiltonian
\begin{equation}
{\cal H}(s)=\int\d A\left[
\tau(s)+\frac{\kappa(s)}{2}H^2+
\overline{\kappa}(s)K+\ldots\right].
\end{equation}
Since the ``order parameter'' in our case is the length $h,$
there is no need of the additional ``wavefunction''
rescaling, which usually appears in critical phenomena.
The relation between the effective parameters and the
renormalized ones is simply
\begin{equation}
\kappa(s)=\kappa_{\rm eff}(s);\qquad \overline{\kappa}(s)=
\overline{\kappa}_{\rm eff}(s);\qquad \tau(s)=s^2\tau_{\rm eff}(s).
\end{equation}

The renormalized parameters are
actually computed by integrating the
renormalization flow equations, which are obtained by
considering $s=1+\epsilon,$ where $\epsilon$ is infinitesimal.
By the definitions of the effective potential
$\Gamma$ and of the effective Wilson hamiltonian
${\cal H}_{\rm eff}(s)$ it is clear that
\begin{equation}
\lim_{s\to\infty}{\cal H}_{\rm eff}(s)=\Gamma.
\end{equation}
Now, if we are considering an ensemble in which the
total area $A$ of the membrane fluctuates and
a frame tension $\tau$ is applied to it,
we have to choose the initial condition $(\gamma_0,\kappa_0,%
\overline{\kappa}_0)$ of the renormalization trajectory
in such a way that
\begin{equation}
\lim_{s\to\infty}\tau_{\rm eff}(s)=\tau.
\end{equation}
We see therefore that the essential step of the
renormalization group calculation is the
evaluation of ${\cal H}_{\rm eff}(s),$
where $s=1+\epsilon.$ This is quite similar
to the calculation of the effective potential, only
that one restricts the wavenumbers $\underline{q}$ of the
fluctuations to be integrated out to
the shell $\Lambda/s<|q|<\Lambda.$
This calculation can be performed perturbatively.
One chooses as a reference configuration
the equilibrium one $\vec{r}_0$ and expands in
the small (normal) deformation field $\vec{h}=\vec{r}-\vec{r}_0.$
The effective Wilson hamiltonian ${\cal H}_{\rm eff}(s)$
can be then expanded in powers of the temperature:
\begin{equation}
{\cal H}_{\rm eff}(s)=\sum_{\ell}{\cal H}_{\rm eff}^{\ell}(s)
\left(k_{\rm \beta}T\right)^{\ell}.
\end{equation}
If one stops at one loop level, it is sufficient
to consider only the second variation $\delta^2$ of the Helfrich
hamiltonian with respect to the deformation field
$\vec{h}=\vec{r}-\vec{r}_0.$
The result reads
\begin{equation}
{\cal H}_{\rm eff}(s)=F[\vec{r}_0]-\kt\log\int_s
{\cal D}\vec{h}\exp\left(-\frac{\delta^2F}{\kt}\right).
\end{equation}
The notation $\int_s$ reminds the constraint on the fluctuations.

The measure factor ${\cal D}\vec{h}$ has to be considered with
some\break care \cite{David89,Guitter89,tension,Cai}, and has
been discussed in the seminar by
Thomas Powers in this School~\cite{Powers}.
In principle it contains two contributions: one (known as the
Faddeev-Popov determinant) which takes into account
the volume occupied by different
reparametrizations of the same surface.
It is defined and discussed in Appendix B.
The other (discussed in particular in refs.~\cite{Cai,Powers}),
which takes into account the fact that the actual number of
degrees of freedom involved in the fluctuating
surface is proportional to the number of molecules, i.e., to the
actual area element $\d A$.
It turns out, however,
that one can simplify the problem (to the level of
not worrying at all about these contributions) if
one works in the {\em normal gauge.}
\begin{figure}
\begin{center}\includegraphics[width=8cm]{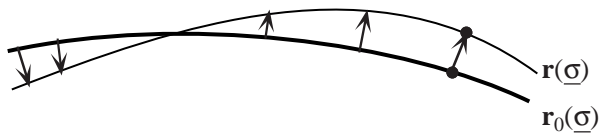}\end{center}
\caption{Definition of the normal gauge. Given
a reference shape represented as
$\vec{r}_0(\underline{\sigma}),$
the deformation is represented by the displacement
$\vec{h}=\vec{r}(\underline{\sigma})-
\vec{r}_0(\underline{\sigma})$ along the normal
to the reference shape.}
\label{normal:fig}
\end{figure}
We consider momentarily membranes embedded
in $d$-dimensional space.
Let the reference shape be parametrically represented
by $\vec{r}_0=\vec{r}_0(\underline{\sigma}),$
where $\vec{r}_0\in{\bf R}^d.$
The instantaneous shape of the membrane is represented
by $\vec{r}=\vec{r}(\underline{\sigma}).$
In the {\em normal gauge\/} one imposes the condition
\begin{equation}
\left(\vec{r}(\underline{\sigma})-\vec{r}_0(\underline{\sigma})
\right)\cdot\partial_i\vec{r}_0(\underline{\sigma})=0,\qquad i=1,2.
\end{equation}
The displacement $\vec{h}(\underline{\sigma})
=\vec{r}(\underline{\sigma})-\vec{r}_0(\underline{\sigma})$
has therefore $d-2$ independent components.
In the normal gauge the Faddeev-Popov determinant $\Delta_{\rm FP}$
turns out to be local and to contribute only a
(divergent) shift to the effective
membrane tension $\gamma.$

The calculation of the renormalization group
flow to one loop is reported in Appendix C.
The result is the following~\cite{Helfrich85,Peliti85,%
Kleinert86,tension}:
\begin{eqnarray}
s\frac{\partial \kappa(s)}{\partial s}&=&
-\frac{3}{4\pi}\frac{k_{\rm B}T}{1+\tau(s)/\kappa\Lambda^2},\\
s\frac{\partial\overline{\kappa}(s)}{\partial s}&=&
\frac{5}{6\pi}\frac{k_{\rm B}T}{1+\tau(s)/\kappa\Lambda^2},\\
s\frac{\partial \tau(s)}{\partial s}&=&2\tau(s)+
\frac{\Lambda^2}{4\pi}k_{\rm B}T\log\left(
\frac{\kappa+\tau(s)/\Lambda^2}{k_{\rm B}T}\right).
\end{eqnarray}
The calculation is valid to one loop,
hence it only holds if the terms of order
$\kt$ are small. This implies either $\kappa>\kt$
or $\tau/\Lambda^2>\kt.$
Let us remark that these equations imply
that the bending rigidity $\kappa$ decreases
when one goes to larger and larger scales:
the membrane becomes more and more crumpled.
On the other hand, the Gaussian rigidity
$\overline{\kappa}$ {\em increases\/}:
although this has no effect
on the behavior of an isolated membrane (whose topology
is fixed), it is important as one considers
ensembles of fluctuating membranes,
as we shall do in the next section.

\begin{figure}
\begin{center}\includegraphics[width=8cm]{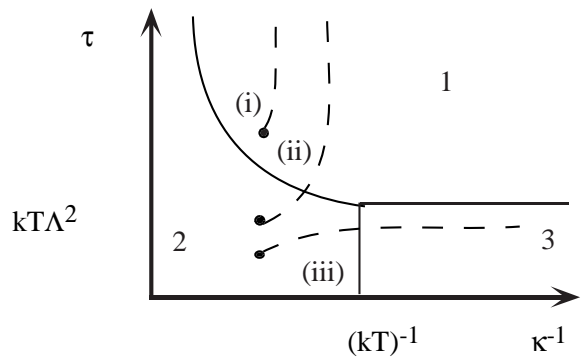}\end{center}
\caption{Different behaviors of fluid membranes with
respect to fluctuations in the $(\tau,\kappa^{-1})$
plane. 1: Tension-dominated region.
2: Rigidity-dominated region. 3: Fluctuation-dominated region.
Also shown are the renormalization
trajectories (i)--(iii) discussed below.}\label{renorm:fig}
\end{figure}
This perturbative result allows us to
define three different regions
characterizing the fluctuations of fluid membranes:
\begin{enumerate}
\item {\em Tension-dominated region:}
\begin{equation}
\tau>\kappa \Lambda^2;\qquad \tau>k_{\rm B}T\Lambda^2.
\end{equation}
In this region the fluctuations are small and
governed by tension $\tau$ (``drumhead
model''). The flow equations can be approximated by
\begin{equation}
s\frac{\partial \kappa}{\partial s}=s\frac{\partial \bar{\kappa}}{\partial s}
=0;\qquad s\frac{\partial \tau}{\partial s}=2 \tau.
\end{equation}
\item {\em Rigidity dominated region:}
\begin{equation}
\kappa>k_{\rm B}T;\qquad \tau<\kappa \Lambda^2.
\end{equation}
In this region fluctuations are small, but are dominated by
the rigidity term. One has the following approximate
flow equations:
\begin{eqnarray}
s\frac{\partial \kappa}{\partial s}&=&-\frac{3k_{\rm B}T}{4\pi},\\
s\frac{\partial \bar{\kappa}}{\partial s}&=&
\frac{5k_{\rm B}T}{6\pi},\\
s\frac{\partial \tau}{\partial s}&=&\tau
\left(2+\frac{k_{\rm B}T}{4\pi\kappa}
\right).
\end{eqnarray}
The renormalization of the surface term can be
attributed to an increase of the ratio {\em total area/projected area\/}
due to thermal fluctuations. Indeed, to leading order in
$k_{\rm B}T$ this ratio is given by
\begin{equation}
\frac{\left<A\right>}{A_{\rm p}}=1+\frac{k_{\rm B}T}{4\pi\kappa}
\log A_{\rm p}.\label{area:eq}
\end{equation}
\item {\em Thermal fluctuations dominated region:}
\begin{equation}
\kappa<k_{\rm B}T;\qquad \tau<k_{\rm B}T\Lambda^2.
\end{equation}
In this domain perturbation theory breaks down: one expects that
steric interactions and topology changes become important.
\end{enumerate}

As one considers the same membrane at increasing length scale
$\ell,$ the renormalization group flows will carry one from
one region to another. One can find different kinds
of trajectories. Starting from a given value $\kappa_0>\kt$
of the rigidity one can have:
\begin{itemize}
\item[(i)] If the surface tension $\tau_0$ is large enough, the
trajectory remains in the tension-dominated region at all scales,
and is described by the simple drumhead model like an ordinary
interface, with rigidity-induced corrections with the fixed
rigidity $\kappa_0$.
\item[(ii)] For smaller surface tensions, one may cross over from the
rigidity-dominated region to the tension-dominated region: it exists
therefore a crossover length $\ell_{\rm c}$ such that
for $\ell>\ell_{\rm c}$ one is in the tension-dominated regime
with some (nonzero) effective tension $\tau_{\rm eff}$
and an effective rigidity $\kappa_{\rm eff}.$ The crossover
length $\ell_{\rm c}$ will be given by
\begin{equation}
\ell_{\rm c}=\sqrt{\frac{\kappa_{\rm eff}}{\tau_{\rm eff}}}.
\end{equation}
For $\ell<\ell_{\rm c}$ one is in the rigidity-dominated
regime: tension effects can be neglected, and the effective
rigidity depends on the scale $\ell$:
\begin{equation}
\kappa(\ell)=\kappa_0-\frac{3k_{\rm B}T}{4\pi}\log\left(\frac{\ell}{\ell_0}
\right),\label{kappa:eq}
\end{equation}
where $\ell_0$ is the molecular size.
\item[(iii)] If the tension is too small, there is a length scale $\ell_{\rm
c}$
in which one crosses over to the fluctuation-dominated regime and the
model breaks down. This length can be defined by the condition
that $\kappa(\ell_{\rm c}),$ as defined by eq.~(\ref{kappa:eq}),
is of order $k_{\rm B}T.$ This length is known as the {\em persistence
length\/}~\cite{deGennes82} and depends exponentially
on the rigidity $\kappa_0$:
\begin{equation}
\ell_{\rm c}=\xi\simeq \ell_0
\exp\left(\frac{4\pi\kappa_0}{3k_{\rm B}T}\right).
\end{equation}
At larger scales the membrane will be crumpled, with
a correlation length for its normals
of the order of $\xi.$
\end{itemize}

Fran\c{c}ois David~and Emmanuel Guitter
\cite{David86,David88} have considered the
model with bending rigidity
in the large $d$ limit. The results can be described as
a phase diagram in the tension-rigidity plane. For sufficiently
large surface tension $\tau,$ the system is in a ``flat''
phase with a finite ratio of the projected area to the total
area. As one approaches a critical line $\tau_{\rm c}(\kappa_0),$
this ratio vanishes: on the other side of this line the surface
is crumpled and the system does not exist---unless one takes into account
steric interactions and topology changes. Along the critical
line the frame tension remains finite. However, even before
reaching the critical line, the system becomes instable with respect
to nonhomogeneous fluctuations. As one reaches the transition line,
the frame tension remains finite.

The effective model describing the membrane at long distances
is the Liouville action model~\cite{Polyakov81},
introduced by Polyakov in the context of strings.
It describes a Gaussian surface (defined by the vector
field $\vec{r}$), coupled to a fluctuating
intrinsic metric $g_{ij}.$ We discuss it briefly in
Appendix D. It is usually believed that this model also
applies to membranes embedded in finite-dimensional spaces.
However, in this case, self-avoiding effect
are probably essential in determining the actual
behavior of the membrane. Moreover, it
is generally believed~\cite{Froelich}
that these membranes should assume
configurations characteristic of
branched polymers.
Since this is a nonperturbative effect,
it is rather difficult to study it
analytically. Cates~\cite{Cates88}
argues that it is related to an
instability of the Liouville field theory
with respect to the formation of ``spikes''.

A way to control this instability in closed vesicles is to introduce
osmotic pressure. Leibler, Singh and Fisher~\cite{LSF}
verified by numerical studies that two-dimensional
self-avoiding vesicles change continuously from
a deflated state with the characteristics of
branched polymers to an inflated state as the
osmotic pressure $p$ is increased. This transition has
been further studied by real-space renormalization
group techniques~\cite{Maritan} and by conformal
group techniques~\cite{Cardy}, which have allowed to determine
its critical exponents in 2-D.
There have been fewer studies of the corresponding
transition in 3-D. In particular, Gompper and Kroll~\cite{GK}
and Baumg\"artner~\cite{Baum} have numerically shown that the
deflated-inflated transition is first-order in 3-D.
A finite-size scaling analysis of the
deflation-inflation transition has been recently
performed by Damman {\em et al.}~\cite{Dammann}.

I close this section by pointing out that the fact that the effective
rigidity decreases at larger and larger length scales can be
interpreted in terms of the Mermin-Wagner theorem~\cite{MW},
which states that a continuous symmetry cannot be spontaneously
broken in two dimensions by a local Hamiltonian. In our case
the continuous symmetry is the Euclidean one, which would be
broken for our two-dimensional
membranes if they exhibited a flat phase.

\section{Interacting fluid membranes}\setcounter{equation}{0}
We now consider an {\em ensemble\/} of fluctuating, self-avoiding
membranes, following Huse and Leibler~\cite{HL}. The effects
of excluded volume and of the topological term $\okappa\int \d A\,
K$ are essential. We shall first
go across the phase diagram, and identify the phases
which may be present. We shall then examine
each of them more closely.

Let us first set
$\okappa=0,$ thus effectively neglecting the topological term
but allowing for changes in topology.

{}From a physical point of view, we are considering a solution
of amphiphile in water. We assume for the time being that
the bending rigidity $\kappa_0$ is quite large,
and neglect momentarily renormalization effects.
The ``bare'' surface tension
$\tau_0$ appearing in the Helfrich hamiltonian is proportional
to the chemical potential of the amphiphile. It can be controlled
by changing the amphiphile concentration $\phi.$
For $\tau_0$ large and positive, i.e., for a small concentration
$\phi$ of the amphiphile, the membrane breaks down into small
isolated vesicles: as a consequence, the volume is separated
into two components: one {\em inside\/} and one {\em outside\/}
of the droplets. We can assign ``spins'' to each component
(e.g., ``down'' inside and ``up'' outside) and we recognize
a spontaneous symmetry breaking between the two components.

In principle, we could imagine to force the spins to be up
on one side of the sample, and down on the other side. In
this case there should be a membrane running across the sample,
costing a free energy proportional to its section.
In this phase, therefore, there is a nonzero surface tension
$\tau$ for the membrane at a {\em macroscopic\/} level: Huse
and Leibler~\cite{HL} call it the {\em tense droplet\/} phase.

Increasing the amphiphile concentration (reducing $\tau_0$)
the volume enclosed by the droplets increases. The membranes
come closer to one another and are likely to form ``necks''
in order to increase entropy. Therefore the region of minority
spins becomes connected over longer and longer distances. Well
before reaching a volume fraction equal to one half, the
connected regions can percolate. We thus have two infinite
connected regions of unequal size. We have obtained
a {\em bicontinuous phase.} The symmetry between inside
and outside is still broken. Therefore there will still be a nonzero
macroscopic surface tension $\tau$ between a prevalently ``up''
and a prevalently ``down'' region. The phase can therefore
be called {\em tense bicontinuous.}

If $\tau_0$ is further reduced, the symmetry should eventually be
restored. We thus obtain
an {\em isotropic random\/} phase. This phase is analogous
to the disordered phase of a ferromagnet,
whereas the tense phase is analogous to the ordered one.
The spin-up spin-down symmetry can be explicitly
broken by introducing a spontaneous curvature $H_0$,
or a ``magnetic field'' $\mu$ coupling to the spins.
The transition from the tense bicontinuous to the random isotropic
phase should be critical at the symmetry point ($H_0=\mu=0$)
and belong to the Ising universality class.
Both phases are also called ``sponge'' phases in the recent
literature. In particular, the
tense bicountinuous phase is called
the asymmetric sponge, whereas the
random isotropic is called
the symmetric sponge. They are
usually identified with the shear
birifringent L${}_3$ phase observed in some ternary
(and also binary) mixtures.

The tense bicountinuous phase is ``paradoxical'': the broken
symmetry is not observable. In principle, we could have two
identical-looking samples belonging to different phases,
and one could notice this fact only by observing the interface
which forms when the two are put in contact.
Something similar happens in antiferromagnets:
the order parameter of an antiferromagnet has a free choice,
but it is not possible to distinguish between
samples corresponding to different values of it. In
principle, if one puts them in contact, one should observe an
interface: in practice, we can probe its order only
indirectly.

Let us now go all the way to a high concentration of amphiphile.
Having to accommodate a large amount of membranes, which refuse
to cross and pay energy to bend, the most likely organization is
to pack them into stacks. The membranes are called lamellae
in this context, and this phase is called the {\em lamellar
phase.} From the symmetry point of view it is a (lyotropic)
smectic~A phase, and it is characterized by quasi-long
range order in the positions of the lamellae, and long-range
order in their orientations. The lamellae lie on parallel
planes, on average, in different regions
of the sample. On the other hand, the concentration-concentration
correlation function of the amphiphile
decays (like a power law) to an average value
instead of keeping the rippled behavior characteristic of
the lamellar organization at short distances.

\begin{figure}
\begin{center}\includegraphics[width=8cm]{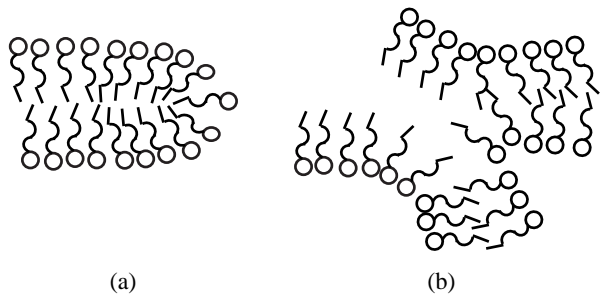}\end{center}
\caption{Examples of defects in lamellar phases:
(a) edges; (b) seams. These defects are assumed
to be forbidden in the simplified model
of interacting fluid membranes.}
\label{seams}\end{figure}

\begin{figure}
\begin{center}\includegraphics[width=8cm]{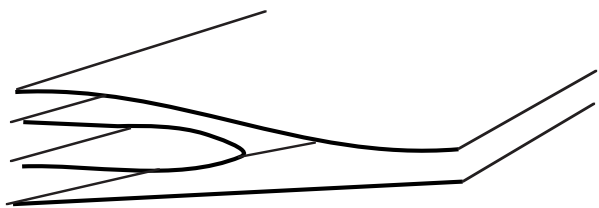}\end{center}
\caption{An edge dislocation which does not
break ``spin'' ordering in lamellar phases.}
\label{edge}\end{figure}

The quasi-long range positional order can be destroyed
by a defect-un\-bind\-ing mechanism not unlike that
underlying the Kosterlitz-Thouless transition. One thus
goes over to a {\em nematic phase,} characterized by
exponential decay of the concentration-concentration
correlations (and therefore macroscopically homogeneous),
but with long-range orientational order. Microscopically,
the difference between the lamellar (smectic) and the nematic
phase lies in the presence of unbound defects.
There may be actually different kinds of defects.
The bilayer
may end at a free edge, or three bilayers meet on one line (seams),
as shown in fig.~\ref{seams}.
In both cases, by going around the dislocation,
the number of lamellae increases or decreases by one.
On the other hand one may have an edge dislocation,
where one lamella bends on itself, as in fig.~\ref{edge},
or a more
complicated screw dislocation, but with a mismatch of
two in lamella number.

These defects may also interfere with the ``spin''
order of the solvent.
In a perfect lamellar phase, it is possible to assign
to regions occupied by the solvent a spin, such that regions separated
by a membrane have opposite spin. This is not
possible if the membrane possesses free edges
or seams.
However, if these defects remain at a microscopical
scale, this is still possible to define an ideal
surface which spans the defect, and the macroscopical ``spin''
order is not disrupted.
With the proliferation of defects, one reaches the
point in which there is no macroscopically acceptable
assignment of spins. In this case, spin symmetry is
restored. On the other hand also the orientational
order characteristic of the nematic phase will eventually
undergo a transition (of the Heisenberg universality class).
These considerations make it plausible that at a sufficiently
high temperature (or rather, small bending rigidity
$\kappa$) the nematic phase will leave the place
to an isotropic phase, which we can identify with the random,
isotropic phase discussed above.

The frontiers between the different phases should correspond to
renormalization group trajectories. In the large rigidity
region, we have seen that the renormalization group
trajectories are given by
\begin{eqnarray}
s\frac{\partial \kappa}{\partial s}&=&-\frac{3\kt}{4\pi},\\
s\frac{\partial \tau}{\partial s}&=&\tau
\left(2+\frac{4\kt}{4\pi\kappa}\right).
\end{eqnarray}
Therefore, defining $\beta=\kt/\kappa,$ we have $\Delta \tau\sim
 s^{2+O(\beta)},$
while $\Delta\beta\sim\log s.$ We thus expect the trajectories to
have the form
\begin{equation}
\tau\propto\exp\left(-\frac{8\pi\kappa}{\kt}\right).
\end{equation}

\begin{figure}
\begin{center}\includegraphics[width=8cm]{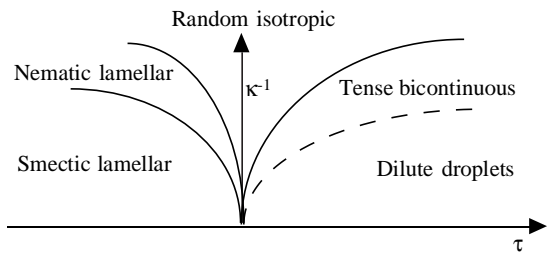}\end{center}
\caption{Phase diagram for fluctuating membranes at $\okappa=0.$
The percolation transition between the dilute
droplets and the tense bicontinuous
phase does not correspond to a thermodynamic singularity.
{}From D.~Huse, S.~Leibler, {\em J.~Phys. France\/} {\bf 49}
605 (1988).}
\label{HL1.fig}\end{figure}

The resulting phase diagram in the $(\kappa^{-1},\tau)$
plane at $\okappa=0$ is drawn in
fig.~\ref{HL1.fig}.

\begin{figure}\begin{center}
\includegraphics[width=8cm]{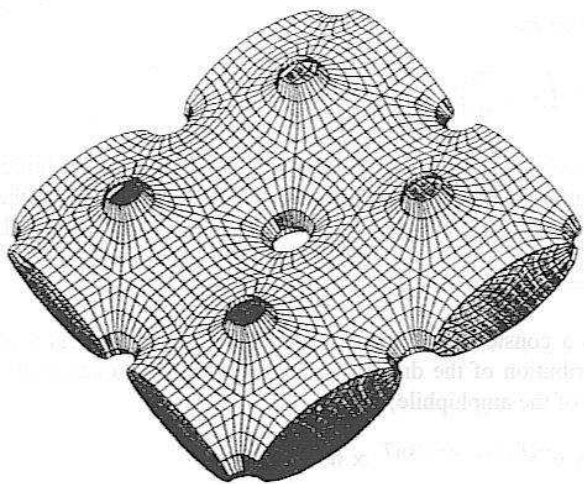}
\end{center}\caption{An example of a plumber's nightmare structure. From
D.~Huse, S.~Leibler, {\em J.~Phys. France\/} {\bf 49}
605 (1988).}\label{plumber}
\end{figure}

Let us now consider the effects of the gaussian rigidity
$\overline{\kappa},$ neglecting for the time being
its renormalization. From the Gauss-Bonnet theorem
we know that the integral over the surface
of the Gaussian curvature $K$
is related to the number $n_{\rm c}$ of connected surface
components and to the number $n_{\rm h}$ of handles:
\begin{equation}
\int\d A K=4\pi (n_{\rm c}-n_{\rm h}).
\end{equation}
Let us keep a large value of the bending rigidity $\kappa.$
If we let $\overline{\kappa}$ become large and positive,
structures with many handles will be favored: these structures
can be built up with minimal surfaces, in which the mean
curvature locally vanishes. Therefore, for both $\overline{\kappa}$
and $\kappa$ large and positive, we expect the formation of regular
structures, made of minimal or almost minimal surfaces,
with a great number of handles. These structures are collectively
known as ``plumber's nightmares'', since they represent a
pipework in which one does not know whether one is in the
interior or in the exterior. An example of such structures is
shown in fig.~\ref{plumber}.
If we go on the other side, towards large {\it negative\/}
values of $\overline{\kappa},$ the formation of disconnected
components will be favored. This can take place in the presence
of a rather large concentration of amphiphile. One will
observe a large concentration of droplets, which, because of
their steric repulsion, will probably tend to organize in a
close-packed droplet crystal.

\begin{figure}
\begin{center}\includegraphics[width=8cm]{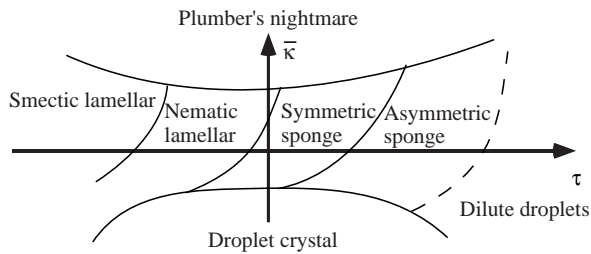}\end{center}
\caption{Phase diagram of an ensemble of nonintersecting membranes
in the $(\tau,\overline{\kappa})$-plane, for fixed $\kappa\gg \kt.$
{}From D.~Huse, S.~Leibler, {\em J.~Phys. France\/} {\bf 49}
605 (1988).}\label{HL2.fig}
\end{figure}

Summing up these considerations, we obtain the
phase diagram in the $(\tau,\overline{\kappa})$-plane,
assuming a large value of the bending rigidity $\kappa,$
and neglecting for the time being the possible occurrence of
free edges and seams, and their effects. The diagram is shown
in fig.~\ref{HL2.fig}. Along the $\overline{\kappa}=0$ axis
we recover the sequence of phases we had described before.

We now discuss each of the phases in detail, and take into
account the controversial consequences of defects and fluctuations.

\smallskip\leftline{\em Droplet phase}\nobreak\noindent
As I have mentioned above, amphiphilic vesicles are
usually formed by sonication and are not assumed to be in
equilibrium in usual conditions. However, there have been
recently observations of droplet phases apparently at equilibrium
in binary~\cite{Cantu91,Cantu94} as well as in more complicated
systems~\cite{Hoffmann92a,Hoffmann92b,Herve93}. This phase
has been named L$_4.$
Experimentally,
it appears in a very narrow region of the phase diagram,
or with a rather special amphiphile in the case of the binary
mixture.

In the equilibrium droplet phase one should observe a
large polydispersity of droplet size. If one neglects the
fluctuations, the elastic energy of a spherical droplet
is given by
\begin{equation}
E_{\rm d}=8\pi\left(\kappa+\frac{\okappa}{2}\right),
\end{equation}
{\em independent\/} of its size. Because of this
scale invariance, logarithmic corrections to
the free energy of the droplet (which are usually
neglected in the thermodynamic limit) become very important.
One obtains an expression of the form
\begin{equation}
F_{\rm d}(n)=\alpha_{\rm d}\kt\ln n+F_0,\label{logcorr}
\end{equation}
where $F_0$ is a constant, and $\alpha_{\rm d}$
is a universal coefficient.
As a consequence, the size distribution of the droplets becomes
(taking into account the chemical potential, $\mu,$ of
the amphiphile)
\begin{equation}
\rho_{\rm d}(n)\propto \e^{-(F_{\rm d}(n)-\mu n)/\kt}\propto
n^{-\alpha_{\rm d}}\e^{-n/n^*},
\end{equation}
where $n^*=\kt/|\mu|$ is a characteristic size, dictated by the amphiphile
concentration. Of course, this form can only hold for
$n>n_{\rm c},$ where $n_{\rm c}$ is the minimum number
of amphiphilic molecules necessary to make a droplet.

The form (\ref{logcorr}) was first proposed by
Helfrich~\cite{Helfrich86}. It has been reexamined by
Huse and Leibler~\cite{HL}, Simons and Cates~\cite{Simons92},
and more recently by Morse and Milner~\cite{Morse94}.
The contributions proportional to $\ln n$ arise from the
renormalization of the elastic constant which we have mentioned
before, but also by the reduction in the translational and
undulational degrees of freedom, which depend on droplet size.
The authors do not agree on the value of $\alpha_{\rm d}$. Part
of the disagreement arises from a controversy over the
coefficients, $\alpha_{\kappa}$ and $\alpha_{\okappa},$
which appear in the renormalization of the bending constants.
In the previous section, I have reported the most widely
accepted results, namely $\alpha_{\kappa}=-3$ and
$\alpha_{\okappa}=10/3.$ In terms of these coefficients,
we may write down the following expressions of $\alpha_{\rm d}$:

\vspace{10pt}
\begin{tabular}{lll}
Helfrich~\cite{Helfrich86},
Simons and Cates~\cite{Simons92}:&$\alpha_{\rm d}$
&$=\alpha_{\kappa}+\alpha_{\okappa}/2=-4/3;$\\
Huse and Leibler~\cite{HL}:&
&$=\alpha_{\kappa}+\alpha_{\okappa}/2+2=1/3;$\\
Morse and Milner~\cite{Morse94}:&
&$=\alpha_{\kappa}+\alpha_{\okappa}/2
+5/2=7/6;$
\end{tabular}

\vspace{10pt}
The behavior of the number density of droplets containing
$n$ surfactant molecules is rather different, depending on
whether $\alpha_{\rm d}$ is negative or positive. If it
is negative, $\rho_{\rm d}(n)$ vanishes at small $n,$
exhibits a peak for $n\simeq n^*,$ and eventually drops
exponentially. On the other hand, if $\alpha_{\rm d}>0,$
$\rho_{\rm d}(n)$ appears to diverge at small $n,$
only to be cutoff (by nonlinear effects we have not explicitly
considered) for $n<n_{\rm c}.$ For $n>n_{\rm c},$
$\rho_{\rm d}(n)$ decreases, first like a power, then
(for $n>n^*$) exponentially.

The free energy density (per unit volume) of the dilute droplet phase
can be obtained by considering it an ideal-gas mixture
of droplets of different sizes:
\begin{equation}
f=\kt \sum_n\rho_{\rm d}(n)\left[\ln\left(\rho_{\rm d}(n)
\lambda^3\right)-1+F_{\rm d}(n)/\kt\right],
\end{equation}
where $\lambda$ is the thermal de~Broglie wavelength
for an amphiphile molecule.
Substituting the equilibrium values of $\rho_{\rm d}(n)$
in this expression, we obtain
\begin{equation}
f=-\kt\left(\rho_{\rm amph}/n^*+\rho_{\rm d}\right),
\end{equation}
where $\rho_{\rm amph}=\sum_n n\rho_{\rm d}(n)$ is the
number density of amphiphilic molecules, and $\rho_{\rm d}$
is the total number density of the droplets.
The number density of amphiphilic molecules,
$\rho_{\rm amph}$ is related to the volume fraction
of the amphiphile, $\phi$ by the obvious relation
$\phi=\rho_{\rm amph} v_{\rm amph},$ where $v_{\rm amph}$
is the volume of a single amphiphilic molecule.
The total amphiphile density is dominated for
any $\alpha_{\rm d}<2$ by contributions from droplets
of size $n\simeq n^*,$ yielding
\begin{equation}
n^*\simeq\left(\rho_{\rm amph}\lambda^3
\e^{F_0/\kt}\right)^{1/(2-\alpha_{\rm d})}.
\end{equation}
This yields $n^*\sim \phi^{\zeta},$ where
\begin{equation}
\zeta=\cases{3/10&for $\alpha_{\rm d}=-4/3;$\cr
3/5&for $\alpha_{\rm d}=1/3;$\cr
6/5&for $\alpha_{\rm d}=7/6.$}
\end{equation}
On the other hand, if $\alpha_{\rm d}<1,$
both $\rho_{\rm d}$ and $f$ are dominated by
droplets of size $n\simeq n^*,$
yielding
\begin{equation}
f\propto -\phi^{(1-\alpha_{\rm d})/(2-\alpha_{\rm d})},
\end{equation}
which decreases as a power law as $\phi$ increases.
However, if $1<\alpha_{\rm d}<2,$
$\rho_{\rm d}$ and $f$ are dominated by
the contributions of small vesicles of size
$n\simeq n_{\rm c},$ yielding a free energy density that is
essentially independent of $\phi.$

In conclusion, I think that a careful reexamination
of the calculation od the free energy of fluctuating
vesicles, in particular taking into account the
constraint of the mean curvature, should be
done in order to clear the controversy.

\smallskip\leftline{\em Sponge phases}\nobreak\noindent
The volume fraction $\Phi$ of the solvent contained
inside the droplets can be estimated by
\begin{equation}
\Phi\propto \sum_n n^{3/2}\rho_{\rm d}(n),
\end{equation}
and is always of order $(n^*)^{5/2}.$ It therefore
increases with increasing amphiphile concentration,
until it becomes larger than a percolation threshold
$\Phi_{\rm p},$ which, for a three-dimensional space,
will be smaller than one half. For $1/2>\Phi>\Phi_{\rm p},$
there will be two infinite connected regions occupied
by the solvent, separated by the membrane, and of unequal volume.
This is the {\em asymmetric sponge\/} phase, also
called tense bicontinuous, because the macroscopic
surface tension between two phases which differ
in the choice of the minority component does not vanish.
On the other hand, as the amphiphile
concentration increases, the symmetry
between the two components should eventually be restored,
leading to the {\em symmetric sponge\/} (random isotropic)
phase. A compact review of sponge phases is contained
in ref.~\cite{sponge}.

Several experimental observation lead to the conclusion that
the shear birifringent L$_3$ phase, which often occurs
in amphiphilic solutions exhibiting a lyotropic smectic A
phase, is indeed a sponge phase.

\begin{itemize}
\item Conductivity measurements~\cite{Gazeau}:
one considers an oil-rich phase, where
the solvent is an insulator and the membrane is an inverted bilayer
containing charges coming from the surfactant counterions.
The membrane acts therefore as a conductor. The measured reduced
conductivity is a fraction of that obtained for the same total amount
of pure water plus charges. This is orders of magnitude larger
than the conductivity observed for a nearby droplet phase.
This indicates that the bilayers are connected and that continuous
water paths are present through the sample.
\item Neutron scattering~\cite{Porte88,Gazeau}: The scattering intensity can be
divided
in two regimes: the large wavevector $q$ regime, where all data
fall on a universal curve corresponding to the scattering
from a flat bilayer, and the small $q$ regime, where correlations
between pieces of the bilayer form a broad peak.
In a lamellar phase the quasi-Bragg peak due to the long-range
order of the bilayer is located at $q=2\pi/d,$ where $d=\delta/\phi$
is the repeat distance of the lamellae ($\delta$ is the membrane
thickness). For a random distribution of connected
bilayers, one expects $d=\gamma\, \delta/\phi,$ where $\gamma$
is a number larger than 1. The peak position for the L$_3$
phase shows indeed a $1/\phi$ behavior, with $\gamma=1.4$--1.6.
\item Freeze fracture~\cite{Strey90}: Direct images obtained by
freeze fracture of the L$_3$ phase show a sponge-like structure,
quite similar to that of bicontinuous microemulsions, with the difference
that, instead of two solvents, there is only one solvent present.
\end{itemize}

Light scattering data also yield detailed information on the
organization of the sponge phase, and will be discussed later.

We can make more precise our description of the sponge phase by
defining a lattice model~\cite{Cates88b,Cates90}, inspired by the
Talmon-Prager~\cite{Talmon} and de~Gennes-Taupin~\cite{deGennes82}
model of microemulsions, or in a less explicit but more general way,
by a Landau-Ginzburg approach~\cite{Roux90,Cates90,Coulon91a,Coulon91b}.
We shall dwell on the second approach. It starts from the consideration
that {\em two\/} order parameters are necessary to describe
the sponge phase: one is obviously the amphiphile concentration
$\phi,$ or rather, its deviation from a reference value $\phi^*$:
\begin{equation}
\rho=\phi-\phi^*.
\end{equation}
The value $\phi^*$ corresponds to the volume fraction at a special
point of high symmetry, in fact a {\em double critical endpoint\/}
(DCE).
The second order parameter describes the breaking of the symmetry
between the two infinite connected components of the space
occupied by the solvent, which we arbitrarily label ``inside''
(I) and ``outside'' (O). Denoting their volume fraction by
square brackets, we define
\begin{equation}
\eta=\frac{[{\rm I}]-[{\rm O}]}{[{\rm I}]+[{\rm O}]}.
\end{equation}
The underlying I/O symmetry of the hamiltonian implies that
$\eta$ can only enter the Landau expansion in even powers.

We can now write down the following expansion of the thermodynamic
potential $\Phi=f-\mu\rho,$ where $f$ is the Helmholtz free
energy density:
\begin{equation}
\Phi=-\mu\rho+\frac{a}{2}\rho^2+\frac{1}{4}\rho^4
+\frac{A}{2}\eta^2+\frac{1}{4}\eta^4+\frac{1}{2}\eta^2.\label{Landau}
\end{equation}
This expansion starts from the
point in which both $\rho$ and $\eta$
vanish, which (as we shall see) is a DCE.
We have used some of the available freedom to rescale the fields
and the thermodynamic potential in order to get rid of
irrelevant coefficients: in this way we are left
with $\mu,$ $a,$ and $A.$
This expression is very similar to that obtained in the Landau
expansion of the Blume-Emery-Griffiths (BEG) model,
which was originally invented to describe the superfluid/normal
fluid transition in He$^3$-He$^4$ mixtures~\cite{BEG}.
It is useful to switch (via a Legendre transformation)
to the $(a,A,\rho)$ space, which is
closer to the experimental situation (in which the amphiphile
concentration is fixed) and also yields a slightly
simpler calculation. In this space the transition lines are
identified by the usual double tangent construction.
We therefore consider the following expression of the free energy:
\begin{equation}
f(\rho,\eta)=\frac{a}{2}\rho^2+\frac{1}{4}\rho^4+\frac{A}{2}\eta^2
+\frac{1}{4}\eta^4+\frac{1}{2}\rho\eta^2.
\end{equation}
If we fix $\rho,$ and minimize this expression with respect
to $\eta,$ we obtain a value $\eta=\eta_{\rm m}(\rho)$ and
a corresponding value $f(\rho)$ of the free energy. The double
tangent construction can now be applied to $f(\rho).$
This construction corresponds to the requirement
of equality both of the chemical potential and of the osmotic
pressure of the amphiphile:
\begin{eqnarray}
\left.\frac{\partial f}{\partial\rho}\right|_1&=&
\left.\frac{\partial f}{\partial\rho}\right|_2,\nonumber\\
f(\rho_1)-\rho_1\,\left.\frac{\partial f}{\partial\rho}\right|_1&=&
f(\rho_2)-\rho_2\,\left.\frac{\partial f}{\partial\rho}\right|_2.
\label{minimization}
\end{eqnarray}
We obtain:
\begin{equation}
\eta_{\rm m}(\rho)=\cases{0,&if $\rho\ge\rho^*,$\cr
(\rho^*-\rho)^{1/2},&if $\rho<\rho^*,$}
\end{equation}
where $\rho^*=-A.$ Therefore
\begin{equation}
f(\rho)=\cases{(a/2)\rho^2+\rho^4/4,&if $\rho\ge\rho^*,$\cr
(a/2)\rho^2+\rho^4/4-(\rho^*-\rho)^2/4,&if $\rho<\rho^*.$}
\end{equation}
This corresponds to a usual symmetry-breaking transition in $\eta,$
located at $\rho=\rho^*.$
For $\rho\ge\rho^*,$ one has $\eta_{\rm m}(\rho)=0,$ corresponding to
a symmetric state, whereas for $\rho<\rho^*,$ one has
$\eta_{\rm m}(\rho)\neq 0.$ The rest of the calculation
involves solving eq.~(\ref{minimization}) where spinodal
lines exist. A good qualitative feeling for the phase diagram
can be obtained just in plotting the function $f(\rho)$
and looking for double tangents.

\begin{figure}
\begin{center}\includegraphics[width=8cm]{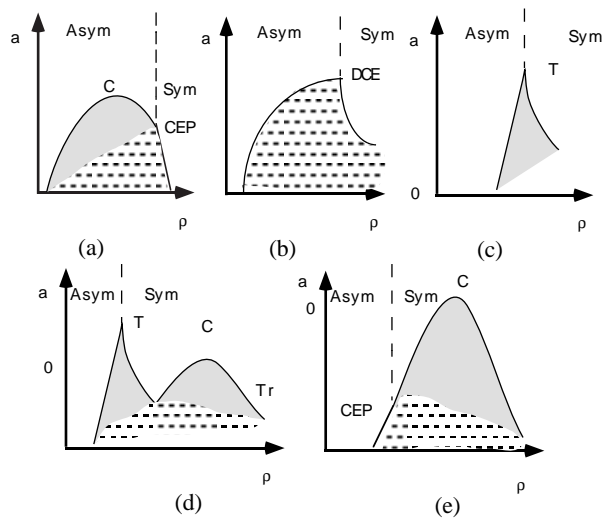}\end{center}
\caption{Schematic phase diagrams obtained from the
minimization of the Landau free energy in
the $(a,\rho)$ plane, corresponding
to the five types discussed in the text.
The transition from symmetric to asymmetric sponge
can be either continuous (dashed line)
or first order (solid line).
T is a tricritical point, Tr a triple point,
CEP a critical endpoint, DCE a double critical
endpoint, and C a liquid-gas type critical point.
{}From D. Roux, C.~Coulon, M.~E.~Cates,
{\em J.~Phys.~Chem.} {\bf 96} 4174 (1992).}
\label{phd1}
\end{figure}

\begin{figure}
\begin{center}\includegraphics[width=8cm]{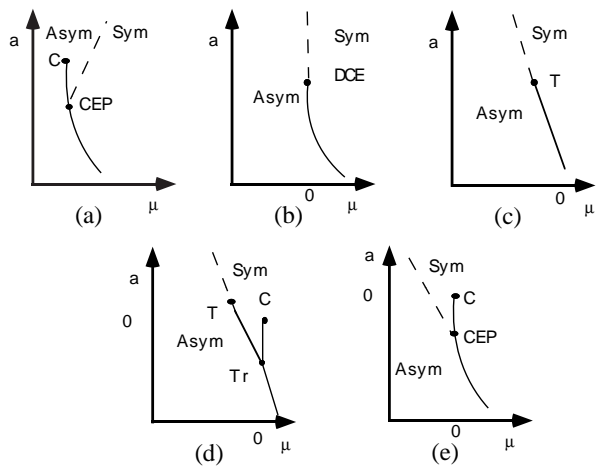}\end{center}
\caption{Phase diagrams (as above) in the $(a,\mu)$ plane.
{}From D. Roux, C.~Coulon, M.~E.~Cates,
{\em J.~Phys.~Chem.} {\bf 96} 4174 (1992).}
\label{phd2}
\end{figure}

\begin{figure}
\begin{center}\includegraphics[width=8cm]{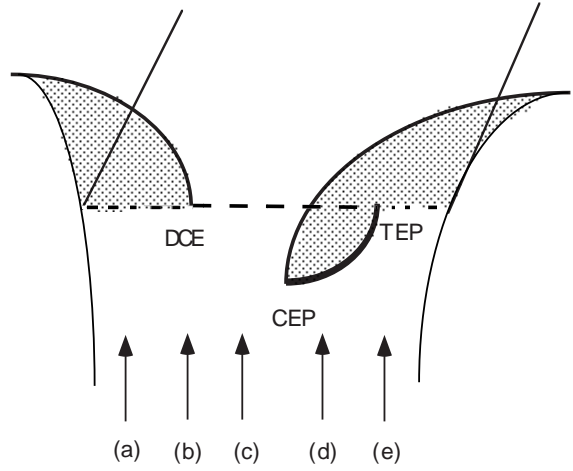}\end{center}
\caption{Representation of the Landau phase
diagram in the $(a,\mu,A)$ space.
Continuous line: line of critical points.
Broken line: line of tricritical points.
Thick line: line of triple points.
Dotted line: line of critical endpoints.
DCE: double critical endpoint; CE: critical endpoint;
TCE: tricritical endpoint.
{}From D. Roux, C.~Coulon, M.~E.~Cates,
{\em J.~Phys.~Chem.} {\bf 96} 4174 (1992).}\label{phd3}
\end{figure}

In total, five types of phase diagram are obtained
in the $(a,\rho)$ plane, depending only on the location
of $\rho^*$ with respect to the spinodal curves.
The phase diagrams are shown in fig.~\ref{phd1}:
one goes from one to another by varying $\rho^*.$
The same phase diagrams are represented in the
$(a,\mu)$ plane in fig.~\ref{phd2}.  The complete
structure of the phase diagram can be represented
in the $(a,\rho^*,\mu)$ space as in fig.~\ref{phd3}.
We now describe in turn the five types of phase
diagram.

\begin{itemize}
\item[(a)] When $\rho^*>0,$ there is a critical line (CL)
for large values of $a$ ($a>a^*_1=1/2-{\rho^*}^2$),
separating the symmetric phase ($\eta_{\rm m}(\rho)=0$)
from the asymmetric one. This line ends in a critical
endpoint (CEP) where the CL meets a first-order line.
This first-order line end at a critical point (CP)
at $\rho=0,$ $a=1/2,$ which corresponds to a demixtion
critical point between two asymmetric sponge phases,
with different amphiphile concentration.
For $a<a^*_1,$ the first order line separates
a symmetric from an asymmetric phase.
\item[(b)] When $\rho^*=0,$ the
line of continuous transition separating
the symmetric from the asymmetric phase
ends at a double critical
endpoint (DCE) at $\rho=0,$ $a=a_{\rm c}=1/2.$
There the shape of the coexistence curve goes
as $a_{\rm c}-a\propto(\rho-\rho^*)^2$ on one
side to $\propto(\rho-\rho^*)^{1/2}$ on the other side.
There is no longer a separate ordinary
critical point.
\item[(c)] When $-1/4<\rho^*<0,$
the symmetric and asymmetric phase are separated
by a second order line for $a>a^*_2=1/2-3{\rho^*}^2,$
and by a first order line for $a<a^*_2.$
These lines meet at a tricritical point.
\item[(d)] For $-1/2<\rho^*<-1/4,$
one also finds a first order line between
two symmetric phases. This line ends at a critical
point at $a=0,$ $\rho=0.$
\item[(e)] When $\rho^*<-1/2,$ the tricritical
point T disappears and the continuous transition
between symmetric and asymmetric phase ends
at a critical endpoint (CEP).
\end{itemize}

Upon looking at this phase diagram it is possible to
identify the starting point of the expansion
as the double critical endpoint (DCE) appearing
in the diagram of Type (b). This high order critical
point has in fact an ``upper critical dimension''
of 8/3. Therefore, the mean field treatment describes
correctly the behavior around this point for
three dimensional systems. On the other hand,
the corrections to mean-field theory are very weak
(logarithmic) around the tricritical point.
One should thus expect the mean field
predictions to be satisfied in a wide region
for systems which happen to be near the DCE or
the tricritical point T.

The occurrence of a line of tricritical points
in the $(a,A,\rho)$ plane implies
that, even in a binary surfactant-solvent system
there is a finite probability of finding a tricritical
point in a $(\phi,T)$ phase diagram, without any special
adjustment of other thermodynamic variables.

Since the Landau expansion obtained above has two
independent parameters (up to rescalings), and
since the elastic description of bilayers has
two independent elastic coefficients $\kappa$
and $\okappa,$ it should be possible to span
the whole of the phase diagram by varying $\kappa$
and $\okappa.$ In practice one can vary these
quantities by adjusting concentrations of
cosurfactants, salt, etc. These systems may therefore
provide a testing ground for theories
of higher order critical behavior.

In order to investigate the behavior
of correlation function, one adds
suitable gradient terms
to the expansion (\ref{Landau}).
They involve the gradients of the two
order parameters $\rho$ and $\eta,$
although only the $\rho$-$\rho$
correlations are directly observable. The most general
form of the gradient terms is
\begin{equation}
\Delta\Phi=\frac{\gamma_{\rho}}{2}|\vec{\nabla} \rho|^2
+\frac{\gamma_{\eta}}{2}|\vec{\nabla} \eta|^2
+\frac{\gamma_{\rm c}}{2}\eta|\vec{\nabla}\eta\cdot
\vec{\nabla}\rho|.\label{grad}
\end{equation}
In the simple case in which $\gamma_{\rho}=\gamma_{\rm c}=0,$
it is possible to calculate exactly the correlations
within the Gaussian approach~\cite{Roux90}.
The third term in eq.~(\ref{grad})  is not
quadratic in the fluctuating fields and
prevents the use of the Gaussian approach when
$\gamma_{\rm c}\neq 0.$
In this case, the probability distribution
for $\rho(\vec{r})$ depends only on the local
value of $\eta(\vec{r}).$ One can then integrate
over the $\rho(\vec{r})$ field, obtaining
an effective potential for the $\eta$ field:
\begin{equation}
\Gamma[\eta]=\int\d\vec{r}\,\left\{\frac{A'}{2}
\eta^2+\frac{\lambda}{4}\eta^4+\frac{\gamma_{\eta}}{2}
|\vec{\nabla}\eta|^2\right\},\end{equation}
where $A'=A+\mu/a,$ and $\lambda=1-a/2.$
This form implies
\begin{equation}
\left<\eta(\vec{r})\eta(0)\right>=
\frac{\e^{-r/\xi_{\eta}}}{4\pi\gamma_{\eta}r},
\end{equation}
where $\xi_{\eta}=\gamma_{\eta}/A'.$

The {\em amphiphile\/} concentration-concentration
correlation function is defined by
\begin{equation}
g(\vec{r})=\left<\delta\rho(\vec{r})\delta\rho(0)\right>,
\end{equation}
where $\delta\rho(\vec{r})=\rho(\vec{r})-\left<\rho\right>.$
It contains two contributions: the direct contribution,
which in our hypotheses is very short-ranged (delta-like),
and an indirect contribution, due to its coupling to
the $\eta$ fluctuations. It is not difficult
to see that one has in fact
\begin{equation}
g(\vec{r})=\frac{1}{a}\delta(\vec{r})+
\frac{1}{4a^2}\left[\left<\eta^2(\vec{r})\eta^2(0)
\right>^2-\left<\eta^2\right>^2\right].
\end{equation}
Since we assume gaussian fluctuations, we have
\begin{equation}
\left<\eta^2(\vec{r})\eta^2(0)\right>=\left<\eta^2\right>^2
+2\left<\eta(\vec{r})\eta(0)\right>^2.
\end{equation}
By using the expression of the correlations of $\eta$
obtained above, we have
\begin{equation}
g(\vec{r})=\frac{\delta(\vec{r})}{a}+
\frac{1}{4a^2}\frac{\e^{-2r/\xi_{\eta}}}{(4\pi\gamma_{\eta}r)^2},
\end{equation}
whose Fourier transform is
\begin{equation}
I(q)=C_1\left[C_2+\frac{\tan^{-1}(q\xi_{\eta}/2)}{q\xi_{\eta}/2}
\right],\label{struc}
\end{equation}
where
\begin{equation}
C_1=\frac{\xi_{\eta}}{16\pi a^2\gamma_{\eta}^2};\qquad
C_2=\frac{16\pi a\gamma_{\eta}^2}{\xi_{\eta}}.\end{equation}

This expression can be compared with the results of
light scattering experiments. It involves
a $1/q+{\rm const.}$ behavior at large $q$
instead of the more usual (Ornstein-Zernike)
one: $1/q^2+{\rm const.}$ The result should hold when
the hypothesis of neglecting $\gamma_{\rho}$ and
$\gamma_{\rm c}$ applies: i.e., near the critical
symmetric-asymmetric line and away from the tricritical point.

A perturbation method~\cite[App.~A]{sponge} can be introduced to
obtain the expression for $I(q)$ in the general case. The result reads
\begin{equation}
I(q)=\left<|\delta\rho_q^2|\right>=
\frac{1}{a+\gamma_{\rho}q^2}+K\frac{
\left(1+(\gamma_{\rm c}q^2/2)\right)^2}{(a+\gamma_{\rho}q^2)^2}
\frac{\tan^{-1}(q\xi_{\eta}/2)}{q\xi_{\eta}/2},
\end{equation}
for the symmetric sponge phase, and
\begin{eqnarray}
I(q)&=&\left<|\delta\rho_q^2|\right>=
\frac{1}{a+\gamma_{\rho}q^2}\nonumber\\
&&+\frac{
\left(1+(\gamma_{\rm c}q^2/2)\right)^2}{(a+\gamma_{\rho}q^2)^2}
\left[\frac{K_1\left<\eta\right>^2}{A+\gamma_{\rho}q^2}
+K_2\frac{\tan^{-1}(q\xi_{\eta}/2)}{q\xi_{\eta}/2}
\right],\end{eqnarray}
for the asymmetric sponge phase.

Two different experimental situations have
been analyzed by light scattering. In the first case~\cite{Roux90},
one has studied an oil-rich sponge phase as a function of
membrane concentration (this experiment
will be denoted as ``oil'' dilution).
In the system under investigation, the sponge phase undergoes
a first-order transition to a phase
of practically pure solvent.
In the second case~\cite{Coulon91a,Coulon91b},
one has studied a water-rich
system which exhibits a second-order phase
transition between a symmetric and an asymmetric
sponge phase.  The transition is made apparent
by a maximum turbidity line (MTL) separating
the two isotropic phases.
This is the line where the structure factor
$I(q)$ (proportional to the
osmotic compressibility) diverges at $q=0,$ corresponding
to very large fluctuations in
the amphiphile concentration. In this way the rather
remarkable feat of exhibiting a symmetry breaking
which cannot be directly observed has been achieved.
This experiment will be denoted as ``water'' dilution.
More recently, there have been careful experiments
on the quaternary system Ma-octylbenzenesulfonate
(OBS)/$n$-pentanol/brine~\cite{Filali}. These
experiments challenge the picture of the asymmetric-symmetric
transition given above, and will be discussed later.

\begin{figure}\begin{center}
\includegraphics[width=10cm]{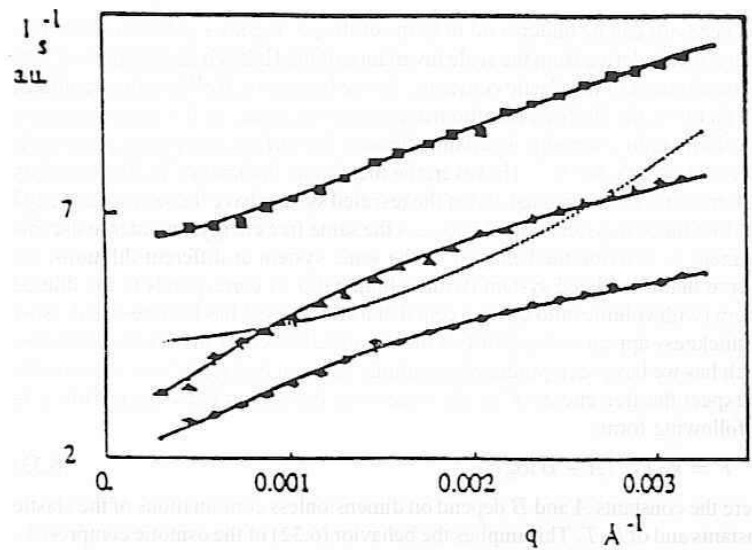}
\end{center}\caption{Plot of the inverse scattering
intensity against $q$ for a sponge
phase at different concentrations.
Dotted line: Ornstein-Zernike form.
Solid lines: eq.~(133). From D.~Roux, C.~Coulon,
M.~E.~Cates, {\em J.~Phys.~Chem.}
{\bf 96} 4174 (1992).}\label{scatt}
\end{figure}

The inverse scattering intensity $1/I(q)$ measured in oil
dilution experiments is shown in fig.~\ref{scatt} as a function
of the wavenumber $q.$ The form is strikingly different
from the Ornstein-Zernike (dotted line), and fits well to the simple
form (\ref{struc}) (solid lines). The fact that the simple
expression works so well indicates that the correlation
length for $\rho$ is rather small. It is possible
to extract from the data the ratio of
the arctan to the constant term
appearing in eq.~(\ref{struc}). This ratio turns out
to be about 2/3, and is a measure of the relative importance of
the fluctuations of $\eta$ with respect to the fluctuations
of amphiphile concentration.

The osmotic compressibility $I(0)$ as a function
of the amphiphile volume fraction $\phi$ varies approximately
as $1/\phi,$ but with logarithmic corrections:
\begin{equation}
I(0)\propto\left(\phi\log(\phi/\phi^*)\right).\label{comp}
\end{equation}
This behavior can be understood in terms of the
scaling laws~\cite{HL,Porte89} of the free energy,
which derive from the scale invariance of the
Helfrich hamiltonian and the renormalization
of the elastic constants. As we have seen,
if all lengths are dilated by a factor $\lambda,$
the Helfrich hamiltonian remains invariant.
In this transformation, the volume ratio $\phi$
remains invariant, although the surface area per
unit volume is decreased by a factor $\lambda^{-1}.$
However, the membrane thickness $\delta$ is also
scaled by $\lambda.$ Microstates
of the original and of the rescaled system
have the same energy, and therefore the two systems
should possess the same free energy. In order to use this
argument to describe the behavior of the same system
at different dilutions, we observe that the dilated
system (with volume ratio $\phi$) corresponds to
the diluted system (with volume ratio $\phi/\lambda$)
except that its thickness $\delta$ has been rescaled.
Now the thickness appears as a cutoff in the renormalization
of the elastic constants, which has we have seen produce
logarithmic terms in the free energy. As a result,
we expect the free energy $F$ of the solution to depend on
the volume ratio $\phi$ in the following form:
\begin{equation}
F=\kt\phi^3(A+B\log\phi),
\end{equation}
where the constants $A$ and $B$ depend on dimensionless
combinations of the elastic constants and of $\kt.$
This implies the behavior (\ref{comp}) of the
osmotic compressibility. The logarithmic corrections
observed in $I(0)$ are therefore an experimental
evidence for the renormalization of the elastic constants.

The water dilution experiments allow to exhibit the divergence
of the correlation length $\xi_{\eta}$ and of the osmotic
compressibility $I(0)$ at the symmetric-asymmetric
transition. Experimentally in particular $I(0)\propto
(\mu-\mu_{\rm c})^{-\zeta},$ where $\zeta\simeq 0.5.$
The Landau-Ginzburg approach predicts $I(0)\propto
(\mu-\mu_{\rm c})^{-1/2}.$ However, the relation
between density and chemical potential is singular
at the transition, and the exponents undergo the
so-called Fisher renormalization~\cite{Fisher}.
One has $I(0)\propto (\rho-\rho_{\rm c})^{-1}$
near the tricritical point. Near the critical point
one should use nonclassical (Ising) exponents,
and the Fisher renormalization has only a small effect. The
result reads $I(0)\propto (\rho-\rho_{\rm c})^{-\alpha/(1-\alpha)},$
where $\alpha\simeq 0.11$ is the specific heat exponent.
This yields $\zeta\simeq 0.12,$ also in contrast with the
measured value. This discrepancy may arise from a crossover
from critical to tricritcal behavior, but has not yet been solved.

\smallskip\leftline{\em Lamellar phases}\nobreak\noindent
Membranes can assume two forms that minimize the
bending energy: as a stack of planar sheets, or as
per\-iod\-ic struc\-tures of min\-imal surfaces (plumber's
nightmare). In both cases
the mean curvature vanishes, and therefore the
free energy of the structure is determined by the
contribution of the gaussian rigidity and by fluctuation
effects. Lamellar phases have been observed
for a long time, and they are usually known
as L$_{\alpha}$ phases~\cite{amphi}.

The stability of the lamellar phases arises
from a compromise between the area coefficient $\tau_0,$
which wants to pack the layers as close as possible,
and the interaction between the lamellae. The interaction
between neutral membranes (for the electrostatic
properties of membranes I refer to the review
by D. Andelman~\cite{Andelman}) arises from the
following sources:
\begin{itemize}
\item[(a)]A strong, short range repulsion due to the
hydration interaction, yielding the interaction
potential per unit area (as a function
of the membrane spacing $d$)
$V_{\rm h}(d)=A_{\rm h}\exp(-d/\lambda_{\rm h}).$
\item[(b)]A long range attraction due to the van der Waals
forces:
\begin{eqnarray}
V_{\rm VW}(d)&=&-\frac{W}{12\pi}\left[\frac{1}{d^2}-\frac{2}{(d+\delta)^2}
+\frac{1}{(d+2\delta)^2}\right],\nonumber\\
&\sim&-\frac{W}{2\pi}\frac{\delta^2}{d^4},\qquad \hbox{for }d\gg\delta.
\end{eqnarray}
Here $\delta$ is the bilayer thickness and
the parameter $W$ is known as the Hamaker constant.
\item[(c)]The long range steric repulsion, due to the occasional
collision between lamellae~\cite{Helfrich78}:
\begin{equation}
V_{\rm st}(d)=c_{\rm H}\frac{(\kt)^2}{\kappa(d-\delta)^2}.
\end{equation}
Here $c_{\rm H}$ is a universal coefficient, over whose
value there is some controversy \cite{Helfrich78,GK89,David90}.
\end{itemize}
However, one cannot simply sum up these contributions to obtain
an effective potential $V_{\rm eff}(d).$ This potential has a
minimum either at very short distances or at infinite
separation: as a consequence one would have a first order
transition between lamellar phases at high amphiphile concentration
and a practically pure solvent. On the other hand
very dilute lamellar phases (where $d$ can be as
large as 1 m$\mu$) have been observed~\cite{Larche}.

Lipowsky and Leibler~\cite{LL} introducend a complicated
functional renormalization scheme to attack the problem,
and obtained a continuous transition between a state
of ``bound'' lamellae and one of ``unbound'' lamellae,
as the Hamaker constant $W$ crossed a critical value $W_{\rm c}.$
In this theory, the mean lamellar spacing $\bar{d}$ diverges
like a power law near the transition: $\bar{d}\sim(W-W_{\rm c})^{-\psi},$
where $\psi=1.00\pm0.03.$ This result did not appear to admit
a simpler description.

\begin{figure}\begin{center}
\includegraphics[width=10cm]{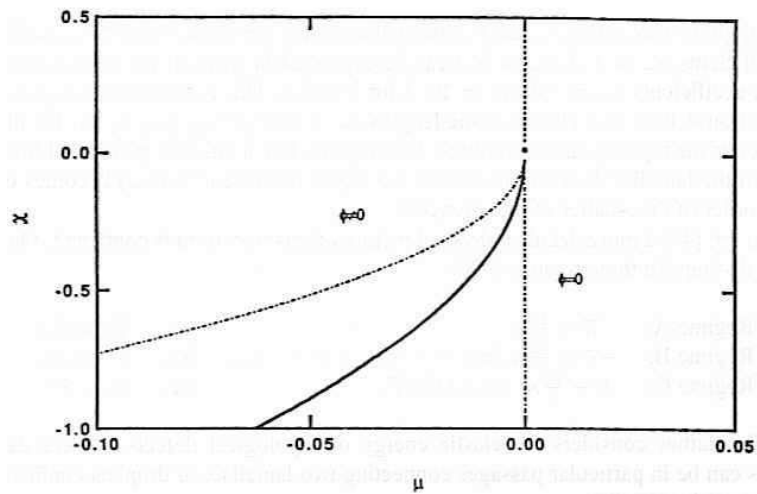}
\end{center}\caption{Phase diagram of a lamellar phase in the $(\mu,\chi)$
plane. Broken line: first order transition. Solid line:
second order transition. Thin lines: spinodals. From
S.~T.~Milner, D.~Roux, {\em J.~Phys.~I France\/} {\bf 2} 1741 (1992).}
\label{lamellar}\end{figure}

Recently, Milner and Roux~\cite{Milner92} as well as
Helfrich~\cite{Helfrich93} have proposed simpler (Flory-like)
explanations of this continuous ``unbinding'' transition. However, the
theory of Milner and Roux applies to the uniform lamellar
phase, whereas the one by Helfrich describes the unbinding
of a finite stack of $n$ layers from a wall. We shall thus
dwell on the theory by Milner and Roux.
First of all, one chooses the volume ratio
$\phi$ of the amphiphile as the
order parameter.
Then the free energy is obtained in analogy with the van der Waals
theory of the liquid-gas transition, by adding to
the Helfrich estimate of the entropy of the stack
(which only takes into account the hard-wall repulsion)
a term representing the correction to the second virial coefficient due
to the other interactions. In analogy with the
Flory-Huggins theory of polymers, the entropy term is estimated
in the absence of all correlations due to the attractive
interactions, and in the correction to the virial coefficient
one neglects the connectivity of the surface.
This enthalpic correction is referred to ``patches''
of the membrane of size $\nu=\ell_0^2\delta,$ where
$\ell_0^2=(\kappa/\kt)\delta^2$ is the shortest
length where the membrane could bend.
The result reads
\begin{equation}
f_{\rm lam}(\phi)=\frac{c_{\rm H}(\kt)^2}{\kappa\delta^3}
\phi^3-\kt\chi\phi^2.\label{Miln}
\end{equation}
The coefficient $\chi$ appearing in the enthalpic correction
can be given as a function of the interaction $U_{\nu}(\vec{r})$
between patches by
\begin{equation}
\chi=-\frac{1}{2\nu^2}\int\d^3\vec{r}\,\left\{1-\exp\left[
-U_{\nu}(\vec{r})/\kt\right]\right\},
\end{equation}
as long as the interaction between patches is sufficiently
weak. We expect that $\chi$ vanishes linearly
as the Hamaker constant reaches some critical value.
The phase diagram can be obtained by minimizing with
respect to $\phi$ the free energy
\begin{equation}
g(\chi,\mu)=f_{\rm lam}(\phi)-\mu\phi,
\end{equation}
where $\mu$ is the chemical potential of the amphiphile.
The resulting phase diagram is shown in fig.~\ref{lamellar}.
The expression (\ref{Miln}) has the form of a Landau expansion
around $\phi=0.$ In this expression, the cubic term in $\phi$
does not necessarily lead to a first order phase transition,
because negative values of $\phi$ are unphysical (a similar
situation arises in the mean field treatment of percolation).
Therefore the order parameter vanishes at the second order
transition with an exponent 1 instead of the usual mean field
exponent 1/2.

\smallskip\leftline{\em Topological instabilities}\nobreak\noindent
We have thus found that it is possible to stabilize lamellar
phases with arbitrarily small amphiphile concentrations,
at least if the Hamaker constant is large enough. However,
we have argued before that as the solution is diluted one should
go over to isotropic or cubic phases. In order to understand this
transition it is necessary to include topology changes
and renormalization effects, which are absent from the previous
considerations. Two recent papers deal with this
problem~\cite{Golubovic94,Morse}. The essential point
is that, while the bending rigidity $\kappa$ decreases as
one goes to longer and longer length scales, the gaussian
rigidity $\okappa$ {\em increases.} Therefore structures
of higher and higher genus are favoured at longer and longer
scales.
\def\lcoll{\ell_{\rm coll}}

In ref.~\cite{Morse} one starts from the expression of the
Helfrich hamiltonian in terms of the principal curvatures
$c_1$ and $c_2$:
\begin{equation}
F=\int\d A\left[\frac{1}{2}\kappa\left(c_1+c_2\right)^2
+\okappa\,c_1c_2\right],
\end{equation}
and rewrites it introducing the {\em topological rigidities\/}
$\kappa_+=\kappa+\okappa/2,$ $\kappa_-=-\okappa/2$:
\begin{equation}
F=\int\d A \left[\frac{1}{2}\kappa_+\left(c_1+c_2\right)^2
+\frac{1}{2}\kappa_-\left(c_1-c_2\right)^2\right].\end{equation}
One has of course $\kappa=\kappa_++\kappa_-$. When
either topological rigidity vanishes, the simply connected
membrane becomes unstable with respect to fluctuations
leading to a different topology:

\smallskip\noindent
\begin{tabular}{ll}
For $\kappa_+<0,$ &towards the formation of many spherical surfaces;\\
for $\kappa_-<0,$ &towards the formation of an infinite minimal surface.
\end{tabular}\smallskip

Since $\kappa=\kappa_++\kappa_-,$ this instability occurs
{\em before\/} the instability towards a continuous deformation
of the membrane, which occurs when $\kappa$ becomes negative,
except in the special case in which $\kappa_+\simeq\kappa_-\simeq 0.$

The topological rigidities renormalize with respect to the
scale $\ell$ according to the equation
\begin{equation}
\kappa_{\pm}(\ell)=\kappa_{\pm}-\frac{\alpha_{\pm}\kt}{4\pi}
\log\left(\frac{\ell}{\ell_0}\right).
\end{equation}
In this equation, $\ell_0$ is a microscopic distance
(of the order of the lateral size of an
amphiphilic molecule), $\kappa_{\pm}$ are the bare values
of the topological rigidites, and the coefficients
$\alpha_+=4/3,$ $\alpha_-=5/3$
can be expressed in terms of the renormalization coefficients
$\alpha_{\kappa}=-3,$ $\alpha_{\okappa}=10/3$ for $\kappa$
and $\okappa.$
This renormalization group equation defines two characteristic lengths
$\xi_{\pm}\propto\exp\left(4\pi\kappa_{\pm}/(\alpha_{\pm}\kt)\right)$
for the onset of the topological instabilities.
One expects that a lamellar phase will melt when the lamellar
distance (or, better, the repeat distance $d=\delta/\phi$) becomes
of the order of the smaller of these lengths.

In ref.~\cite{Golubovic94} a more detailed physical picture
of this transition is contained.
One can distinguish three regimes~\cite{Golubovic94}:

\smallskip
\begin{tabular}{llll}
Regime A:& $-\okappa<\frac{10}{9}\kappa,$ &i.e.,
 &$\xi_{\okappa}\ll\xi_{\kappa};$\\
Regime B:& $-\okappa>\frac{10}{9}\kappa,$ but $|\okappa+\frac{10}{9}
\kappa|>O(1)\,\kt,$ &i.e., &$\xi_{\kappa}\ll\xi_{\okappa};$\\
Regime C:& $|\okappa+\frac{10}{9}|\kappa<O(1)\,\kt,$&i.e.,&$\xi_{\kappa}
\sim\xi_{\okappa}.$
\end{tabular}\smallskip

The author considers the
elastic energy of topological defects. These
defects can be in particular passages
connecting two lamellae, or
droplets confined between the lamellae.

Let us first consider a passage connecting
two lamellae a distance $d$ apart. Its curvature
energy can be estimated as
\begin{equation}
E_{\rm p}=O(1)\,\kappa(\ell)\left(\frac{r}{\ell}\right)^2
-4\pi\okappa(r),\label{passage}\end{equation}
where $r$ is the size of the passage neck, and $\ell$ is the
lateral size of the passage deformation region.
At distances from the neck smaller than $\ell,$ the presence
of the passage curves the membrane giving rise to the
first term in eq.~(\ref{passage}). The second term in this
equation is the gaussian curvature contribution, and
originates essentially from the region of the neck.
Therefore the two rigidities are evaluated at different
scales. (This effect is not taken
into account in ref.~\cite{Morse}).
The passage deformation region is approximately a catenoid,
yielding the condition $d\simeq 2r\log(\ell/r),$ relating
$r$ and $\ell.$ We let this constraint in eq.~(\ref{passage})
and obtain the equilibrium energy and size of the
passage, as a function of $d.$ Assuming $\kappa(d)\gg\kt,$
and neglecting slowly varying logarithmic factors, we obtain
\begin{eqnarray}
E_{\rm p}&=&-\alpha_{\okappa}\kt\log\frac{d}{\xi_{\okappa}},\\
\ell_{\rm eq}&=&O(1)\left(\frac{\kappa(d)}{\kt}\right)^{1/2}d,\\
r_{\rm eq}&=&O(1)d.
\end{eqnarray}
The elastic energy of the passage decreases as $d$ increases,
and becomes negative as $d$ becomes of order $\xi_{\okappa}.$
On the other hand, the deformation region $\ell_{\rm eq}$
increases with $d.$

The interaction among passages can be represented by a hard
core repulsion, with effective hard core size $\ell_{\rm eq}.$
When the area density of $n_{\rm p}$ of passages is
smaller than $\ell_{\rm eq}^{-2},$ a simple grand canonical
calculation yields $n_{\rm p}(d)\simeq n_0(d),$ with
\begin{equation}
n_0(d)=\left(\frac{\kappa(d)}{\kt}\right)^2\frac{1}{d^2}
\left(\frac{d}{\xi_{\okappa}}\right)^{\alpha_{\okappa}}
\propto d^{\alpha_{\okappa}-2}=d^{4/3},\end{equation}
which increases with increasing $d.$ Such a growth
of the passage density was observed by Harbich et al.\
in egg-lecithin membrane stacks~\cite{Harbich78}.
The free energy of the ``passage fluid'' per unit membrane
area is given by $F_{\rm p}(d)=-\kt n_0(d)\sim -d^{4/3}.$
The passages begin to overlap when $d$ becomes of order
$d^*,$ where
\begin{equation}
d^*=\left(\frac{\kt}{\kappa(\xi_{\okappa})}
\right)^{3/\alpha_{\okappa}}\xi_{\okappa}\sim \xi_{\okappa}.
\end{equation}
In this situation, the free energy can be estimated
by minimizing with respect to the passage density $n$
a trial free energy containing the contribution
of the hard core repulsion:\begin{equation}
F_{\rm p}(n,d)=\kt n\log\left(\frac{n}{\e n_0 (d)}\right)
-\kt n\log\left(1-n\ell_{\rm eq}^2\right).\end{equation}
The first term reproduces the dilute result $n=n_0(d),$
while the second represents the free energy increase due
to the hard-core repulsion.
We obtain therefore $n_{\rm p}\simeq \ell_{\rm eq}^{-2}$
for $d\gg d^*,$ with a free energy
\begin{eqnarray}
F_{\rm p}(d)&\simeq& -O(1)\left(\frac{\kt}{\ell_{\rm eq}^2}
\right)\log\left(n_0(d)\ell_{\rm eq}^2\right)\nonumber\\
&\simeq& -O(1)\frac{(\kt)^2}{\kappa(d)d ^2}\log\left(1+
(d/d^*)^{\alpha_{\okappa}}\right).\end{eqnarray}
This free energy corresponds to an {\em attractive\/}
interaction between membrane pairs, which tends to keep
them at a distance $d\sim \xi_{\okappa}.$ This attraction
actually dominates over the Helfrich steric repulsion
\begin{equation}
V_{\rm st}(d)=-c_{\rm H}\frac{(\kt)^2}{d^2}.\end{equation}
for $d>\xi_{\okappa}.$

The free energy of the lamellar phase per unit
volume is given by
\begin{equation}
F_{\rm lam}(\tau_0)=\min_d\Phi(d,\tau_0),\label{Flam}
\end{equation}
where $\tau_0$ is the area coefficient (proportional to
the chemical potential of surfactant) and
\begin{equation}
\Phi(d,\tau_0)=\frac{1}{d}\left[\tau_0+F_{\rm p}(d)+
V_{\rm st}(d)\right].\end{equation}
This expression does not include the
van der Waals interaction. In the absence of
the passage term, one has a single minimum
for $\tau_0<0,$ at $d=d_{\rm eq}\sim |\tau_0|^{-1/2}.$
One would thus expect a swelling of the lamellar
phase until $d\sim\xi_{\kappa},$ where the lamellae crumple.
But the inclusion of the passage interaction changes the behavior,
and $d_{\rm eq}$ remains essentially fixed at $\xi_{\okappa}$
also for {\em positive\/} $\tau_0,$ where one would expect a
diluted droplet phase.
The free energy of the droplet
phase can be shown to be essentially zero compared to
that of eq.~(\ref{Flam}). The lamellar phase therefore
gives way to the droplet phase when the free energy (\ref{Flam})
vanishes, what happens for
$\tau_0\sim (\kt)^2/(\kappa(\xi_{\okappa})\xi_{\okappa})^2.$
This only holds
in regime A, where $\xi_{\okappa}\ll\xi_{\kappa}.$
In regime A one therefore expects the lamellar phase (with
a repeat distance $d\sim d^*$) to remain stable also for a positive
area coefficient. One would thus have a strong first-order phase
transition between a passage-rich, but still
orientationally ordered, lamellar phase and a dilute droplet phase.

In regime B, on the other hand, passages will be
very dilute and the most important role
is played by the droplets. The length scale
of droplets is fixed by $\xi_+,$ which is much
smaller than $\xi_{\kappa}$ in this regime. In ref.~\cite{Golubovic94}
it is assumed that the
elastic energy of a droplet of scale $R$ is given by
$E_{\rm d}(R)=8\pi\kappa_+(R)=2\alpha_+\kt\log(\xi_+/R).$
(The results would probably have to be modified
in view of the results of ref.~\cite{Morse94}.)
By means of this expression, it is possible
to compute the free energy, per unit area of lamella,
of a polydisperse solution of droplets confined
between two lamellae. The result reads
\begin{equation}
F_{\rm d}(d)=d\,{\cal F}_{\rm d}+f_{\rm d}(d),
\end{equation}
where ${\cal F}_{\rm d}$ is the free energy per unit
volume of the droplet phase, which in our hypotheses is given
by
\begin{equation}
{\cal F}_{\rm d}=-O(1)\frac{\kappa^2}{\kt}\left(
\frac{\ell_0}{\xi_+}\right)^{2\alpha_+}\ell_0^{-3}.
\end{equation}
The effects of the confinement are contained in
$f_{\rm d}(d),$ which is given by
\begin{equation}
f_{\rm d}(d)=O(1)\frac{(\kappa(d))^Ò}{\kt}
\left(\frac{d}{\xi_+}\right)^{2\alpha_+}d^{-2}\sim d^{2/3},
\end{equation}
corresponding to an attractive interaction between the lamellae.
Therefore the {\em difference\/} between the free energy
density of the lamellar and the droplet phase exhibits a minimum
at a value $d^*=d^*(\tau_0).$ For a critical value $\tau_{\rm c}
\sim -(\kt)^2/(\kappa(d_{\rm max})d_{\rm max}^2)$
of $\tau_0$ the free energy difference vanishes and
the lamellar phase melts into the droplet
phase. The maximal distance $d_{\rm max}$ between the lamellae
is given by
\begin{equation}
d_{\rm max}=d^*(\tau_{\rm c})\sim
\left(\frac{\kt}{\kappa(\xi_{\rm d})}\right)^{3/\alpha_+}
\xi_{\rm d}\sim\xi_{\rm d}.
\end{equation}
This is also the radius $R_{\rm max}$ of the largest droplets.

In this regime it should therefore be possible to
observe a weak first-order
transition to a dense droplet phase. This transition takes place
when $d$ reaches the value $\xi_+.$ A transition to an
isotropic sponge phase can only occur in regime C
(corresponding to $d_{\rm max}\sim
R_{\rm max}\sim \xi_{\okappa}\sim\xi_{\kappa}\sim\xi_+$)
where the creation of passages between the largest
droplets is favored.

These expectations agree with those of ref.~\cite{Morse} in regime B
and regime C. The predictions for regime A are different. In particular
they contradict the prediction made in ref.~\cite{Morse} of
a transition from the lamellar phase into a L$_3$ phase of
multiply-connected surfaces with small mean curvature.

\smallskip\leftline{\em Free edges and gauge theories}\nobreak\noindent
I discuss in this subsection
the elegant connection made by Huse and\break Leibler~\cite{HL91}
between the sponge phases of membranes and the $Z_2$ gauge-Higgs
models on a lattice.

When one considers real membranes, one should also take
into account the possibility of free edges. Free edges
(see fig.~\ref{seams}(a)) involve a region with strong
curvature in the amphiphilic arrangement, and where contact
between water and the tails of the amphiphile is eased.
It is likely therefore that these defects cost
a relatively high price in terms of energy per unit length.
However, this price is not infinite, and it is interesting to ask
how the phase diagram of the membranes is affected by
introducing them.

One can stard by adding to the Helfrich hamiltonian a term
\begin{equation}
F_1=\lambda_0\int\d \ell,
\end{equation}
where the integral is over all edges, and $\lambda_0$
is the chemical potential per unit length, of the
edges ``living'' on the membrane. For infinite $\lambda_0$
edges are suppressed and the previous results hold.
When $\lambda_0$ is finite, we start running into difficulties
in separating without ambiguities ``inside'' (I) from ``outside''
(O) in the sponge and droplet phases. Moreover, it is no more
possible to define an effective surface tension, as we had done
in the {\em gedankenexperiment\/} mentioned in the discussion
of the asymmetric sponge phase.
Indeed, if we try to force a macroscopic interface between a
prevalently ``up'' and a prevalently ``down''
spin phase, the system will prefer to introduce a macroscopic
hole in the membrane, and the free energy will be higher
by a term of order $L$ (instead of order $L^2$)
with respect to a homogeneous system.

\begin{figure}
\begin{center}\includegraphics[width=8cm]{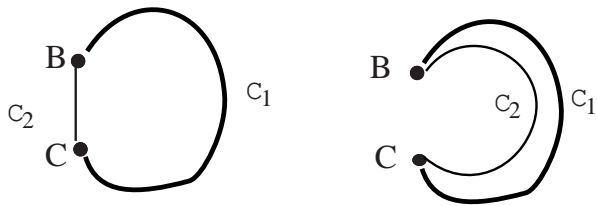}\end{center}
\caption{Theoretical construction for the
definition of the macroscopic line tension.
An edge (thick line) is forced in the sample,
connecting points B and C along the curve ${\cal C}_1.$
The return path ${\cal C}_2$ may either be a straight line
(a) or a curved line close to ${\cal C}_1,$
depending on the energy cost of the creation of membrane
patches. From D. H. Huse and S. Leibler,
{\em Phys. Rev. Lett.} {\bf 66} 437 (1991).}
\label{edges}\end{figure}

Instead, one may consider the effects of forcing
a macroscopic {\em edge\/} through the system.
Imagine to introduce an edge along a curved
line ${\cal C}_1$ connecting points B and C, set a distance
$\ell$ apart, as in fig.~\ref{edges}.
The length of the curved line is equal to $L.$
The system must choose the return path ${\cal C}_2$
connecting C to B. If the free energy cost needed to create
a macroscopic patch of the membrane is small,
nothing prevents the system from choosing
the straight line BC as ${\cal C}_2$ (fig.~\ref{edges}(a)).
In this case, the free energy price of the
free edge BC is proportional to $\ell,$
and the {\em macroscopic\/} line tension
is nonzero. On the other hand, if the
free energy cost of the membrane is large,
the system prefers to return along the curve ${\cal C}_2$
situated very close to ${\cal C}_1$ (fig.~\ref{edges}(b)),
in order to avoid creating extra surface. In this
regime, the macroscopic line tension is formally zero.
The first regime is expected to hold in
the symmetric sponge phase. We expect the macroscopic
line tension to vanish in the  asymmetric sponge
phase and also in the phase corresponding
to the droplet phase, but in which small membrane
patches with free edges are also allowed.
This is the {\em droplet-and-disc\/} phase.

In order to understand more closely the phase diagram,
let us fix the value of the microscopic line tension
$\lambda_0,$ of the rigidity $\kappa_0,$ and let us
start from a large value of the area coefficient
$\tau_0.$ We shall be in the droplet-and-disc
phase, and the edges appear as closed rings,
spanned by the membrane.
As $\tau_0$ decreases, an infinite
piece of connected surface appears,
although (with sufficiently large
$\lambda_0$) the size of the edge loops
remain finite: this is the asymmetric
sponge regime. If we now reduce $\lambda_0,$
one reaches a regime in which connected
edge paths of infinite length are present:
this is the {\em sponge-with-free edges\/}
regime. One can picture this regime as
a melt of ring polymers (the edges)
of arbitrary length. These polymers are
Gaussian at large lengths, and therefore
have only a finite probability
of coming close to themselves. However,
there are an infinite number of these
polymers, and different polymers will come
close to each other infinitely many times.
So the surface attached to one edge will almost
certainly be connected to the surface
attached to the other edge, and therefore
one expects only one connected piece
of membrane to be present.

We have therefore four different
isotropic regimes: droplet-and-disc,
asymmetric sponge, sponge with free edges
(all three of them with a vanishing macroscopic
line tension), and the symmetric sponge regime
(with a nonzero line tension). Only
the symmetric sponge regime is separated from
the others by a {\sl bona fide\/}
phase transition.

\begin{figure}\begin{center}
\includegraphics[width=10cm]{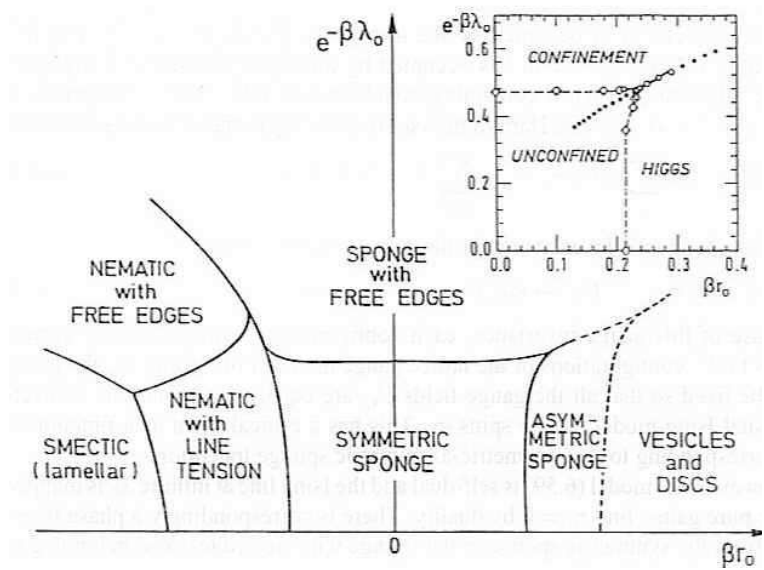}
\end{center}\caption{Schematic phase diagram for an ensemble
of membranes with edges, according
to D. H. Huse and S. Leibler,
{\em Phys. Rev. Lett.} {\bf 66} 437 (1991).
The dashed lines correspond to geometric
(percolation) transitions,
while the solid lines show the
thermodynamic phase transitions.
The $\tau_0>0$ of the phase diagram corresponds
to the phase diagram of the $Z_2$
gauge Higgs model shown in the inset.
In the inset the dotted line is the self-duality line;
the unconfined-confined transition on the
vertical line has its dual equivalent
in the unconfined-Higgs transition
on the horizontal line.}\label{HLphd}\end{figure}

For sufficiently large and negative $\tau_0$
one goes over to the orientationally ordered
phases. Here, the presence of free edges is
incompatible with smectic (positional) order,
as we have discussed before: but we may have,
or not have, a macroscopic line tension.
We have therefore two {\em nematic\/}
phases, depending on the value of
the macroscopic line tension:
nematic with line tension, and nematic with free
edges. The resulting phase diagram in the
$(\tau_0,\lambda_0)$ plane is sketched in fig.~\ref{HLphd}.

The part of the phase diagram involving the sponge phases
is in fact equivalent to the phase
diagram of the $Z_2$ gauge-Higgs
system (for a review, see, e.g.,
\cite{Drouffe83}).
 The connection can be made explicit by
considering random surfaces on a lattice. The
lattice length should correspond
to the persistence length $\xi_{\kappa}$:
at longer lengths, the bending rigidity
can be neglected. Thus, let us consider
a random surface living on the plaquettes
of a simple cubic lattice. Label the
elementary cubes by indices $i.$
One introduces Ising spins $\sigma_i=\pm 1$
on each elementary cube, and a gauge
field $U_{ij}=\pm 1$ on each plaquette $ij$
(separating cube $i$ and cube $j$).

The plaquette $ij$ is occupied by the
membrane if $\sigma_iU_{ij}\sigma_j=-1.$
The link shared by cubes $i,$ $j,$
$k,$ and $l$ is occupied by an edge (or
a seam) if the number of adjacent plaquettes containing
membranes is odd. This
corresponds to $U_{ij}U_{jk}U_{kl}U_{li}=-1.$
The Hamiltonian corresponding to
these fields is therefore
\begin{equation}
H=-\tau_0\sum_{\left<ij\right>}
\sigma_iU_{ij}\sigma_j-\lambda_0\sum_{\left<ijkl\right>}
U_{ij}U_{jk}U_{kl}U_{li}.\end{equation}
This hamiltonian is invariant under
the gauge transformation
\begin{equation}
\sigma_i\to\epsilon_i\sigma_i;\qquad
U_{ij}\to \epsilon_i\epsilon_j\,U_{ij};\qquad
\epsilon_i=\pm 1.\label{gauge}\end{equation}
Because of this gauge invariance,
each configuration of the membrane corresponds
to $2^N$ configurations of the lattice
gauge model. For infinite $\lambda_0,$
the gauge may be fixed so that all the
gauge fields $U_{ij}$ are equal to $+1,$
and one recovers the usual Ising model
for the spins $\sigma_i.$ This has
a critical point
as a function of $\tau_0,$ corresponding
to the symmetric-asymmetric sponge
transition.

However, the model (\ref{gauge}) is self-dual
and the Ising line at infinite $\lambda_0$
is mapped to the pure gauge line $\tau_0=0$
by duality. There is correspondingly
a phase transition from the symmetric sponge
to the sponge with free edges, also
belonging to the Ising universality class.

Away from the $\tau_0=0$ or $\lambda_0=\infty$
lines one may exploit Monte-Carlo simulations
of the $Z_2$ gauge system~\cite{Jongeward}.
The result is that these lines of continuous
transition become first-order line near the triple
point where the three phases meet. A line of first-order
transition also extends along the self-dual line, separating
for a while the asymmetric sponge from the
sponge with free edges phase, and terminating in a critical point,
beyond which there is no thermodynamic singularity
separating the two regimes.

As I have mentioned above, the experimental results
of ref.~\cite{Filali} on the quaternary
OBS/$n$-pentanol/brine system could not be interpreted
without difficulties in the framework of the
phenomenological theory of the asymmetric/symmetric
transition of ref.~\cite{sponge}. In this system,
one observes a sharp maximum of the osmotic compressibility
$I(0)$ on the maximum turbidity
line (MTL) far inside a one-phase region; however,
the light scattering function $I(q)$ has a Ornstein-Zernike $q^{-2}$
form on the $n$-pentanol rich side of the MTL
(while it deviates from this form on the $n$-pentanol
poor side).

This behavior is not compatible with the predictions
of ref.~\cite{sponge}, which would imply a
(weak) divergence of $I(0),$ and a different form of the scattering
function, nor with the hypothesis of a tricritical behavior,
because one is deep inside a one phase region (while a tricritical
point is necessarily close to a two-phase region).
The authors suggest to introduce a different phenomenological
model, introducing a new order parameter, $m,$
linearly (and not quadratically) coupled to
the amphiphile density $\rho.$ In this way one would
explain both the strong divergence of $I(0)$
and the Ornstein-Zernike form of  $I(q).$
The divergence is then ``rounded off'' because
the amphiphile concentration acts as a ``magnetic
field'' for $m,$ and the critical point is reached
only for that particular value of the concentration
for which the ``magnetic field'' vanishes.

The authors then go on to suggest that the MTL in this
system can be related to the symmetric sponge/sponge-with-free-edges
(S/SFE) transition considered by Huse and Leibler~\cite{HL91}.
The idea is that the (small) alcohol molecules
essentially act by reducing the line tension
$\lambda_0$ for the formation of free
edges and seams. However, I do not see how this
suggestion helps in understanding the asymmetry
in the form of the scattering function
$I(q),$ or the apparent lack of divergence
in $I(0).$ Concerning the last point the authors
argue that the actual behavior of the membrane
should be described by
a more general model, in which edges and seams
have different line tensions. However,
the critical behavior of this model is still
unknown.

\smallskip\leftline{\em Plumber's nightmares}\nobreak
\noindent I close by a very brief discussion of plumber's
nightmare phases~\cite{HL}. The free energy per unit
volume of this phase as a function of the
lattice constant $s$ is given by
\begin{equation}
{\cal F}_{\rm pn}(s)\simeq C_1\frac{\tau_0}{s}
+\frac{C_2\kt-C_3\okappa(s)}{s^3}.\end{equation}
The density of handles is $C_3/(4\pi s^3),$ and
the constants $C_i$ are of order unity. The
lattice constant will presumably be of the
order of the short-distance cutoff $\ell_0,$
at least if the ``bare'' gaussian rigidity $\okappa_0$
is positive. In this case, by
comparing this free energy to that of the
droplet phase (which is zero on the scale
$\tau_0/\ell_0$) the transition to the
plumber's nightmare phase will take place
when $C_3\okappa_0\simeq C_2\kt+C_1\tau_0\ell_0^2.$
We expect the transition from the droplet phase
to be strongly first-order. On the other hand
one can view the transition from the lamellar
side to be analogous to a ``crystallization''
of the passage fluid considered, e.g., in
ref.~\cite{Golubovic94}. In this case it should
be weakly first order.

\section*{Conclusions}\setcounter{equation}{0}
Unfortunately I have not been able
to discuss in these notes some fascinating
arguments, like the behavior of membranes with
reduced symmetry (hexatic and cristalline),
or made of several components. But I had presumed
too much on my speed in writing and I could not
abuse of the editors' patience. I hope however
that I have conveyed to you part of my
fascination for this delicate field of Statistical
Physics.

\section*{Acknowledgments}
I am grateful to Fran\c cois David and Paul Ginsparg
for inviting me to give this course and for their patience
in waiting for my contribution to these
Proceedings. I also thank
all those who have helped me during the preparation
and the writing-up of these lectures, and in
particular R.~Cohen, T.~Di~Palma, G.~Esposito,
M.~Falcioni, U.~Marini-Bettolo, T.~Powers.

\appendix
\renewcommand{\theequation}{\Alph{section}.\arabic{equation}}
\section{Differential equations for vesicle shapes}\setcounter{equation}{0}

The Euler-Lagrange equations for
vesicle shapes cannot
be analytically solved in general.
We show in this Appendix how,
limiting oneself to the consideration of
{\em axisymmetric shapes},
one can transform
them into a system of first-order ordinary differential
equations, which can be solved numerically
(\cite{Deuling,Peterson}).

\begin{figure}
\begin{center}\includegraphics[width=6cm]{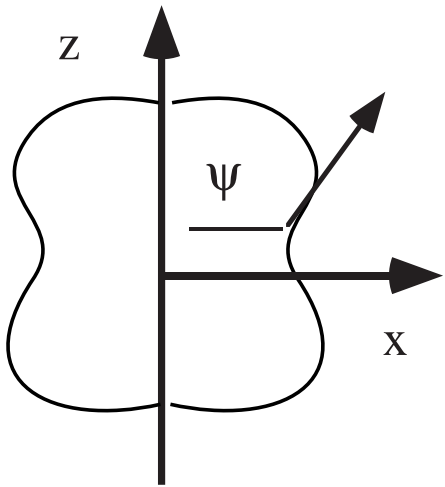}\end{center}
\caption{Axisymmetric vesicle shape: the $z$
axis is the axis of rotational symmetry.}
\label{azim-fig}
\end{figure}

The geometry is defined in fig.~(\ref{azim-fig}).
Because of rotational symmetry, one has
\begin{equation}
H=c_{\rm m}+c_{\rm p},
\end{equation}
where $c_{\rm m,p}$ are the principal curvatures
(the eigenvalues of the curvature matrix $\Omega$),
corresponding respectively to the meridians (m)
and the parallels (p). Denoting by $\psi$ the angle
between the surface normal and the $z$-axis,
we have
\begin{equation}
c_{\rm p}=\frac{\sin\psi(x)}{x},\qquad c_{\rm m}=\cos\psi(x)
\frac{\d\psi}{\d x}.
\end{equation}
The angle $\psi(x)$ is related to the contour $z(x)$ by
\begin{equation}
\frac{\d z}{\d x}=-\tan \psi(x).
\end{equation}
On the other hand, the contour $z(x)$
and the volume $V$ can be expressed
in terms of $x$ and $c_{\rm p}$:
\begin{eqnarray}
z(x)&=&z(0)-\int\d x x c_{\rm p}\left[1-(xc_{\rm p})^2\right]^{-\frac{1}{2}},
\\
\d V&=&\pi x^3 c_{\rm p}\left[1-(xc_{\rm p})^2\right]^{-\frac{1}{2}}\d x,\\
\d S&=&2\pi x\left[1-(xc_{\rm p})^2\right]^{-\frac{1}{2}}\d x.
\end{eqnarray}
It is useful to replace the independent variable
$x$ by the normalized surface area $s$ using the
relation
\begin{equation}
\d x=\pm \frac{2}{x}\left[1-(xc_{\rm p})^2\right]^{\frac{1}{2}}.
\end{equation}
The variable $s$ ranges from $s=0$ at the
upper pole to a value $s_{\rm m}$ at the
dividing parallel of latitude, and then
to $s=1$ at the lower pole.
One has
\begin{equation}
\frac{\d x}{\d s}\cases{\ge 0,&for $0\le s\le s_{\rm m}$;\cr
\le 0,&for $s_{\rm m}\le s\le 1.$}
\end{equation}
Introducing the
notation $f=x^2$ the equations
become:
\begin{eqnarray}
\frac{\d c_{\rm m}}{\d s}&=&\pm\left(1-fc^2_{\rm p}\right)^{-\frac{1}{2}}
\times \left\{c_{\rm p}\left[\left(c_{\rm p}-c_0\right)^2-c^2_{\rm m}
\right]+\right.\nonumber\\
&&+\left.\frac{2\gamma}{\kappa}c_{\rm p}
+\frac{p}{\kappa}-2\left(1-fc_{\rm p}^2\right)
\frac{c_{\rm m}-c_{\rm p}}{f}\right\};\\\label{eq:cm}
\frac{\d c_{\rm p}}{\d s}&=&\pm 2\left(1-fc_{\rm p}^2\right)^{\frac{1}{2}}
\frac{c_{\rm m}-c_{\rm p}}{f},\\
\frac{\d f}{\d s}&=&\pm 4\left(1-fc_{\rm p}^2\right)^{\frac{1}{2}}.
\end{eqnarray}

These equation have singularities for either $f(s)\to 0$ or
$f(s)c_{\rm p}^2\to 1$. At the poles of the vesicles we have
$f=0$, but at the same time the difference $c_{\rm m}-c_{\rm p}$
tends to zero. This yields the correct boundary condition
at the poles:
\begin{equation}
\left.\frac{\d c_{\rm p}}{\d s}\right|_{\rm poles}=\frac{1}{3}
\left.\frac{\d c_{\rm m}}{\d s}\right|_{\rm poles}.
\end{equation}
The singularity for $f(s)c^2_{\rm p}\to 1$ is approached
when $f(s)$ reaches an extremum, and should be removed
by imposing that the factor in curly brackets in eq.~(\ref{eq:cm})
vanishes. This allows one to join the $\pm$ branches
of the equations by expanding in a Taylor series around
the value $s=s_m$ where $f(s)$ reaches the extremum.

One can then obtain the three curves $c_{\rm m}(s),$
$c_{\rm p}(s),$ $f(s),$ and the matching at those
intermediate points $s_{\rm m}$ where $f(s)c^2_{\rm p}\to 1$
gives three nonlinear equations for the unknown
boundary conditions $c_{\rm m}(s{=}0),$ $f(s{=}s_{\rm m}),$
and $c_{\rm m}(s{=}s_{\rm m}).$ A further integration
yields the actual vesicle shape $z(x).$

\section{The Faddeev-Popov determinant}\setcounter{equation}{0}
Let us consider the integral
\begin{equation}
{\cal I}=\int\d^2\vec{x}\,f(\vec{x}),
\end{equation}
where the function $f(x,y)$ is invariant upon rotations
in the $(x,y)$ plane:
\begin{equation}
\vec{x}\to\vec{x}_{\theta}=\Omega(\theta)\cdot \vec{x},
\end{equation}
where $\Omega(\theta)$ is a rotation matrix:
\begin{equation}
\Omega=\left(\begin{array}{cc}\cos\theta&-\sin\theta\\
\sin\theta&\cos\theta\end{array}\right).\end{equation}
It is then convenient to integrate the function over
one line, whose images upon rotation sweep the whole
two-dimensional plane. This line can be represented by
an equation in $\vec{x}$:
\begin{equation}
\psi(\vec{x})=0.
\end{equation}
Given any vector $\vec{x}$ there will be a value $\theta^{*}$
such that $\psi(\vec{x}_{\theta^{*}})=0.$
Using the properties of the delta function, we obtain
\begin{equation}
\int_0^{2\pi}\d\theta\delta\left(\psi(\vec{x}_{\theta})\right)
=\left|\frac{\partial \psi(\vec{x}_{\theta})}{\partial \theta}
\right|_{\theta^{*}}^{-1}.\end{equation}
Therefore
\begin{eqnarray}
{\cal I}&=&\int_0^{2\pi}\d\theta\int\d^2\vec{x}\,
\delta\left(\psi(\vec{r}_{\theta})\right)
\left|\frac{\partial \psi(\vec{x}_{\theta})}{\partial\theta}\right|
f(\vec{r})\nonumber\\
&=&2\pi\int\d^2\vec{x}\,
\delta\left(\psi(\vec{r})\right)
\left|\frac{\partial \psi(\vec{x}_{\theta})}{\partial\theta}
\right|_{\theta=0}f(\vec{r}).
\end{eqnarray}
The advantage of this expression is that all the quantities
need only to be evaluated on (or near) the line $\psi(\vec{x})=0.$

The Faddeev-Popov determinant is the generalization of
the factor introduced above to correct the effect of the
integration over $\theta.$ We consider a general integral
\begin{equation}
{\cal I}=\int\d\mu(x)\,f(x),
\end{equation}
where the measure $\d\mu(x)$ and the function $f(x)$
are both invariant with respect to a $N$-dimensional group $G,$
whose elements $g$ can be
represented via its $N$ generators $t_{\alpha}$:
\begin{equation}
g=\exp\left[\sum_{\alpha}\theta^{\alpha}t_{\alpha}\right].\end{equation}
We ``fix the gauge'', introducing $N$ constraints
$\psi_{\alpha}(x)=0,$ $i=1,\ldots,N.$ We then have
\begin{equation}
{\cal I}=\int\d\mu(x)f(x)\,\prod_{\alpha=1}^N\delta\left(\psi_{\alpha}
(x)\right)\,|\Det(J_{\alpha\beta}(x))|\,{\rm vol}\,G,\end{equation}
where we have introduced the volume of the group $G,$ and
\begin{equation}
J_{\alpha\beta}(x)=\left.\frac{\partial \psi_{\alpha}(x)}{\partial
\theta^{\beta}}\right|_{\theta=0}.
\end{equation}

In our case, the relevant group is generated by the infinitesimal
diffeomorphisms of the form
\begin{equation}
\usigma\to\usigma+\underline{\epsilon}(\usigma).
\end{equation}
The corresponding transformations on the $\vec{r}$ read
\begin{equation}
\vec{r}\to\vec{r}+\epsilon^i\partial_i \vec{r}.
\end{equation}
Consider now a deformation $\vec{h}$ with respect to
a background configuration $\vec{r}_0$:
\begin{equation}
\vec{r}(\usigma)=\vec{r}_0(\usigma)+\vec{h}(\usigma).
\end{equation}
The {\em normal gauge\/} is realized by imposing the constraints
\begin{equation}
\psi_i(\usigma)=\vec{h}(\usigma)\cdot\partial_i\vec{r}_0(\usigma)=0,
\qquad i=1,2.\end{equation}
Let us compute how the constraints are affected by an infinitesimal
diffeomorphism:
\begin{eqnarray}
\delta \psi_i(\usigma)&=&\delta\left(\vec{r}-\vec{r}_0\right)
\cdot\partial_i\vec{r}_0=\delta\vec{h}\cdot\partial_i\vec{r}_0\nonumber\\
&=&\epsilon^j\partial_j\vec{r}\cdot\partial_i\vec{r}_0\nonumber\\
&=&\epsilon^j\left(\partial_j\vec{r}_0+\partial_j\vec{h}\right)
\cdot\partial_i\vec{r}_0\nonumber\\
&=&\epsilon^j\left(g_{ij}+\partial_j\vec{h}
\cdot\partial^i\vec{r}_0\right).
\end{eqnarray}
We see that this expression is purely local.
Therefore, in this gauge, the Faddeev-Popov determinant reads
\begin{equation}
\Delta_{\rm FP}=\prod_{\usigma}\Det\left[g_{ij}(\usigma)+
\partial_i\vec{h}(\usigma)\cdot\partial_j\vec{r}_0(\usigma)\right].
\end{equation}
In the particular case of the Monge representation,
$\partial_i\vec{h}(\usigma)\cdot\partial_j\vec{r}_0(\usigma)=0,$
so that the Faddeev-Popov determinant is trivial.

\section{One-loop calculation of the renormalization group\setcounter{equation}{0}
flow}
In this Appendix I report the calculation of the renormalization group
flow to one loop, following essentially \cite{tension,David89,Guitter89}.
We use the method of the background field. We take the reference
configuration to be parametrically described
by $\vec{r}_0(\usigma)$ and we sum over all configurations
$\vec{r}(\usigma)$ such that: (i) they are sufficiently
``close'' to the reference one (so that the deformation
$\vec{h}(\usigma)$ is small); (ii) the wavenumbers contained
in the Fourer transform of $\vec{h}(\usigma)$ belong to the
infinitesimal shell $\Lambda/s<|\underline{q}|<\Lambda.$
We write the effective Wilson hamiltonian ${\cal H}_{\rm eff}(s)$
as follows:
\begin{equation}
{\cal H}_{\rm eff}(s)=-\kt\log\int_s{\cal D}\vec{h}
\exp\left\{-\frac{F[\vec{r}_0+\vec{h}]}{\kt}\right\}
+\int\d A\,\vec{\lambda}\cdot\vec{r}_0,\label{Legendre}
\end{equation}
where $\int_s$ is a reminder on the constraint on the fluctuation
wavenumber, and $\vec{\lambda}$ is the external field needed to set
$\left<\vec{r}\right>_{\lambda}=0.$
We can expand $F[\vec{r}_0+\vec{h}]$ in powers of $\vec{h}$
up to second order. The contribution of the first
order will cancel with the second term of the rhs of
eq.~(\ref{Legendre}). We obtain
\begin{eqnarray}
{\cal H}_{\rm eff}(s)&=&F[\vec{r}_0]-\log\int_s{\cal D}\vec{h}
\exp\left\{-\frac{\delta^2 F[\vec{r}_0+\vec{h}]}{\kt}\right\}\nonumber\\
&=&F[\vec{r}_0]+\frac{\kt}{2}\Tr_s\log\left\{
\frac{\delta^2F}{\delta h\delta h}\right\},\label{Trlog}
\end{eqnarray}
where $\delta^2F$ is the second variation of the Helfrich hamiltonian
with respect to $\vec{h}.$

We now introduce the metric tensor $g_{ij}$ of the reference configuration
$\vec{r}_0$:
\begin{equation}
g_{ij}(\usigma)=\partial_i\vec{r}_0(\usigma)\cdot\partial_j\vec{r}_0
(\usigma).\end{equation}
On the basis of $g_{ij}$ we can define the covariant derivative
$D_i.$ We can then express the curvature tensor $\vec{\Omega}_{ij}$ as
follows:
\begin{equation}
\vec{\Omega}_{ij}=D_iD_j\vec{r}_0.
\end{equation}
One can check that $\Omega_{ij}$ is normal to the surface for
any $i,$ $j.$ It can therefore represented
by $\vec{\Omega}_{ij}=\Omega_{ij}\vec{n},$
where $\vec{n}$ is the normal unit vector.
The Gaussian curvature $K$ and the square of
the mean curvature $H^2$ are then given by
\begin{eqnarray}
H^2&=&\Omega_i{}^i\Omega_j{}^j;\\
2K&=&\Omega_i{}^i\Omega_j{}^j-\Omega_i{}^j\Omega^i{}_j.
\end{eqnarray}
We use the normal gauge, and represent the deformation $\vec{h}$
by the scalar $\nu,$ defined by $\vec{h}=\nu\vec{n}.$
We then have:
\begin{eqnarray}
\delta^2F&=&\int\d A\,\nu\left\{\kappa\left[\Delta^2+\frac{1}{2}
(H^2-4K)+2 H\Omega^{ij}D_iD_j+\ldots\right]\right.\nonumber\\
&&\left.+\tau\left[-\Delta+2K\right]\right\}\nu,
\end{eqnarray}
where we have introduced the Laplacian (with respect to the
metric tensor $g_{ij}$):
\begin{equation}
\Delta=D_iD^i.
\end{equation}
The gaussian curvature
term drops out of the variation.
We have neglected terms which are irrelevant in our
approximations, being of higher order in the curvatures.
When integrating upon the fluctuations, the $\Tr_s\log$
term in eq.~(\ref{Trlog}) can be replaced by
$\int\d\usigma\int_s\d^2\underline{q}(2\pi)^{-2},$
and the operator $\Delta$ by $q^2.$
The result reads
\begin{eqnarray}
{\cal H}_{\rm eff}(s)&=&F+\frac{\kt}{2}\int\d A
\frac{\Lambda^2\epsilon}{2\pi}\nonumber\\
&&\times\log\left\{
\frac{1}{\kt\Lambda^4}\left[\kappa\Lambda^4+\tau
\Lambda^2-\frac{3}{2}\kappa H^2\Lambda^2+\frac{10}{3}\kappa K
\right]\right\}.\end{eqnarray}
We can read off this formula the effective parameters:
\begin{eqnarray}
\tau_{\rm eff}(s)&=&\tau+\epsilon\Lambda^2
\frac{\kt}{4\pi}\log\left(\frac{\kappa}{\kt}+\frac{\tau}{\kt \Lambda^2}
\right),\\
\kappa_{\rm eff}(s)&=&\kappa-\epsilon
\frac{3\kt}{4\pi}\frac{1}{1+(\tau/\Lambda^2)},\\
\overline{\kappa}_{\rm eff}(s)&=&\overline{\kappa}
+\epsilon\frac{5\kt}{3\pi}\frac{1}{1+(\tau/\Lambda^2)}.
\end{eqnarray}
However, this formula does not consider the
contributions of the measure terms \cite{Cai}.
Their effect is a shift in $\tau_{\rm eff}$:
\begin{equation}
\tau_{\rm eff}\to\tau_{\rm eff}'=\tau_{\rm eff}-\epsilon
\frac{\Lambda^2}{4\pi}\kt\log\left(\frac{\kappa}{\kt}\right).
\end{equation}
Taking into account this contribution and
rescaling back the lengths we obtain the renormalized
couplings, which obey the renormalization group equations
\begin{eqnarray}
s\frac{\partial \tau}{\partial s}&=&2\tau+
\frac{\Lambda^2}{4\pi}\kt\log\left(1+
\frac{\tau}{\kappa\Lambda^2}\right),\\
s\frac{\partial \kappa}{\partial s}&=&-\frac{3\kt}{4\pi}
\frac{1}{1+(\tau/\Lambda^2)},\\
s\frac{\partial \overline{\kappa}}{\partial s}&=&
\frac{5\kt}{3\pi}\frac{1}{1+(\tau/\Lambda^2)}.
\end{eqnarray}

\section{The Liouville model}\setcounter{equation}{0}
The Liouville model is defined by the hamiltonian
\begin{equation}
{\cal H}[\vec{r},g_{ij}]=\gamma_0\int\d^2\usigma\sqrt{g}+
\lambda\int\d^2\usigma\sqrt{g}\,g^{ij}\partial_i\vec{r}\cdot
\partial_j\vec{r}.\label{Liouvmod}
\end{equation}
The metric tensor $g_{ij}$
fluctuates in this model, and is not induced by the
membrane configuration described by $\vec{r}.$

The Liouville model is
best studied in the {\em conformal gauge\/} where the
internal metric tensor $g_{ij}$ takes the form
\begin{equation}
g_{ij}=\e^{\phi}\delta_{ij}.
\end{equation}
It is always possible to construct locally
such a coordinate system (see, e.g., \cite[Sec.~2.8]{David89}).
In this system, the gaussian curvature reads
\begin{equation}
K(\usigma)=-\frac{1}{2}\e^{-\phi(\usigma)}
\left(\frac{\partial^2}{\partial\sigma_1^2}+
\frac{\partial^2}{\partial\sigma_2^2}\right)\phi(\usigma).
\end{equation}

If we fix $\phi,$ we can integrate over the Gaussian field
$\vec{r},$ obtaining an effective hamiltonian for the
metric field $\phi.$
This integration can be performed via the
{\em conformal anomaly\/}~\cite{Polyakov81,David89}.

Consider a scalar field $\psi(\usigma),$ defined on a closed
surface $S,$ and associate
to it a hamiltonian of the form
\begin{equation}
{\cal H}_0[\psi]=\frac{1}{2}\int_S\d^2\usigma\,\sqrt{g}
g^{ij}\partial_i\psi\partial_j\psi.
\end{equation}
Integrating by parts we obtain
\begin{equation}
{\cal H}_0[\psi]=\frac{1}{2}\int_S\d^2\usigma\,\sqrt{g}
\psi(-\Delta)\psi,
\end{equation}
where $\Delta$ is the scalar Laplacian in the metric $g_{ij}$:
$\Delta=g^{ij}D_iD_j.$
$D_i$ is the covariant derivative with
respect to $\sigma^i.$
In the conformal gauge, the Laplacian is simply given by
\begin{equation}
\Delta=\e^{-\phi(\usigma)}\left(\frac{\partial^2}{\partial\sigma_1^2}+
\frac{\partial^2}{\partial\sigma_2^2}\right).
\end{equation}

Integrating over the Gaussian field $\psi$ yields formally
\begin{equation}
Z=\int{\cal D}\psi\,\e^{-{\cal H}_0}[\psi]
\propto\left[\Det(-\Delta)\right]^{-\frac{1}{2}}.
\end{equation}
One can express it formally in terms of the eigenvalues
$\{-\lambda_k\}$ of the Laplacian:
\begin{equation}
\Det(-\Delta)=\prod_k\lambda_k=\exp\left(\sum_k\log\lambda_k\right).
\end{equation}
However, this expression is ill-defined. On the one hand, the
constant function belongs to the eigenvalue zero (yielding
an infinite contribution). On the other hand,
the sum diverges at large $k.$ One has to introduce an
ultraviolet (large wavenumber) regulator $\Lambda\simeq
\pi/a_0,$ to suppress this divergence. This regulator should
be introduced in a reparametrization-invariant way,
which is obtained via the {\em heat-kernel\/} regularization.

The ``heat kernel'' $G(\usigma,\usigma';t)$ is the solution
of the differential equation
\begin{equation}
\frac{\partial}{\partial t}G(\usigma,\usigma';t)=\Delta_{\sigma}G(\usigma,
\usigma';t),
\end{equation}
satisfying the initial condition
\begin{equation}
G(\usigma,\usigma';t{=}0)=\frac{1}{\sqrt{g(\usigma})}
\delta(\usigma-\usigma').\label{kernel:eq}
\end{equation}
The extra factor on the rhs is introduced to maintain
reparametrization invariance. At short times, $G(\usigma,
\usigma';t)$ looks locally like a Gaussian, with a width
proportional to $\sqrt{t},$ like a diffusion kernel.
However, this diffusion takes place in a curved space.
It will therefore be slower in a space of positive curvature and
faster in a space of negative curvature. The asymptotic expansion
of $G(\usigma,\usigma;t)$ for $t\to 0^+$
is indeed given by~\cite{Greiner}
\begin{equation}
G(\usigma,\usigma;t)\sim\frac{1}{4\pi}\left[\frac{1}{t}+
\frac{K(\usigma)}{3}+O(t)\right].
\end{equation}

We can now define the regularized determinant of the Laplacian
as follows:
\begin{eqnarray}
\log\Det'(-\Delta_{\epsilon})&=&\Tr'\log(-\Delta_{\epsilon})
=-\Tr'\left[\int_{\epsilon}^{\infty}
\frac{\d t}{t}\e^{t\Delta}\right]\nonumber\\
&=&-\int_{\epsilon}^{\infty}\frac{\d t}{t}\Tr'\left(\e^{t\Delta}\right).
\end{eqnarray}
The parameter $\epsilon\sim\Lambda^2$ acts as a cutoff. The
prime indicates that the zero eigenvalue
has been suppressed. By using the expression (\ref{kernel:eq})
we obtain
\begin{equation}
\Tr'\log(-\Delta_{\epsilon})\simeq
-\frac{1}{4\pi\epsilon}\int_S\d^2\usigma\sqrt{g}
-\frac{\log\epsilon}{12\pi}\int_S\d^2\sqrt{g}K
+{\rm finite\ terms}.\label{determ:eq}
\end{equation}
Note that the coefficient of $1/\epsilon$ depends
on the cutoff procedure used, while the coefficient of
$\log\epsilon$ does not.

The finite contribution can be computed in the following
way. Consider a local conformal rescaling
\begin{equation}
g_{ij}(\usigma)\to g_{ij}(\usigma)\e^{\varphi(\usigma)}=g'_{ij}.
\end{equation}
Under such a transformation, the Laplacian changes as
\begin{equation}
\Delta(g)\to\Delta(g')=\e^{-\varphi(\usigma)}\Delta(g).
\end{equation}
Therefore
\begin{eqnarray}
\frac{\delta}{\delta\varphi(\usigma)}
\Tr'\log(-\Delta_{\epsilon})&=&
-\int_{\epsilon}^{\infty}\frac{\d t}{t}\Tr\left[t
\frac{\delta\Delta}{\delta\varphi}\e^{t\Delta}\right]\nonumber\\
&=&\int_{\epsilon}^{\infty}\frac{\d t}{t}\sqrt{g(\usigma)}
[t\Delta\e^{t\Delta}]_{\usigma\usigma}\nonumber\\
&=&\int_{\epsilon}^{\infty}\d t\frac{\d}{\d t}
[\e^{t\Delta}]_{\usigma\usigma}\sqrt{g(\usigma)}\nonumber\\
&=&-[\e^{\epsilon\Delta}]_{\usigma\usigma}\sqrt{g(\usigma)}
=-\sqrt{g(\usigma)}G(\usigma,\usigma;\epsilon)\nonumber\\
&=-&\sqrt{g(\usigma)}\left[\frac{1}{4\pi\epsilon}+
\frac{K(\usigma)}{12\pi}+O(\epsilon)\right].
\end{eqnarray}
This is the conformal anomaly~\cite{Polyakov81}. This result
can also be written in the remarkable form:
\begin{eqnarray}
\frac{\delta}{\delta\varphi(\usigma)}
\Tr'\log(-\Delta_{\epsilon})&=&
\frac{\delta}{\delta\varphi(\usigma)}\left[\frac{1}{4\pi\epsilon}
\int_S\d^2\usigma\sqrt{g(\usigma)}
-\frac{1}{12\pi}\int_S\d^2\usigma\int_S\d^2\usigma'\right.\nonumber\\
&&\left.
\sqrt{g(\usigma)}K(\usigma)G(\usigma,\usigma')
\sqrt{g(\usigma')}K(\usigma')\right],
\end{eqnarray}
where $G(\usigma,\usigma')$ is the inverse of the Laplacian
in the metric $g_{ij}.$
Therefore one can integrate out the variation in $\varphi(\usigma),$
starting, e.g., from a constant metric $g^0_{ij},$
conformally equivalent to $g_{ij},$
and obtain
\begin{eqnarray}
\Tr\log(-\Delta_{\epsilon}(g))&=&\frac{1}{4\pi\epsilon}
\int_S\d^2\usigma\sqrt{g(\usigma)}\nonumber\\
&&-\frac{1}{12\pi}\int_S\d^2\usigma\int_S\d^2\usigma'
\sqrt{g(\usigma)}K(\usigma)G(\usigma,\usigma')
\sqrt{g(\usigma')}K(\usigma')\nonumber\\
&&+{\cal F}[g^0]+O(\epsilon),
\label{Liouville}
\end{eqnarray}
where ${\cal F}[g^0]$ depends only on the conformal class of the
metric $g,$ and contains the logarithmically divergent
part obtained in~(\ref{determ:eq}).
In the conformal gauge, this expression becomes simply
\begin{equation}
{\cal H}_{\rm Liouville}=-\frac{1}{48\pi}\int_S\d^2
\usigma\left[\frac{1}{2}(\partial\phi)^2-\frac{12}{\epsilon}\e^{\phi}
\right].\end{equation}

The finite part in eq.~(\ref{Liouville}) is called the {\em
Liouville action\/}~\cite{Polyakov81}. The Liouville model
(\ref{Liouvmod}) involves $d$ scalar fields, corresponding
to the components of the field $\vec{r}.$ On the other hand, the
Faddeev-Popov determinant (App.~B) is nontrivial in the conformal gauge.
Its contribution turns out to be also proportional to the
Liouville action, but with a different coefficient.
The effective hamiltonian for $\phi$ reads
\begin{equation}
{\cal H}_{\rm eff}=\frac{26-d}{48\pi}\int_S\d^2\usigma
\left[\frac{1}{2}(\partial\phi)^2+\mu^2\e^{\phi}\right],
\end{equation}
where $\mu^2$ is a ``mass'' which represents the effect of the
surface tension.

\end{document}